\documentclass[12pt]{iopart}

\usepackage{times}

\usepackage{graphicx}

\begin{document}
\pagestyle{myheadings}

% by PG MESSO A POSTO !!!
\def\lesssim{\mathrel{\hbox{\rlap{\hbox{\lower4pt\hbox{$\sim$}}}\hbox{$<$}}}}
\def\gtrsim{\mathrel{\hbox{\rlap{\hbox{\lower4pt\hbox{$\sim$}}}\hbox{$>$}}}}
\newcommand{\ffrac}[2]{\left(\frac{#1}{#2} \right)}

\title{Ultra High Energy Particle Astronomy, \\ Neutrino Masses  and Tau Airshowers }

\author{D. Fargion$^{1}$$^{,2}$, M.Khlopov $^{1}$$^{,3}$, R.Konoplich $^{4}$, P.G. De Sanctis Lucentini,\\ M. De Santis, B. Mele$^{2}$}

%$^{***}$Physics Department, Universita' degli studi "La
%Sapienza",5, Piazzale Aldo Moro - I 00185 Roma, Italy.}
\address{$^{1}$Physics Department, Rome University "La Sapienza",Ple.A.Moro 2,00185, Rome, Italy}
\address{$^{2}$INFN, Sezione di Roma, Rome, Italy}
\address{$^{3}$Center for Cosmo-particle Physics "Cosmion"
and M.V.Keldysh Institute of Applied Mathematics RAS, 125047,
Moscow, Russia and Moscow Engineering Physics Institute Technical
University, Moscow, Russia.}
\address{$^{4}$ Department of Physics, New York University, New York, NY $10003$
and Manhattan College, Riverdale, New York, NY $10471$}

%Dept. of Physics, Manhattan College,Riverdale,New York,NY $10471$}
\begin{abstract}
Ultra High Energy (UHE) neutrino Astronomy and relic Neutrino
masses may play a key role in  solving the Ultra High Energy
Cosmic Ray (UHECR) puzzle within the Z-Showering model.  Tau
air-shower originated by UHE neutrino in matter may probe and
amplify this expected UHE neutrino Astronomy. The discover of
upcoming and horizontal $\tau$ air-showers (UPTAUs, HORTAUs),
born by  UHE, $\nu_{\tau}$ interacting inside mount chains or
along  Earth Crust crown masses at the Horizons edges, will
greatly amplify the single UHE $\nu_{\tau}$ track.   Observing
UPTAUs and HORTAUs along their line of showering, from higher
balloons or satellites, at Greisen-Zatsepin-Kuzmin (GZK) energies
($\geq 10^{19}$ eV), may test a calorimeter mass exceeding $150$
$km^3$ (water equivalent). Observing Horizontal Shower by EUSO
will probe even larger ring areas generated in huge terrestrial
volumes larger than $\simeq 2360$ $km^3$. These highest GZK
energies $\nu_{\tau}$ astronomy are well tuned to test the needed
abundant $\nu$ fluence in GZK Z-Showering model where ZeV ($\geq
10^{21}$ eV) Ultra High Energy neutrinos hit on relic light
($\sim 0.1-3$ eV masses) anti-neutrinos, in hot dark matter
halos, creating boosted UHE Z bosons. Their nucleon decay in
flight may born secondaries observed on Earth atmosphere as
UHECR. The wide EUSO acceptance may  test easily Z-showering
fluxes as well as  GZK neutrinos at their minimal granted
fluence. The rare ($78$) observed upcoming terrestrial gamma
flashes observed by BATSE satellite during last decade may be
also be traces of such UPTAUs and HORTAUs.
%%%%%%%%%%%%%%%%%%%%%%%%%%%%%%%%%%%%%%%%%%%%%%%%%%%%%%%%%%%%%%%%%%%%%%
%High Energy Neutrino signals from PeV up to  GZK, $E_{\nu}\geq
%10^{19}$ eV, energies  amplified as Upward and Horizontal Tau air
%Showers,\textbf{UPTAUS}, and \textbf{HORTAUS}, may flash toward
%mountain peaks, airplanes, balloons  and satellites in future Tau
%$air-showering experiments

%Observing from  high mountains the
%detectable crown masses at UHE $\nu$ energies around EeV  are
%comparable to a few $km^3$, while observing from higher balloons
%or satellites at orbit altitudes, at GZK  energies  , the
%Horizontal Crown effective Masses  may even exceed $150$ $km^3$.
%The consequent event rate by expected GZK or Z-Showering
%neutrinos models must arise at large rate by present scheduled
%(EUSO-like) or future (Crown Array Detector) satellite in Space.
%%%%%%%%%%%%%%%%%%%%%%%%%%%%%%%%%%%%%%%%%%%%%%%%%%%%%%%%%%%%%%%%%%%%%%%%

\end{abstract}
\newpage
% Leave the next line commented out!

% \maketitle
\tableofcontents

\newpage

\section{Introduction: The rise of  Ultra High Energy (UHE) Particle   astronomy}

 Astronomy is based on the
direct, undeflected radiation from astronomical objects to the
observer. The neutral messengers which we did use in last three
centuries were mainly photons in the optical range. During last
century the whole electromagnetic spectrum  opened to Astronomy.
%%%%%%%%%%%%%%%%%%%%%%%%%%%%%%%%%%%%%%%%%%%%%%%%%%%%%%%%%%%%%%%%%%
Moreover in the last $30$ years also neutrino astronomy arose and
proved solar and supernova neutrino astronomy. Ultra High Energy
Cosmic Ray (UHECR)  may offer a New Particle Astronomy because,
even charged, they are almost un-bent by galactic and
extra-galactic magnetic fields. This UHECR astronomy is bounded
(by primordial photon drag, the well known Greisen, Zatsepin,
Kuzmin (GZK) cut-off
\cite{Greisen:1966jv},\cite{Zatsepin:1966jv}. ) in a very narrow
(almost local) Universe (few tens Mpc). However we did't find yet
in present UHECR arrival maps  any nearby known galactic
structures or clear super-galactic imprint. Observed UHECR
isotropy call for a cosmic link,  well above the narrow GZK
radius. Observed UHECR  clustering in groups suggest compact
 sources (as AGN or BL Lacs beaming Jets ) respect to any homogeneous and isotropic
halo of primordial topological defects.  Recent evidence for a
dozen or more BL Lacs sources correlation (some at medium
redshift $z \simeq 0.3 \gg z_{GZK}$) with  UHECR clustered
events are giving support (with the isotropy) to a cosmic
 origin for UHECR sources \cite{Kalashev:2002kx}. Incidentally UHECR must produce by pion decays GZK
neutrino secondaries at a fluence at least comparable with
observed UHECR. Therefore there must be a minimal  underlying
 neutrino astronomy at observed GZK cosmic ray level. In order to solve the GZK
puzzles  it may be necessary to consider the traces of UHE
neutrinos either primary (at much higher fluxes)  of UHECR by
Z-Showering or at least as secondaries. Indeed  UHE Relic
neutrinos with a light mass may play a role as calorimeter
 for UHE neutrinos from cosmic distances at ZeV energies. Their scattering may
solve the GZK paradox, \cite{Fargion Salis 1997}, \cite{Fargion
Mele Salis 1999}, \cite{Weiler 1999}, \cite{Yoshida  et all
1998}\cite{Fargion et all. 2001b}, \cite{Fodor Katz Ringwald
2002}; in this scenario UHE $\nu$ Astronomy is not just a
consequence but itself the cause of UHECR signals.

To test this idea we need to open an independent UHE neutrino
Astronomy. Different muon track detectors, cubic $km^3$ in
underground may
 be reveal them at PeVs energies. Nevertheless UHE neutrino tau may interact on
Mountains or better on Earth crust, at a huge concrete
calorimeter, leading to Tau Air-Showers, probing easily both
(secondary and/or primary)  UHE neutrino astronomy  at GZK
energies. Present article review  the problems and the
 experimental prospects on Mountains, planes detectors quota;
  we also reconsidered events measured by satellites,
 like  past BATSE, present gamma satellite INTEGRAL,
  future EUSO  experiments. The recent
signals  in BATSE terrestrial gamma flashes maybe indeed the
first evidence
 for these new UHE neutrino Astronomy upcoming from Earth crust at PeV-EeV energies, leading to UPTAUs and HORTAUs showers.
  In last figure we
 summirized the two consequent Flux signal derived by the TGF event rate data in
BATSE (1991-2000) experiment, normalized for the estimated BATSE
thresholds. These result may be a useful reference estimate for
PeV-EeV neutrino fluxes and apparently are well consistent with
Z-Showering model for a relic mass within the expected values ($m
\simeq 0.04-0.4 eV$).

%%%%%%%%%%%%%%%%%%%%%%%%%%%%%%%%%%%%%%%%%%%%%%%%%%%%%%%%%%%%%%%%%%
 Let us remind that the restrictions on any
astronomy are related to a the messenger interactions  with the
surrounding  medium on the way to the observer. While Cosmic Rays
astronomy is severely blurred by random terrestrial, solar,
galactic and extragalactic  magnetic lenses, the highest $\gamma$
ray astronomy (above tens TeV) became more or less  blind because
of photon-photon opacity (due to electron pair production) at
different energy windows. Indeed the Infrared- TeV opacity as
well as a more severe Black Body Radiation, BBR,( at $2.75
K$)-PeV cut-off are bounding the TeV -PeV $\gamma$ ray astronomy
in  very nearby cosmic ( or even galactic) volumes. Therefore
rarest TeV gamma signals are at present the most extreme  trace
of High Energy Astronomy. We observe copious cosmic rays at
higher ($\gg 10^{15}$eV) energies almost isotropically spread by
galactic  and cosmic magnetic fields in the sky.

 Let us remind,
among  the $\gamma$ TeV discoveries, the signals of power-full
Jets blazing to us from Galactic (Micro-Quasars) or extragalactic
edges (BL Lacs). At PeV energies astrophysical $\gamma$ cosmic
rays should also be presented, but, excluding a very rare and
elusive Cyg$X3$ event, they have not being up date observed; only
upper bounds are known at PeV energies.  The missing $\gamma$ PeV
astronomy, as we mentioned, are very probably absorbed because of
their own photon interactions (electron pairs creation) at the
source environment and/or along the photon propagation into the
cosmic Black Body Radiation (BBR) and/or into other diffused
background radiation. Unfortunately PeV charged cosmic rays,
easily bend and bounded in a random walk by Galactic magnetic
fields, loose their original directionality and their
astronomical relevance; their tangled trajectory resident time in
the Galaxy is much longer ($\geq 10^{3}$ - $10^{5}$) than any
linear neutral trajectory, as in the case of gamma rays, making
the charged cosmic rays  more probable to be observed by nearly a
comparable length ratio. However astrophysical UHE neutrino
signals in the wide range $10^{13}$eV-$10^{19}$eV (or even higher
GZK energies) are unaffected by any radiation cosmic opacity and
may easily open a very new exciting window toward Highest Energy
sources. Being weakly interacting the neutrinos are an ideal
microscope to deeply observe  in their accelerator (Jet,SN,GRB,
Mini Black Hole) cores they do not experience any strong self
opacity as the case of photon. Other astrophysical $\nu$ sources
at lower energies ($10^{8}$ eV - $10^{12}$ eV) should also be
present, at least at EGRET fluence level, but their signals are
very weak and probably drowned by the dominant diffused
atmospheric $\nu$, secondaries of muon secondaries, produced as
pion decays by the same charged (and smeared) UHE cosmic rays
(while hitting terrestrial atmosphere): the so called diffused
atmospheric neutrinos. Indeed a modulation of the atmospheric
neutrinos signal has inferred the first conclusive evidence for a
neutrino mass and for a neutrino flavor mixing.  At lowest (MeVs)
$\nu$ energy windows, the abundant and steady solar neutrino
flux, (as well as the prompt, but rare, neutrino burst from a
nearby Super-Novae (SN 1987A)), has been, in last twenty years,
successfully explored, giving support to neutrino flavour mixing
and to the neutrino mass reality. More recent additional probes
of solar neutrino flavour mixing and reactor neutrino
disappearance are giving a robust ground to the neutrino mass
existence, at least at minimal $\sim 0.05$ eV level. Let us
mention that Stellar evolution, Supernova explosion but in
particular Early Universe had over-produced and kept in thermal
equilibrium neutrinos whose relic presence here today pollute the
cosmic spaces either smoothly (lightest $m_ {\nu} \ll 0.001 eV$
relativistic $\nu$) or in denser and even possibly clustered Hot
Halos ($\simeq eVs$ relic $\nu$ masses). A minimal tiny
atmospheric neutrino (above $0.05$ eV) $\nu$ mass, beyond the
Standard Model, are already making their cosmic energy density
component almost two order of magnitude larger than the
corresponding $2.75 K$ (Black Body Radiation) BBR radiation
density.   Here we discuss their role in Z-Shower model and we
concentrate on the possibility to detect such a component of the
associated  UHE neutrino flux astronomy (above PeV-EeV up to GZK
energies) by UHE $\nu_{\tau}$ interactions in Mountain chains or
in Earth Crust leading to Horizontal or Upward Tau air-showers
\cite{Fargion 2000-2002}, \cite{Fargion 2002b},\cite{Feng et al
2002} \cite{Fargion 2002c}, \cite{Bertou et all 2002}. The
article will discuss in next Chapter $2$ The Z-Shower scenario
and , in next chapter $3$ the Z-Shower tail spectra, while in
chapter $4$ we discuss the idea that UHE $\tau$ neutrino may
amplify its signal by peculiar $\tau$ air-showers, either upward
and horizontal. In next Chapter $5$ we discuss the possible
connection between observed upward Gamma Flashes and such
expected Upward Tau Air-Showers (UPTAUs). In Chapter $6$ it has
been evaluated the terrestrial skin crown volumes, surrounding
each observer at high quota, where UHE neutrino $\tau$ may hit
and give life to Earth-Skimming Tau nearly Horizontal, whose
decay in flight may be source of Horizontal Tau air-showers
(HORTAUs).In Chapter $7$ we consider the HORTAUs signals
observable from next generation of UHECR detectors as EUSO and or
OWL experiments. In the Conclusion we summirized the tau
astronomy and the resulting  Terrestrial Gamma Flash
interpretation as neutrino UPTAUs and HORTAUs fluxes at ($\sim 100
eV cm^{-2}s^{-1}sr^{-1}$) just an order of magnitude below present
neutrino telescope detector and cosmic ray arrays.

%%%%%%%%%%%%%%%%%%%% section 2 %%%%%%%%%%%%%%%%%%%%%%%%%

\subsection{The Underground $km^3$ detectors for single $\mu$ tracks.}

The UHE $10^{13}$eV - $10^{16}$eV $\nu$ 's , being weakly
interacting and rare, may  be detected mainly inside huge volumes,
bigger than Super-Kamiokande ones; at present  most popular
detectors consider underground ones (Cubic Kilometer Size like
AMANDA-NESTOR) or (at higher energy $10^{19}$eV - $10^{20}$eV) the
widest Terrestrial atmospheric sheet volumes (Auger Array
Telescope or EUSO atmospheric Detectors).
 Underground $km^3$ detection is based mainly on $\nu_{\mu}$ tracks above hundred
 TeV
energies, because of their high penetration in matter, leading to
$\mu$ kilometer size lepton tails \cite{Gandhi et al 1998}. Rarest
atmospheric horizontal showers are also expected due to $\nu$
interactions in air (and, as we shall discuss, in the Earth Crust)
with more secondary tails. While $km^3$ detectors are optimal for
PeV neutrino muons, the Atmospheric Detectors (AUGER-EUSO like)
exhibit a minimal threshold at highest ($\geq 10^{19} eV$)
energies. The $km^3$ sensibility is more tuned to Tens TeV -PeV
astronomy while AUGER has wider acceptance above GZK energies. As
we shall discuss $\tau$ air-shower detectors exhibit also huge
acceptance at both energy windows  being competitive both at PeV
as well as  at EeV energy range as well as GZK ones; we shall not
discuss here the $Km^{3}$ detector as the ICECUBE project.

%%%%%%%%%%%%%%%%%%%% end section 2 %%%%%%%%%%%%%%%%%%%%%%%%%

%%%%%%%%%%%%%%%%%%%% section 3 %%%%%%%%%%%%%%%%%%%%%%%%%

 \subsection{Ultra Relativistic Neutron Astronomy at EeV and secondary neutrinos}

 Incidentally just around EeV ($10^{18} eV$) energies
an associated Galactic Ultra High Energy Neutron Astronomy  might
be already observed in anisotropic clustering of UHECR data
because of the relativistic neutrons boosted lifetime. Therefore
UHE  neutrons at EeV may be a source candidate of the observed
tiny EeV anisotropy in UHECR data. Indeed a $4\%$ galactic
anisotropy and clustering in EeV cosmic rays has been recently
emerged by AGASA\cite{AGASA 1999} along our nearby galactic
spiral arm. These data have been confirmed by a South (Australia)
detector(SUGAR) \cite{Bellido et all 2001}. Therefore AGASA might
have already experienced a first UHECR-Neutron astronomy (UHENA)
at a very relevant energy flux ($\sim 10 eV cm^{-2}s^{-1}$).
These EeV-UHENA signals  must also be a source of at least
comparable parasite ($10^{17}-10^{16}$ eV) secondary tails of
UHE  neutrino $\bar{\nu_{e}}$  from the same neutron beta decay
in flight. After and even more abundant UHE $\tau$ fluxes may be
also produced by a larger pion pair production near the same
accelerating source of UHE EeV neutrons. Their flavor oscillations
and mixing in galactic or extragalactic flights (analogous to
atmospheric and solar ones) must guarantee the presence of all
lepton flavors fluxes $\phi$ nearly at equal foot:
$\phi_{\bar\nu_{e}}$, $\phi_{\bar\nu_{\mu}}$,
$\phi_{\bar\nu_{\tau}}$ $= 1:1:1$\cite{Fargion 2000-2002}.
 The latter UHE $\bar{\nu_{\tau}}$ imprint (added to other local astrophysical UHE $\nu$
production) could be already recorded \cite{Fargion 2000-2002} as
Upward and Horizontal Tau air-showers  Terrestrial Gamma Flash
(considered as secondaries $\gamma$ of Upward Tau air-showers and
Horizontal Tau air-showers): UPTAUs and HORTAUs.

%%%%%%%%%%%%%%%%%%%% section 3 %%%%%%%%%%%%%%%%%%%%%%%%%

%%%%%%%%%%%%%%%%%%%% section 4 %%%%%%%%%%%%%%%%%%%%%%%%%
%%%%%%%%%%%%%%%%%%%%%%%%%%%%%%%%%%%%%%%%%%%%%%%%%%%%%%%%%%
 \section{UHE neutrino scattering on  relic $\nu_r$ neutrinos in Hot Dark Matter Halos}

 At highest energy edges ($\geq 10^{19}-10^{20} eV$), a somehow correlated  new UHE Astronomy is
also expected for charged Cosmic Rays; indeed these UHECR have
such a large rigidity to avoid any bending by random galactic or
extragalactic magnetic fields; being nearly non deflected UHECR
should point toward the original sources showing in the sky a new
astronomical map. Moreover such UHECR astronomy is bounded by the
ubiquitous cosmic $2.75 K^{o}$ BBR screening (the well known
Greisen, Zat'sepin, Kuzmin GZK cut-off) limiting its origination
inside a very local ($\leq 20 Mpc$) cosmic volume. Surprisingly,
these UHECR above GZK (already up to day above $60$ events) are
not pointing toward any known nearby candidate source. Moreover
their nearly isotropic arrival distributions underlines and
testify a very possible cosmic origination, in disagreement with
any local (Galactic plane or Halo, Local Group) expected footprint
by GZK cut-off. A very weak Super-Galactic imprint seems to be
present but at low level and already above GZK volume.  This
opened a very hot debate in modern astrophysics known as the GZK
paradox. Possible solutions has been found recently beyond the
Standard Model assuming a non-vanishing neutrino mass. Indeed at
such Ultra-High energies, neutrino at ZeV energies ($\geq
10^{21}eVs$) hitting onto  relic cosmological light ($0.1-4 eV$
masses) neutrinos \cite{Dolgov2002} nearly at rest in  Hot Dark
Matter (HDM) Halos (galactic or in Local Group) has the unique
possibility to produce UHE resonant Z bosons (the so called
Z-burst or better Z-Showering scenario). The gauge $Z$, $W^+$,
$W^-$, boson decay may shower  the same nucleon secondaries
responsible for observed UHECR the different channel
cross-sections in Z-WW-ZZ-Shower by scattering on relic neutrinos
are shown in Fig.\ref{fig:fig1} while the boosted Z-Shower
secondary chains are in
Fig.2,Fig.\ref{fig:fig2},Fig.\ref{fig:fig3},Fig.\ref{fig:fig4},Fig.\ref{fig:fig5},Fig.\ref{fig:fig6}:
\cite{Fargion Salis 1997}, \cite{Fargion Mele Salis 1999},
\cite{Yoshida  et all 1998}, \cite{Weiler 1999}; for a more
updated scenario see \cite{Fargion et all. 2001b}, \cite{Fodor
Katz Ringwald 2002}.
%%%%%%%%%%%%%%%%%%%%%%%%%%%%%%%%%%%%

\subsection{Solving the GZK puzzle by Z-Burst or ZZ-WW Shower models }

\begin{figure}
\centering
\includegraphics[width=.77\textwidth]{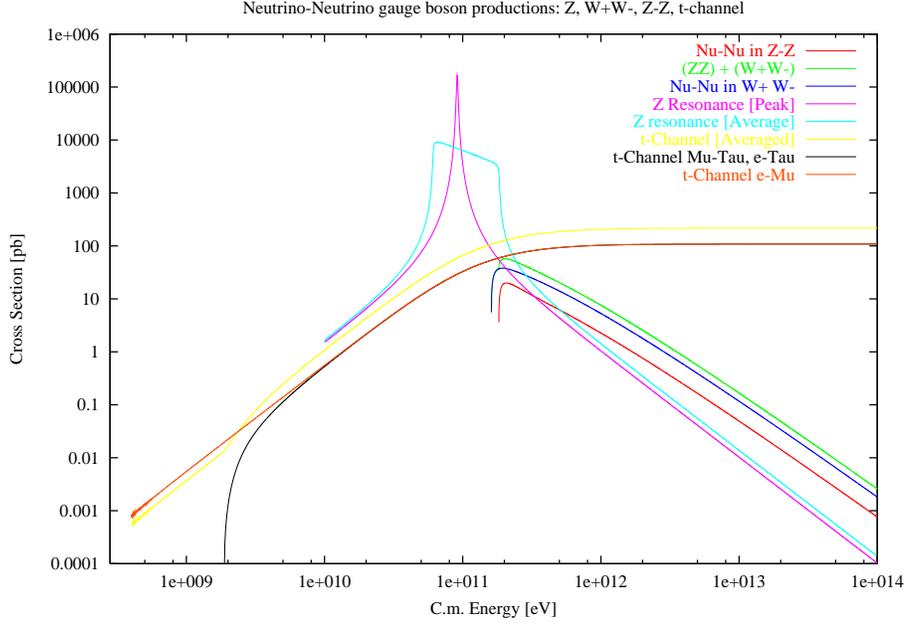}
\caption {The neutrino-relic neutrino cross-sections at center of
mass energy. The Z-peak energy will be smoothed into the
inclined-tower curve, while the WW and ZZ channel will guarantee a
Showering also above a $2 eV$ neutrino masses. The presence of
t-channel plays a role in electromagnetic showering at all
energies above the Z-peak.}\label{fig:fig1}
\end{figure}

If relic neutrinos have  a mass (as they should by recent neutrino
mass splitting evidences \cite{Barger 2002}) larger than their
maximal thermal energy (which is $\sim 5 \cdot 10^{-4} eV$ for the
temperature $1.9 K^o $, predicted in Big Bang universe for
primordial thermal gas of massless two component neutrinos) they
may cluster in Local Group or galactic halos; near eVs masses the
clustering seems very plausible and it may play a role in HDM
cosmology \cite{Fargion83} \cite{Khlopov81}. Their scattering
with incoming extra-galactic UHE neutrinos determine high energy
particle cascades which could contribute or dominate the observed
UHECR flux at $GZK$ edges. The competitive UHE $\nu$ scattering
on terrestrial atmosphere is much less ($\ll 10^{-3}$) effective
and it has an angular spread (mostly horizontal) not in agreement
width the observed UHECRs. Indeed the possibility that neutrino
share a little mass has been reinforced by Super-Kamiokande
evidence for atmospheric neutrino anomaly via $\nu_{\mu}
\leftrightarrow \nu_{\tau}$ oscillation. An additional evidence
of neutral lepton flavor mixing has been very recently reported
also by Solar neutrino experiments (SNO, Gallex), accelerator
experiment K2k and reactor experiment KamLAND. It should be noted
that very recent  indirect cosmological bound by WMAP
\cite{pierce} experiment are constraining neutrino mass into a
very narrow window : $0.05 eV  \geq m_{\nu} \geq 0.23 eV $; for
this reason we shall discuss a very definite low neutrino mass
$m_{\nu}$ in next table and figures. Consequently there are at
least two main extreme scenario for hot dark halos: either
$\nu_{\mu}\, , \, \nu_{\tau}$ are both extremely light
($m_{\nu_{\mu}} \sim m_{\nu_{\tau}} \sim \sqrt{(\Delta m)^2} \sim
0.05 \, eV$) and therefore hot dark neutrino halo is very wide
and spread out to local group clustering sizes (increasing the
radius but loosing in the neutrino density clustering contrast),
or $\nu_{\mu}, \nu_{\tau}$ may share degenerated (around $eV$
masses, ignoring more severe WMAP bound \cite{pierce}) split by a
very tiny different values.  In the latter fine-tuned neutrino
mass case ($m_{\nu}\sim 0.4 eV-1.2 eV$) (see Fig.$2$ and
Fig.\ref{fig:fig2} the Z peak $\nu \bar{\nu}_r$ interaction will
be the favorite one; in the second case (for heavier non
constrained neutrino mass ($m_{\nu} \gtrsim 1.5 \, eV$)) only a
$\nu \bar{\nu}_r \rightarrow W^+W^-$ and the additional $\nu
\bar{\nu}_r \rightarrow ZZ$ interactions, (see the cross-section
in Fig.\ref{fig:fig1})\cite{Fargion Mele Salis 1999} considered
here will be the only ones able to solve the GZK puzzle. Indeed
the relic neutrino mass within HDM models in galactic halo near
$m_{\nu}\sim 4 eV$, corresponds to a lower $Z$ resonant incoming
energy

%\begin{subeqnarray}
\begin{equation}
%\sqrt{(\Delta m)^2}
 {{E_{\nu} =  {\left(
\frac{4eV} {\sqrt{{{m_{\nu}}^2+{p_{\nu}^2}}}} \right)} \cdot
10^{21} \,eV.} \nonumber}
\end{equation}
%\end{subeqnarray}
%%%%%%%%%%%%%%%%%%%%%%%%%%%%%%%%%%%%%%%%%%%%%%%%%%%%%%%%%%%%%%%%%%%%%%%%%%%%%%%%%%%%%%%%%%%%%%

%%%%%%%%%%%%%%%%%%%%%end section 4 %%%%%%%%%%%%%%%%%%%%%%%%%

%%%%%%%%%%%%%%%%%%%%%%%%%%%%%%%%%%%%%%%%%%%%%%%%%%%%%%%%%%%%%%%%%%%%%%%%%%%%%%%%%%%%%%%%%%%%

   This resonant incoming neutrino energy is unable to overcome GZK energies while it is
     showering mainly a  small energy fraction into nucleons ($p,\bar{p}, n, \bar{n}$),
    at energies $E_{p}$ quite below (See Tab.\ref{tab:table1}).

\begin{equation}
%\sqrt{(\Delta m)^2}
 {{E_{p} =  2.2 {\left(
\frac{4eV} {\sqrt{{{m_{\nu}}^2+{p_{\nu}^2}}}} \right)} \cdot
10^{19} \,eV.} \nonumber}
\end{equation}

   Therefore too heavy ($> 1.5 eV$) neutrino mass are not fit to
   solve GZK by Z-resonance while WW,ZZ showering as well as t-channel showering
   may naturally keep open the solution.
   In particular the overlapping of both the Z and the WW, ZZ
   channels described in Fig.\ref{fig:fig1}, for $m_{\nu} \simeq 2.3 eV$ while
   solving the UHECR above GZK they must pile up (by Z-resonance
   peak activity) events at $ 5 \cdot 10^{19} eV$, leading to a correlation with the observed bump in
   AGASA data at this energy. There is indeed a first marginal evidence of such a UHECR bump
    in AGASA and Yakutsk data that may stand for this interpretation.
     More detailed data are  needed to verify such conclusive  possibility.

   Most of us  consider cosmological light relic neutrinos in Standard Model
   at non relativistic regime   neglecting any relic neutrino momentum ${p_{\nu}} $ term.
     However, at lightest mass values
   the momentum may be comparable to the relic mass;
   the spectra may reflect additional relic neutrino-energy injection which are feeding
   standard cosmic relic neutrino   at energies much above the same neutrino mass.
    It can be easily estimated that neutrino background  due to stellar,
   Super Nova, GRBs, AGN past activities,,
    presently  red-shifted into a  KeV-eV $\nu$ spectra
     while piling into a relic neutrino grey-body  spectra, cannot exceed $ 0.01\%$
    of the thermal cosmological neutrinos.
    However in cosmological models of unstable neutrinos  and primordial Black Hole neutrino evaporation such
    background may appear quite naturally in $\simeq eVs$ ranges  with or without
    leading to a present radiation dominated Universe.
     Therefore  it is worth-full to keep the most general
      mass and momentum term in the target relic neutrino spectra.
      In this windy ultra-relativistic neutrino  cosmology, there
      is no clustering halos and the unique  size to be considered is nearly coincident with the GZK one, defined by
      the energy loss length for UHECR nucleons ($\sim 20 Mpcs$).
       Therefore the isotropic UHECR behavior
      is guaranteed. The puzzle related to uniform source
      distribution spectra  seems to persist. Nevertheless the UHE neutrino-
      relic neutrino scattering \textit{do not} follow a flat
      spectrum (as well as any hypothetical $\nu$ grey
      body spectra). This leaves  open the opportunity to have a
      relic relativistic neutrino component at eVs energies as
      well as the observed  non uniform UHECR spectra. This case is similar to the
       case of a very light neutrino mass much below $0.1$ eV. \\
%%%%%%%%%%%%%%%%%%%%%%%%%%%%%%%%%%%%%%%%%%%%%%%%%%%%%%%%%%%%%%%%%%%%%%%%%%%%
%%%%%%%%%%%%%%%%%%%%%%%%%%%%%%%Anisotropy%%%%%%%%%%%%%%%%%%%%

\subsection{Relic Neutrino masses, Hot Halos and UHECRs  Anisotropy and Clustering}

In the simplest case of neutrino dominated cosmology  the neutrino
mass plays a role in defining its HDM Halos size and the
consequent enhancement of UHECR arrival directions due to our
peculiar position in the HDM halo. Indeed for a heavy mass case
$\geq 2 eV$ HDM neutrino halo are mainly galactic and/or local,
reflecting an isotropic or a diffused amplification toward nearby
$M31$ HDM halo. In the lighter case the HDM  might include the
Local Cluster up to Virgo. To each size corresponds also a
different role of UHECR arrival time. The larger the HDM size the
longer the UHECR random-walk travel time (in extra-galactic
random magnetic fields) and the wider the arrival rate lag
between doublets or triplets. The smaller is the neutrino halo
the earlier the UHE neutron secondaries by Z shower will play a
role: indeed at $E_{n}= 10^{20}eV$ UHE neutron are flying a Mpc
and their directional arrival (or their late decayed proton
arrival) are more on-line toward the source. This may explain the
high self collimation and auto-correlation of UHECR discovered
very recently \cite{Tinyakov}. The UHE neutrons Z-Showering fits
with the harder spectra observed in clustered events in AGASA
\cite{Takeda}. The same UHECR alignment may explain the quite
short (2-3 years)\cite{Takeda2} lapse of time observed in AGASA
doublets. Indeed the most conservative scenario where UHECR are
just primary proton from nearby sources at GZK distances (tens of
Mpcs) is no longer acceptable either because the absence of such
nearby sources and because of the observed stringent UHECR
clustering ($2^o - 2.5^o$) \cite{Takeda} in arrival direction, as
well as because of the short ($\sim3$ years) characteristic time
lag between clustered events. Finally the same growth with energy
of UHECR neutron (and anti-neutron) life-lengths (while being
marginal or meaning-less in tens Mpcs GZK flight distances) may
naturally explain, within a  Mpc Z Showering Neutrino Halo, the
arising harder spectra revealed in doublets-triplet spectra
\cite{Aoki}. Nevertheless in the modern multi-component Cold Dark
Matter (CDM) dominated cosmology the size and density of neutrino
halo may loose a direct relationship with neutrino mass and the
final scenario should be much more complicated, taken
into account  the evidences of different types of CDM particles.\\
For instance, EGRET data on diffuse galactic gamma ray background
above 1 GeV together with the results of DAMA direct experimental
searches for cosmic WIMPs might be considered \cite{a9}
\cite{Grossi},\cite{Konoplich2} as the evidence for the existence
of massive stable neutrino of 4th generation with the mass about
50 GeV. This hypothesis can provide explanation for anomalies in
cosmic positron spectrum, claimed by HEAT, and is accessible for
testing in the special analysis of underground neutrino data
\cite{Damour} and in precise measurements of gamma ray background
and cosmic ray fluxes.

%\subsection{Tinyakov-Glushkov  Paradox }

 UHE neutron secondaries from the same Z showering in HDM halo may also solve
an emerging puzzle: the  correlations of arrival directions of
UHECRs found recently \cite{Glushkov} in Yakutsk data at energy
$E= 8\cdot 10^{18} eV$ toward the Super Galactic Plane  are to be
compared with the compelling evidence of UHECR events ($E= 3\cdot
10^{19} eV$ above GZK) clustering toward well defined BL Lacs at
cosmic distances (redshift $z> 0.1-0.2$)
\cite{Tinyakov,Tinyakov2}. Where is the real UHECR sources
location? At Super-galactic disk (50 Mpcs wide, within GZK range)
or at cosmic ($\geq 300Mpcs$) edges? It should be noted that even
for the Super Galactic hypothesis \cite{Glushkov} the common
protons are unable to justify the high collimation of the UHECR
events. Of course both results (or just one of them) maybe a
statistical fluctuation. But both studies seem statistically
significant (4.6-5 sigma) and they seem in obvious disagreement.
There may be still open the  possibility of $two$ new categories
of UHECR sources both of them located at different distances above
GZK ones (the harder the most distant BL Lac sources). But it
seems quite unnatural  the UHECR propagation by direct nucleons
where the most distant are the harder. However our Z-Showering
scenario offers different solutions: (1) The Relic Neutrino Masses
define different Hierarchical Dark Halos and privileged arrival
direction correlated to Hot Relic Neutrino Halos. The real sources
are at (isotropic) cosmic edges \cite{Tinyakov}, \cite{Tinyakov2},
but their crossing along a longer anisotropic relic neutrino cloud
enhances the interaction probability in the Super Galactic Plane.
(2) The nearest SG sources are weaker while the collimated BL Lacs
are harder: both sources need a Neutrino Halo to induce the
Z-Showering UHECR. More data will clarify better the real
scenario.

%%%%%%%%%%%%%%%%%%%%%%%%%%%%%%%%end Anisotropy%%%%%%%%%%%%%%%%%%%%%%%%%%%%%%%%%%%%%%%%%%%%%
%%%%%%%%%%%%%%%%%%%%%%%%%%%%%%%%%%%%%%%%%%%%%%%%%%%%%%%%%%%%%%%%%%%%%%%%%%%%
As we noticed above, relic neutrino masses above a few eVs in  HDM
halo \textit{are not} consistent with naive Z peak; higher
energies interactions ruled by WW, ZZ cross-sections
 may nevertheless solve the GZK cut-off. In this
regime there will be also possible to produce  t-channel $UHE$
lepton pairs by $\nu_i \bar{\nu}_j\rightarrow l_i\bar{l}_j$
through a virtual W exchange, leading to an additional
electro-magnetic showers injection. As we shall see this important
and underestimated signal will produce UHE electrons whose final
traces are TeV synchrotron photons.
 The hadronic tail of the Z or $W^+ W^-$ cascade may be
 the source of final  nucleons $p,\bar{p}, n, \bar{n}$ able to explain UHECR events observed by
Fly's Eye and AGASA  and other detectors. The same $\nu
\bar{\nu}_r$ interactions are a source of Z and W that decay in a
rich shower ramification.
%%%%%%%%%%%%%%%%%%%%%%%%%%%%%%%%%%%%%%%%%%%%%%%%%%%%%%%%
%%%%%%%%%%%%%%%%%%%%%%%%%%%%%%%%%%%%%%%%%%%%%%%%
\subsection{UHECR  Nucleons  from Z showers}
 Although protons (or anti-protons)
  are the most favorite candidates to explain the highest
  energy air shower observed, one doesn't
have to neglect the signature of final neutrons and anti-neutrons
as well as electrons and photons. Indeed the UHECR neutrons are
produced in Z-WW showering at nearly same rate as the charged
nucleons. Above GZK cut-off energies UHE $n$,$ \bar{n}$, share a
life length comparable with the Hot Galactic Dark Neutrino Halo.
Therefore they may be an important component in UHECRs. Moreover
prompt UHE electron (positron) interactions with the galactic or
extra-galactic magnetic fields or soft radiative backgrounds may
lead to gamma cascades from PeV to TeV energies.\\
Gamma photons at energies $E_{\gamma} \simeq 10^{19}$ - $10^{20}
\,eV$ may freely propagate through galactic or local halo
distances (hundreds of kpc to few Mpc) and could also contribute
to the extreme edges of cosmic ray spectrum  and clustering
(see also \cite{Yoshida  et all 1998}\cite{Fargion 2000}). \\
The ratio of the final energy flux of nucleons near the Z peak
resonance, $\Phi_p$ over the corresponding electro-magnetic energy
flux $\Phi_{em}$ ratio is, as in Tab.1 $e^+ e^-,\gamma$ entry,
nearly $\sim \frac{1}{8}$.  Moreover if one considers at higher
$E_{\nu}$ energies, the opening of WW, ZZ channels and the six
pairs $\nu_e \bar{\nu_{\mu}}$, \, $\nu_{\mu} \bar{\nu_{\tau}}$, \,
$\nu_e \bar{\nu_{\tau}}$ (and their anti-particle pairs) t-channel
interactions leading to  highest energy leptons with no nucleonic
relics (as $p, \bar{p}$), this additional injection contributes a
factor $\sim 1.6$ leading to $\frac{\Phi_p}{\Phi_{em}} \sim
\frac{1}{13}$. This ratio is valid at energies corresponding to
the $WW,ZZ$ masses. Since the overall cross section is energy
dependent at the center of mass energies above these values, the
$\frac{\Phi_p}{\Phi_{em}}$ decreases more because of the dominant
role of t-channel (Fig.\ref{fig:fig1}). We focus here  on Z, and
WW,ZZ channels showering in hadrons for GZK events.
The important role of UHE electron showering into TeV radiation is discussed below.\\
%%%%%%%%%%%%%%%%%%%%%%%%%%%%%%%%%%%%%%%%%%%%%%%%%%%%%%%%%%%%%%%%%%%%%%%%%%%%

\subsection{UHE $\nu$ - $\nu_{relic}$ Cross Sections }
Extragalactic neutrino cosmic rays are free to move on cosmic
distances up our galactic halo without constraint on their mean
free path, because the interaction length with cosmic background
neutrinos is greater than the actual Hubble distance. A Hot Dark
Matter galactic or local group halo model with relic light
neutrinos (primarily the heaviest $\nu_{\tau}$ or $ \nu_{\mu} $) ,
acts as a target for the high- energy neutrino beams. The relic
number density and the halo size are large enough to allow the
$\nu \nu_{relic}$ interaction. As a consequence high energy
particle showers are produced in the galactic or local group halo,
overcoming the GZK cut-off.  There is an upper bound density
clustering for very light Dirac fermions due to the maximal Fermi
degenerancy whose  number density contrast is $\delta n \propto
m_{\nu}^3$,  while one finds
 that  the neutrino free-streaming halo grows
only as $\propto m_{\nu}^{-1}$. Therefore the overall interaction
probability grows $ \propto m_{\nu}^{2} $, favoring heavier non
relativistic (eVs) neutrino masses. In this frame above few eV
neutrino masses only WW-ZZ channel are operative. Nevertheless
the same lightest relic neutrinos may share higher Local Group
velocities (thousands $\frac{Km}{s}$) or even nearly relativistic
speeds and it may therefore compensate the common density bound:

%%%%%%%%%%%%%%%%%%%%%%%%%%%%%%%%%%%%%%Tokio%%%%%%%
\begin{equation}
 n_{\nu_{i}}\leq 1.9\cdot10^{3}%\left(\frac{n_{\nu_{cosmic}}}{54cm^{-3}}\right)
\left( \frac{m_{i}}{0.1eV}\right)  ^{3}\left(
\frac{v_{\nu_{i}}}{2\cdot10^{3}\frac{Km}{s} }\right)  ^{3}
\end{equation}

%\begin{equation} { {n_{\nu_{\i}}}= {10^3 {\left(
%\frac{n_{\nu_i}}{54\,{\rm cm}^{-3}}\right)} \;
%\left(\frac{m_{\nu_i}}{{\rm 0.1eV}}\right)^3 \left( \frac{\
%v_{\nu_i}}{2000{\rm km/s}}\right)^3\,\right}} \end{equation}

 From the cross section side there are three main interaction processes that
 have to be considered  leading to nucleons in the
UHE neutrino scattering with relic neutrinos .

 {\bf channel 1.} The UHE neutrino  anti-neutrino (and charge conjugation) relic scattering
$\nu_{e} \bar{\nu_{er}},$ $\nu_{\mu}
\bar{\nu_{{\mu}{r}}}$,$\nu_{\tau} \bar{\nu_{{\tau}{r}}}$
$\rightarrow Z \rightarrow \bar{f} f,$ fermion pair decay at the
Z resonance.

{\bf channel 2.}The UHE neutrino  anti-neutrino (and charge
conjugation) relic scattering $\nu_{e} \bar{\nu_{er}},$ $\nu_{\mu}
\bar{\nu_{{\mu}{r}}}$,$\nu_{\tau} \bar{\nu_{{\tau}{r}}}$
$\rightarrow W^+ W^-$ or $ \rightarrow Z Z$ leading to hadrons,
electrons, photons, through W and Z decay.

 {\bf channel 3.} The $\nu_e$ - $\bar{\nu_{\mu}}$, $\nu_e$ -
$\bar{\nu_{\tau}}$, $\nu_{\mu}$ - $\bar{\nu_{\tau}}$ and
antiparticle conjugate interactions of different flavor neutrinos
mediated in the $t$-channel by the W exchange (i.e. $\nu_{\mu}
\bar{\nu_{\tau_r}} \rightarrow \mu^- \tau^+ $). These reactions
are sources of prompt UHE electrons pairs as well as   secondary
electron pairs ( by muon or tau pair decays) as well as photons
resulting by hadronic (${\pi}^o$) $\tau$ decay secondaries. Most
of these UHE electron energies are soon converted in
electromagnetic radiation (photons) by inverse Compton scattering
on BBR or synchrotron radiation (by galactic or extragalactic
magnetic fields).

\subsection{The UHE neutrino scattering $\nu_{e} \bar{\nu_{er}},$ $\nu_{\mu}
\bar{\nu_{{\mu}{r}}}$,$\nu_{\tau} \bar{\nu_{{\tau}{r}}}$
$\rightarrow Z \rightarrow \bar{f} f,$  }
 The interaction of neutrinos of the
same flavor can occur via a Z exchange in the $s$-channel
($\nu_i\bar{\nu}_{i_r}$ and charge conjugated). The cross section
for hadron production in $\nu_i\bar{\nu}_i\rightarrow
Z^*\rightarrow hadrons$ is
\begin{equation}
\sigma_Z(s)=\frac{8\pi s}{M_Z^2}\frac{\Gamma(Z^o\rightarrow
invis.) \Gamma(Z^o\rightarrow hadr.)}{(s-M_Z^2)^2+M_Z^2
\Gamma_Z^2}
\end{equation}
where $\Gamma(Z^o\rightarrow invis.)\simeq 0.5~GeV$,
$\Gamma(Z^o\rightarrow hadr.)\simeq 1.74~GeV$ and $\Gamma_Z\simeq
2.49~GeV$ are respectively the experimental Z width into invisible
products, the Z width into hadrons and the Z full width
\cite{pdg}.

%%%%%%%%%%%%%%%%%%%%%%%%%%%%%%%%%%%%%%%%%%%%
A $\nu\nu_r$ interaction mediated in the $s$-channel by the Z
exchange, shows a peculiar peak in the cross section due to the
resonant Z production at $s= M_Z^2$. However, this occurs for a
very narrow and fine-tuned windows of  arrival neutrino energies
${\nu}_{i}$  (for  fixed  target neutrino masses and momentum $
\bar{\nu}_{i}$):
\begin{equation}
{E_{\nu_{i}}} =  {\left( \frac{4eV} {\sqrt {{{m_{\nu_
{i}}}^2+{p_{\nu_ {i}}}^2}}} \right)} \cdot 10^{21} \,eV.
\end{equation}

%%%%%%%%%%%%%%%%%%%%%%%%%%%%%%%%%%%%%%%%%%%%%%%
The effective peak cross section  reaches the value ($< \sigma_Z
> = 4.2 \cdot 10^{-32} \, cm^2$).  We assumed here for a more
general case (non relativistic and nearly relativistic relic
neutrinos)  that the averaged cross section has to be extended
over an energy window comparable to a half of the center of mass
energy. The consequent effective
averaged cross-section is described in Fig.\ref{fig:fig1} as a lower truncated hill curve.\\

So in this mechanism the energy of the UHE neutrino cosmic ray is
related to the mass of the relic neutrinos, and for an initial
neutrino energy fixed at $E_{\nu} \simeq 10^{22} \, eV$, the Z
resonance requires a mass for the heavier neutral lepton around
$m_{\nu} \simeq 0.4 \, eV$. Apart from this narrow
 resonance peak at $\sqrt{s}= M_Z$, the
asymptotic behavior of the cross section is proportional to $1/s$
for $s\gg M_Z^2$.
\\

The $\nu \bar{\nu} \rightarrow Z \rightarrow hadrons$ reactions
have been proposed by \cite{Fargion Salis 1997}\cite{Fargion Mele
Salis 1999} \cite{Yoshida et all 1998} \cite{Weiler 1999} with a
neutrino clustering on Supercluster, cluster, Local Group, and
galactic halo scale within the few tens of Mpc limit fixed by the
GZK cut-off. Due to the enhanced annihilation cross-section in
the Z pole, the probability of a neutrino collision is reasonable
even for a
 neutrino density contrast as modest as $\delta n_{\nu} /
n_{\nu} \geq 10^2$. The potential wells of such structures might
enhance the neutrino local group density with an efficiency at
comparable with observed baryonic clustering discussed above. In
this range the presence of extended local group halo should be
reflected into anisotropy (higher abundance) toward Andromeda,
while a much lighter neutrino mass may correspond to a huge halo
containing even Virgo maybe Coma and the Super Galactic Plane.

%%%%%%%%%%%%%%%%%%%% fig 2 %%%%%%%%%%%%%%%%%%%%%%%%%%%%%%%%%%%%%%%%%%%%%

\subsection{The $W^+ W^-$ and $ Z Z $ Channels}

%The Z resonant neutrino annihilation is about three orders of
%magnitude higher than W pair channel, but it occurs for a very narrow and
%fine-tuned window of energy

The  reactions $\nu_{\tau} \bar{\nu_{\tau}} \rightarrow W^+ W^-$,%
 $\nu_{\mu} \bar{\nu_{\mu}} \rightarrow W^+ W^- $,%
 $\nu_{e}  \bar{\nu_{e}} \rightarrow W^+ W^-$, %
 \cite{Enq}, have been
previously introduced in order to explain UHECR as the Fly's Eye
event at ($3 \cdot 10^{20} eV$) detected in $1991$ and last AGASA
data for neutrino mass a few eVs clustered in galactic or local
hot dark halos\cite{Fargion Mele Salis 1999}. The cross section is
then given by

$$\sigma_{WW}(s)=\sigma_{asym}\frac{\beta_W}{2s}\frac{1}{(s-M_Z^2)}\ \left\{4 L(s) \cdot C(s)+D(s)\right\}.$$

where $\beta_W=(1-4 M_W^2/s)^{1/2}$,
$\sigma_{asym}=\frac{\pi\alpha^2}{2\sin^4\theta_W
M_W^2}\simeq~108.5~pb$, and the functions $L(s)$, $C(s)$, $D(s)$
are defined as
\begin{displaymath}
L(s)=\frac{M_W^2}{2\beta_W s}\ln\Big(\frac{s+\beta_W s-2 M_W^2}{s-
\beta_W s-2 M_W^2}\Big)
\end{displaymath}
\begin{equation}
C(s)=s^2+s(2 M_W^2-M_Z^2)+2 M_W^2(M_Z^2+M_W^2)
\end{equation}

$$
D(s)=\frac{\Big[ s^2(M_Z^4-60 M_W^4-4 M_Z^2 M_W^2) + 20 M_Z^2
M_W^2 s (M_Z^2+2 M_W^2)-48 M_Z^2 M_W^4(M_Z^2 + M_W^2) \Big]}{12
M_W^2 (s-M_Z^2)}
$$
This result should be extended with the additional new  ZZ
interaction channel considered in \cite{Fargion 2000}:

$$\label{4}
\sigma_{ZZ} = \frac{G^2M^2_Z}{4 \pi} y \frac{(1 +
\frac{y^2}{4})}{(1 - \frac{y}{2})} \Big\{  \ln \Big[ \frac{2}{y}
(1 - \frac{y}{2} + \sqrt{1 - y}) \Big] -\sqrt{1 - y} \Big\}
$$

where $y = \frac{4M^2_Z}{s}$ and $\frac{G^2M^2_Z}{4 \pi} = 35.2
\,pb$.\\

These cross-section values are plotted in Fig.\ref{fig:fig1}.
 The asymptotic behavior of
this function is proportional to
$\sim(\frac{M_W^2}{s})\ln{(\frac{s}{M_W^2})}$ for $s\gg M_Z^2$.\\
A nucleon arising from WW and ZZ hadronic decay could provide a
reasonable solution to the UHECR events above GZK. We assume that
the fraction of pions and nucleons related to the total number of
particles from the W boson decay is  almost the same as  in the Z
decay boson. So each W hadronic decay (Probability  $P \sim
0.68$) leads on average to about 37 particles, where $<n_{\pi^0}>
\sim 9.19$, $<n_{\pi^{\pm}} > \sim 17$, and $<n_{p,\bar{p}, n,
\bar{n}}> \sim 2.7$. In addition we have to expect by the
subsequent decays of $\pi$'s (charged and neutral), kaons and
resonances ($\rho$, $\omega$, $\eta$) produced, a flux of
secondary UHE photons and electrons. On average it results
\cite{pdg} that the energy in the bosons decay is not uniformly
distributed among the particles. Each charged pion will give an
electron (or positron) and three neutrinos, that will have less
than one per cent of the initial W boson energy, while each
$\pi^0$ decays in two photons, each with 1 per cent of the
initial W energy. In the Tab.\ref{tab:table1} we show all the
channel reactions leading from single Z  in nuclear and
electro-magnetic components up to the observed UHECR; in the same
Tab.\ref{tab:table2} we follow the chain reaction and the
probability for the scattering event assuming a low relic
neutrino mass $m_{\nu}\simeq 0.154 eV$ and a neutrino number
density contrast in local group halo $\delta n_{\nu}/  n_{\nu}$
$\simeq 40$.
%Their energies and corresponding fluence are summirized in Table 2.\\
%At the same time a pure W leptonic decay $W \rightarrow l \nu$
%could occur for each flavor with a probability $P \sim 0.11$
%(see Table 4).\\

\subsection{The t-channel process $\nu_{i}\bar{\nu_{jr}}\rightarrow l_i \bar{l_j}$}

Processes $\nu_{i} \bar{\nu_{jr}} \rightarrow l_i \bar{l_j}$
$(i\neq j)$ (like $\nu_{\mu} \bar{\nu_{\tau}} \rightarrow \mu
\bar{\tau} $ for example)
 occur through the W boson exchange in the t-channel.
The cross-section has been derived in \cite{Fargion Mele Salis
1999}, while the energy threshold depends on the mass of the
heavier lepton produced,\\ $E_{\nu_{th}} = 3.95 \cdot10^{18} eV
(m_{\nu} / 0.4 \, eV)^{-1} (m_{\tau , \mu , e} /m_{\tau} )^2$,
with the term $(m_{\tau} / m_{\tau ,\mu , e})$ including the
different thresholds in all the possible interactions: $\nu_{\tau}
\bar{\nu_{\mu}} $ (or $\nu_{\tau}  \bar{\nu_{e}})$ , $\nu_{\mu}
 \bar{\nu_{e}}$, and $\nu_{e} \bar{\nu_{e}}$.

We could consider the reactions $\nu_{i} \bar{\nu_{j}}
\rightarrow l_i \bar{l_{j}} $,($i=j$), keeping both  $s$ channel
and $t$ channel, while in reactions $\nu_{i} \bar{\nu_{j}}
\rightarrow l_i \bar{l_{j}} $,($i\neq j$) do occur only via $t$
channel.

 In the ultra-relativistic limit ($s \simeq
2E_{\nu} m_{\nu_r} \gg M^2_W$ where $\nu_r$ refers to relic
clustered neutrinos)  the cross-section tends to the
asymptotic value $\sigma_{\nu \bar{\nu_r}} \simeq 108.5 \,pb$.\\

$$
 \sigma_W(s)= \sigma_{asym}\frac{A(s)}{s}
 \bigg\{1+\frac{M_W^2}{s}
 \cdot  \Bigg[2-\frac{s+B(s)}{A(s)}\ln\Bigg(\frac{B(s)+A(s)}{B(s)-
A(s)}\Bigg)\Bigg]\bigg\}
$$
where  $\sigma_{asym}$ is the asymptotic behaviour of the cross
section in the ultra-relativistic limit

 and where $\sqrt{s}$ is the center of mass energy, the
functions A(s), B(s) are defined as
$$
%\begin{split}
A(s)= \sqrt{[s-(m_\tau+m_\mu)^2] [s-(m_\tau-m_\mu)^2]}$$
\\
$$
B(s)=s+2M_W^2-m_\tau^2-m_\mu^2 ~~~~~~~~
%\end{split}
$$
and

%%%%%%%%%%%%%%%%%%%%%%%%%%%%%%%%%%%%%%%%%%%%%%%%%%%%%%%%%%%%%%%
\begin{equation}
s\simeq 2 E_\nu m_\nu= 2\cdot 10^{23} \frac {E_\nu}{10^{22}~eV}
\frac {m_\nu} {10~eV}~eV^2 \gg M_W^2~.
\end{equation}
%%%%%%%%%%%%%%%   Figure 1 Ex 01  %%%%%%%%%%%%%%%%%   FFFFFFFFFFFFFFFFFFFFFFFFFFFFFFFFFFFFFF

%%%%%%%%%%%%%%%   Figure 1 END   %%
These  t-channel interactions lead to electro-magnetic showers
and do not offer any nuclear secondary able to explain UHECR
events.

\section{Boosted Z-UHECR  spectra }

Let us examine the destiny of UHE primary particles (nucleons,
electrons and photons) ($E_e \lesssim 10^{21}\,eV$) produced after
hadronic or leptonic W decay. As we have already noticed in the
introduction, we'll assume that the nucleons, electrons and
photons spectra (coming from W or Z decay) after $\nu \nu$
scattering in the halo, follow a power law that in the center of
mass system is $\frac{dN^*}{dE^* dt^*} \simeq E^{* - \alpha}$
where $\alpha \sim 1.5$. This assumption is based on detailed
Monte Carlo simulation of a heavy fourth generation neutrino
annihilations \cite{Konoplich} \cite{Konoplich2} \cite{Grossi} and
with the model of quark - hadron fragmentation spectrum suggested
by Hill \cite{Hill}.

In order to determine the shape of the particle spectrum in the
laboratory frame, we have to introduce the Lorentz relativistic
transformations from the center of mass system to the laboratory
system.
 The number of particles is clearly a relativistic invariant $dN_{lab} = dN^*$,
while the relation between the two time intervals is $dt_{lab} =
\gamma dt^*$, the energy changes like $ \epsilon_{lab} = \gamma
\epsilon^* (1 + \beta \cos \theta^*) = \epsilon^* \gamma^{-1}(1 -
\beta \cos \theta)^{-1}$, and finally the solid angle in the
laboratory frame of reference becomes $d\Omega_{lab} =\gamma^{2}
d\Omega^*  (1 - \beta \cos \theta )^2$. Substituting these
relations one obtains

%\begin{subeqnarray}
%\left(\frac{dN}{d\epsilon dt d\Omega} \right)_{lab} =
%\frac{dN_{*}}{d\epsilon_{*} dt_{*} d\Omega_{*}} \gamma^{-2}
% (1 - \beta \cos \theta)^{-1} =
% \frac{\epsilon^{-\alpha}_{*} \; \gamma^{-2}} {4 \pi}} \cdot (1 - \beta \cos
% \theta)^{-1} \nonumber \\
%\left( \frac{dN}{d\epsilon dt d\Omega} \right)_{lab} =
% \frac{\epsilon^{-\alpha} \; \gamma^{-\alpha-2}} {4 \pi}} (1 - \beta \cos \theta)^{-\alpha-1}\setcounter{eqsubcnt}{0}
%\end{subeqnarray}

$$
\left(\frac{dN}{d\epsilon dt d\Omega} \right)_{lab} =
\frac{dN_{*}}{d\epsilon_{*} dt_{*} d\Omega_{*}} \gamma^{-2}
 (1 - \beta \cos \theta)^{-1} \\
=\frac{\epsilon^{-\alpha}_{*} \; \gamma^{-2}} {4 \pi} \cdot (1 -
\beta \cos
 \theta)^{-1} \\
= \frac{\epsilon^{-\alpha} \; \gamma^{-\alpha-2}} {4 \pi} (1 -
\beta \cos \theta)^{-\alpha-1}
$$

and integrating over $\theta$ (omitting the lab notation) one
loses the spectrum dependence on the angle.

%\begin{equation}
%\left( \frac{dN}{d\epsilon dt d\Omega} \right)_{lab} \propto
%\epsilon^{-\alpha} \gamma_{Z (W)}^{ \alpha} \sim \epsilon^{-
%\frac{\alpha}{2}} \sim \epsilon^{- 0.75}.
%\end{equation}

The consequent fluence derived by the solid angle integral is:
% Long version
%\begin{equation}
% \frac{dN}{d\epsilon dt} \epsilon^{2}=
% \frac{\epsilon^{-\alpha+2} \; \gamma^{\alpha-2}} {2 \beta \alpha}}
% [(1 + \beta)^{\alpha} -
%  \frac{1} {[(1 + \beta)\gamma^2]^{\alpha}}}] \simeq
% \frac{\epsilon^{-\alpha+2} \; \gamma^{\alpha-2}} {\alpha}}
%\end{equation}

$$
%\begin{split}
\frac{dN}{d\epsilon dt} \epsilon^{2}= \\
 \frac{\epsilon^{-\alpha+2} \; \gamma^{\alpha-2}} {2 \beta \alpha}
 [(1 + \beta)^{\alpha} - (1 - \beta)^{\alpha}] \simeq \\
 \frac{2^{\alpha-1}\epsilon^{-\alpha+2} \; \gamma^{\alpha-2}} {\alpha}
%\end{split}
$$

\begin{figure}
\centering
\includegraphics[width=.77\textwidth]{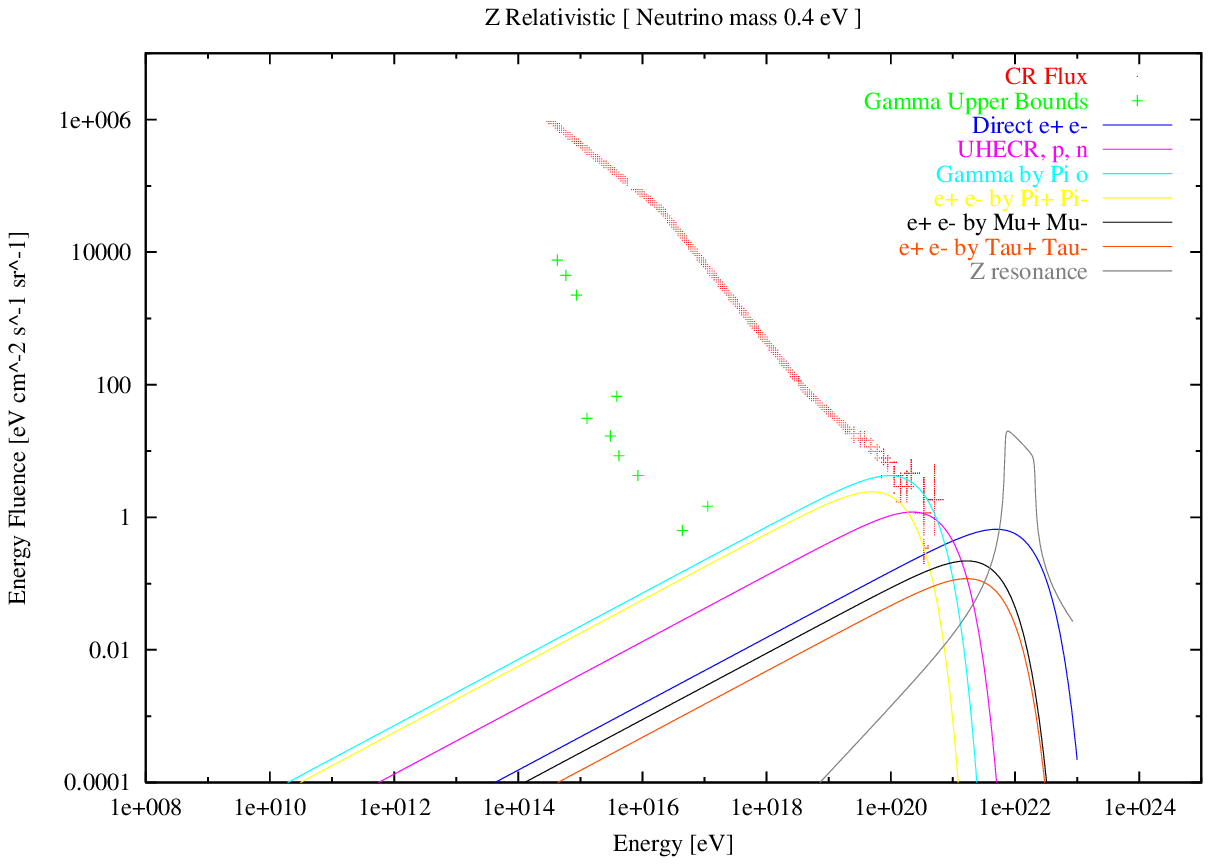}
\caption{Z-Showering Energy Flux distribution for different
    channels assuming a light (fine tuned)  relic neutrino mass
    $m_{\nu} = 0.4 eV$ \cite{Fargion 2000}, \cite{Fargion et all. 2001b},\cite{Fodor
Katz Ringwald 2002}. The total detailed energy percentage
distribution  into neutrino, protons, neutral and charged pions
and consequent gamma, electron pair particles both from hadronic
and leptonic Z, $WW,ZZ$ channels. We also calculated the
electro-magnetic contribution due to the t-channel $\nu_i \nu_j$
interactions as described in previous Fig.2. Most of the $\gamma$
radiation will be degraded around PeV energies by $\gamma \gamma$
pair production with cosmic 2.75 K BBR, or with cosmic radio
background. The electron pairs instead, are mainly relics of
charged pions and will rapidly lose energies into synchrotron
radiation.  Note that the nucleon    injection energy fits the
present AGASA data as well as the most   recent evidence of a
corresponding tiny Majorana neutrino mass $m_{\nu}\simeq 0.4 eV$
\cite{Klapdor-Kleingrothaus:2002ke}.    Lighter neutrino masses
are able to modulate UHECR at higher energies \cite{Fargion et
all. 2001b}.} \label{fig:fig2}
\end{figure}

\begin{figure}
\centering
\includegraphics[width=.77\textwidth]{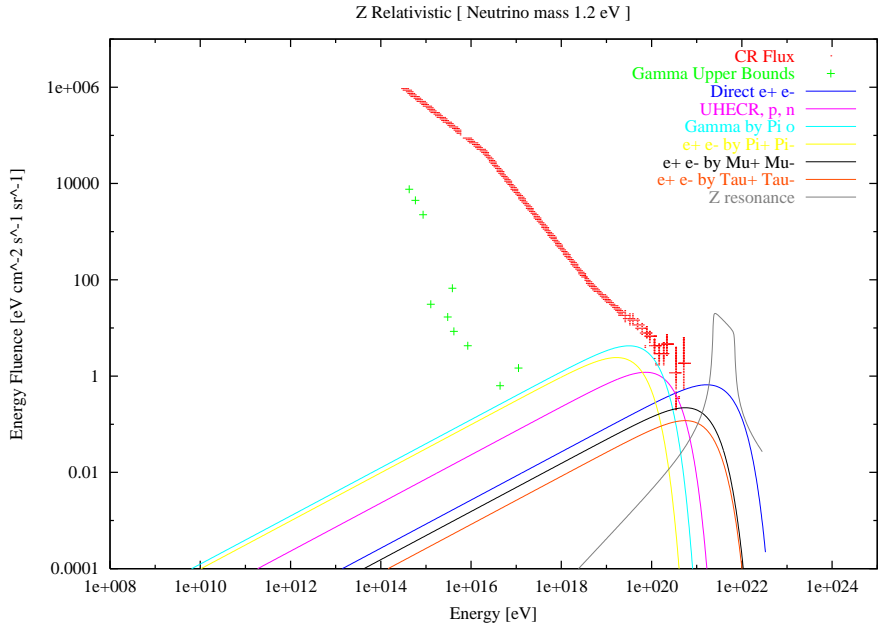}
\caption{Z-Showering Energy Flux distribution for different
channels
    assuming a light (fine tuned)  relic neutrino mass $m_{\nu} = 1.2
    eV$ able to partially fill the highest $10^{20} eV$ cosmic ray
    edges. Note that this value leads to a Z-Knee cut-off, above the
    GZK one, well tuned to present Hires data
\cite{Fargion et all. 2001b}, \cite{Kalashev:2002kx}. }
\label{fig:fig3}
\end{figure}

\begin{figure}
%\begin{figure}
\centering
\includegraphics[width=.77\textwidth]{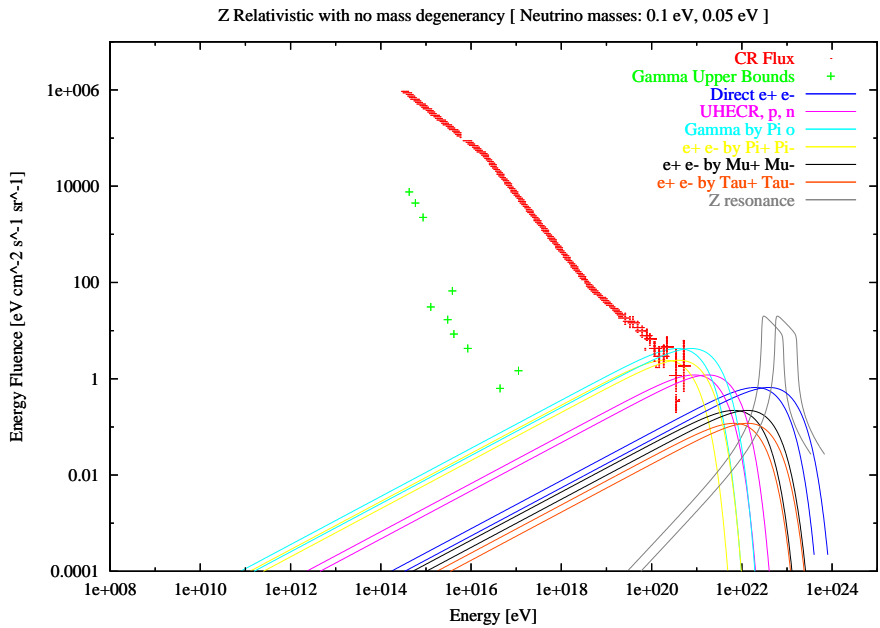}
\caption{Z-Showering Energy Flux distribution for different
channels    assuming a non degenerated twin light  relic neutrino
masses near atmospheric splitting mass values $m_{\nu} = 0.1
eV$,$m_{\nu} = 0.05 eV$ able to partially fill the highest
$10^{20} eV$ cosmic ray   edges. \cite{Fargion et all. 2001b},
\cite{Kalashev:2002kx}. } \label{fig:fig4}
\end{figure}

\begin{figure}
%\begin{figure}
\centering
\includegraphics[width=.77\textwidth]{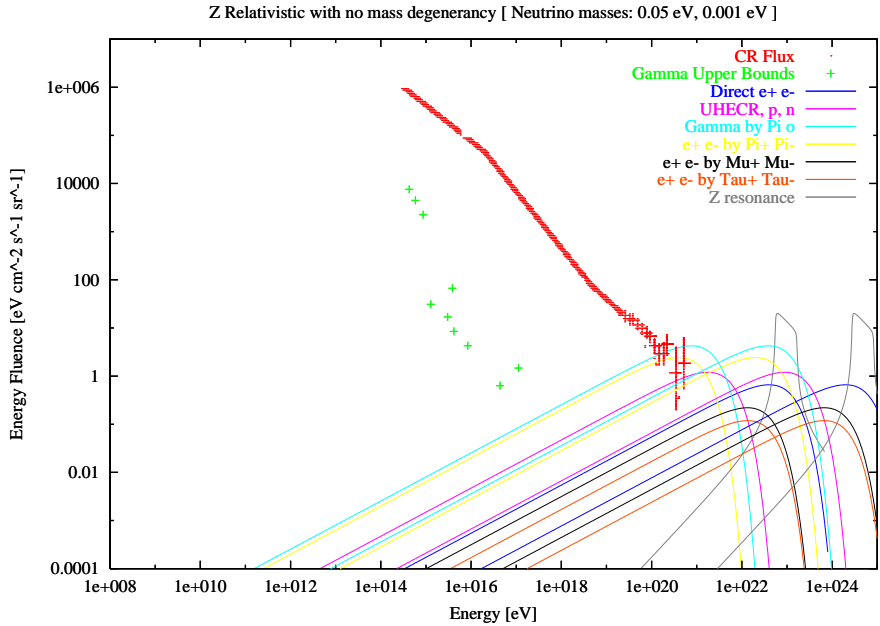}
\caption{Z-Showering Energy Flux distribution for different
channels    assuming a lightest  relic neutrino mass $m_{\nu} =
0.05 eV$ (atmospheric neutrino mass),  $m_{\nu} = 0.001  eV$, just
a small fraction (a seventh part) of the minimal solar neutrino
mass split, able to partially fill the highest $10^{20} eV$ cosmic
ray   edges; \cite{Gallex92},\cite{Fukuda:1998mi},\cite{SNO2002}
and most recent claims by KamLAND \cite{Kamland2002}. Note that
no suppression for neutrino density has been    assumed
here.\cite{Fargion et all. 2001b}, \cite{Kalashev:2002kx}. }
\label{fig:fig5}

\end{figure}

\begin{figure}
\centering
\includegraphics[width=.77\textwidth]{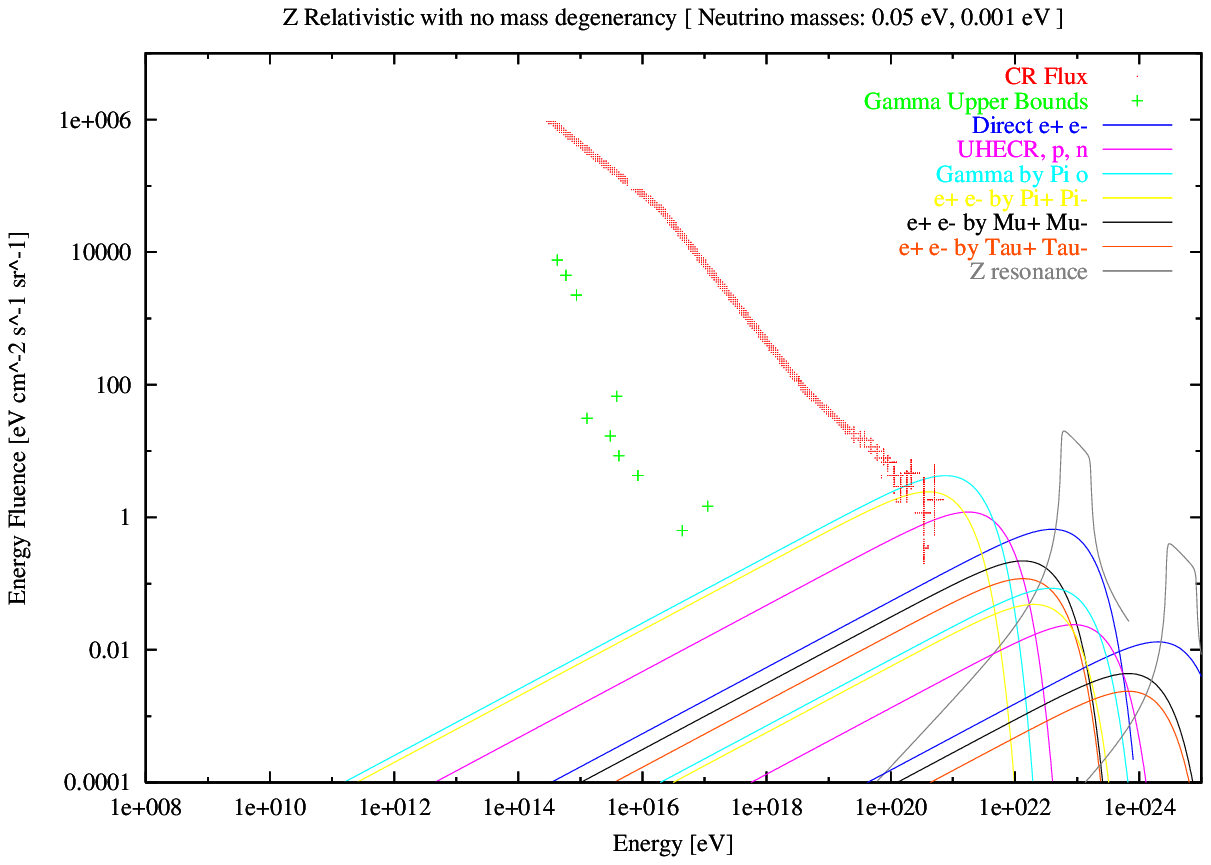}
\caption{Z-Showering Energy Flux distribution for different
channels    assuming a lightest  relic neutrino mass $m_{\nu} =
0.05 eV$ (atmospheric neutrino mass),  $m_{\nu} = 0.001 eV$ ( just
a small fraction (a seventh part) of the minimal solar neutrino
mass split) able to partially fill the highest $10^{20} eV$
cosmic ray  edges. Note that here a realistic suppression for
lighest neutrino density has been assumed. \cite{Fargion et all.
2001b}, \cite{Kalashev:2002kx}. } \label{fig:fig6}

\end{figure}

There are two extreme cases to be considered: the case where the
interaction occurs at Z peak resonance and therefore the center of
mass Lorentz factor $\gamma$ is frozen at a given value (eq.1) and
the case (WW,ZZ pair channel) where all energies are allowable and
$\gamma$ is proportional to $\epsilon^{1/2}$.
%; the latter case will be discussed in detail elsewhere.
 Here we focus only on Z peak resonance. The consequent fluence
 spectrum
 $\frac{dN}{d\epsilon dt}\epsilon^{2}$, as above, is proportional
 to $\epsilon^{-\alpha +2}$. Because $\alpha$ is
nearly $1.5$ all the consequent secondary particles will also show
a spectra proportional to $\epsilon^{1/2}$ following normalized
energies in Tab.\ref{tab:table1} as shown in
Fig.(\ref{fig:fig2}-\ref{fig:fig6}). In the latter case (WW,ZZ
pair channel), the relativistic boost reflects on the spectrum of
the secondary particles, and the spectra power law becomes
$\propto \epsilon^{\alpha/2 +1}=\epsilon^{0.25}$. These channels
will be studied in details elsewhere. In Fig.(\ref{fig:fig2}
-\ref{fig:fig6}) we show the spectra of protons, photons and
electrons coming from Z hadronic and leptonic decay assuming a
nominal primary CR energy flux $\sim 20~eV s^{-1} sr^{-1}
cm^{-2}$, due to the total $\nu \bar{\nu}$ scattering at GZK
energies as shown in Fig.(\ref{fig:fig2}-\ref{fig:fig6}). We
assume an interaction probability $P \equiv \sigma_{\nu-\nu}
n_{\nu} l_{halo}$, for a relic halo neutrino density contrast
$n_{\nu}/n_{\nu-cosmic}$ $\sim 40 $, and integral distance $
l_{halo} \sim 3 Mpc$ whose peak value is $P \sim 8\cdot 10^{-3}$
and a corresponding UHE incoming neutrino energy flux $\sim
2200~eV s^{-1} sr^{-1} cm^{-2}$ near but below present $UHE$
neutrino flux bound from AMANDA,MACRO, Baikal as well as
Goldstone data. The same probability result $P \sim 8\cdot
10^{-3}$ may be obtained assuming a lower neutrino number
contrast but a longer (up to factor ten) integral halo distance:
$n_{\nu}/n_{\nu-cosmic}$ $\sim 4$, $l_{halo} \sim 30 Mpc$, just
within the GZK cut-off maximal distance. The product of such
primary UHE neutrino fluxes with the above probability leads to
the primary Z-Showering peak at energies $\sim 20~eV cm^{-2}s^{-1}
sr^{-1} cm^{-2}$ at $\sim 10^{22} eV$ observed in UHECR nucleon
flux peak $\sim 1.2~eV cm^{-2}s^{-1} sr^{-1} cm^{-2}$ at $\sim
10^{20} eV$ in AGASA data. On the contrary for the much lower and
last HIRES reports the observed flux of UHECR are an order of
magnitude below $\sim 0.1-0.2~eV cm^{-2}s^{-1} sr^{-1} cm^{-2}$,
and they may simply require either a lower relic neutrino density
contrast (within $0.4 $eV neutrino mass model) or (within the
same density contrast), or (and) a lower incoming neutrino flux,
and/or  just a larger ($m_{\nu} \sim 1.2 eV$) relic neutrino mass
(see Fig.\ref{fig:fig3}).
%%%%%%%%%%%%%%%%%%%%%%%%%%%%%%%%%%%%%%%%%%%%%%%%%%%%%%%%%%%
%%%%%%%%%%%%%%%%%%%% fig 1 %%%%%%%%%%%%%%%%%%%%%%%%%
%%%%%%%%%%%%%%%%%%%% fig 1 %%%%%%%%%%%%%%%%%%%%%%%%%

\begin{table}[h]

\begin{center}\footnotesize
\begin{tabular}{|c|c|c|c|c|c|}
 \hline
\multicolumn{6}{|c|}{Secondaries by $\nu \nu \rightarrow Z$
Interactions:
$E_{\nu} = 10^{22} eV$, Fluence $F_{\nu}= 2000 eV cm^{-2}s^{-1}$, ($m_{\nu}=0.4$ $eV$)} \\
\hline\hline & Multiplicity & Energy ($\%$)  & $\sum
E_{CM}(GeV)$& Peak Energy (EeV) & $\frac{dN}{dE}E^2$ (eV)  \\
\hline
 $p$ &2.7 & 6 \% & 5.4 & 2.2$\cdot 10^2$ & 1.2  \\ \hline
 $\pi_0$ & 13 & 21.4 \%  &  19.25 & 1.9$\cdot 10^2$ & 4.25  \\ \hline
 ${\gamma}_{\pi^0}$ & 26 & 21.4 \% & 19.25 & 95 & 4.25  \\ \hline
 $\pi^{\pm}$ & 26 & 42.8 \% & 38.5 & 1.9$\cdot 10^2$  & 4.25 \\ \hline
 $(e^+ e^-)_{\pi}$ & 26 & 12 \% & 11  & 50 & 2.3  \\ \hline
 $(e^+ e^-)_{prompt}$ & 2 & 3.3 \% & 2.7 & 5$\cdot 10^3$ &  1.32 \\ \hline
 $(e^+ e^-)_{\mu}$ & 2 & 1.1 \% & 0.9 & 1.6$\cdot 10^3$ & 0.45 \\ \hline
 $(e^+ e^-)_{\tau}$ & 2 & 1.5 \% & 1.3 &  1.2$\cdot 10^3$  & 0.6 \\ \hline

\end{tabular}
\end{center}
  \caption{The total detailed energy
percentage distribution  into neutrinos, protons, neutral and
charged pions and consequent gamma, electron pair particles both
from hadronic and leptonic channels. We used LEP data for Z
decay. We assumed that on average number of 37 particles produced
during a Z (W) hadronic decay. The number of prompt pions both
charged (18) and neutral (9), in the hadronic decay is increased
by 8 and 4 respectively due to decays of $K^o$, $K^{\pm}$,
$\rho$, $\omega$, and $\eta$. We assumed that the most energetic
neutrinos produced in the hadronic decay mainly come from charged
pion decay. So their number is roughly three times the number of
$\pi$'s. UHE photons are mainly relics of neutral pions.}%%%%%%%%%%
%%%%%%%%%%%%%%%%%%%%%%%%%%%%%%%%%%%%%%%%%%%%%%%%
%Most of the $\gamma$ radiation will be degraded
%around PeV energies by $\gamma \gamma$ pair production with
%cosmic 2.75 K BBR, or with cosmic radio background. The electron
%pairs instead, are mainly relics of charged pions and will
%rapidly lose energies by synchrotron radiation. The contribution
%of leptonic Z (W) decay is also considered and calculated in the
%Table 1B-1B.}%%%%%%%%%%%%%%%
%%%%%%%%%%%%%%%%%%%%%%%%%%%%%%%%%%%%%%%%%%%%
\label{tab:table1}
\end{table}

\begin{table}[h]

\begin{center}\scriptsize
\begin{tabular}{cccc}
\hline \hline \multicolumn {4}{c}{}\\
\multicolumn {4}{c}{}\\
\multicolumn {4}{c}{\scshape{Main Z channel reactions chains for
final proton production}} \\
\multicolumn {4}{c}{} \\
\multicolumn {4}{c}{} \\
 \hline\hline
 \multicolumn {4}{c}{}\\
 {Reaction} & {Probability} & {Multiplicity} &
{Secondary energy} \\
& & & \\
 \hline
& & & \\
& & & \\
\multicolumn {1}{l}{1d)} & & &  \\
\multicolumn {1}{l}{$p+\gamma\rightarrow (p,n)+9\pi$} &
$P_{1d}\simeq 1$ & $M_{1d}= 6$
& $E_{\pi}\sim\frac{E_p}{10}$\\
& & & \\
& & & \\

\multicolumn {1}{l}{2d)} & & & \\
\multicolumn {1}{l}{$\pi^+\rightarrow \mu^++\nu_{\mu}$} &
$P_{2d}\simeq 1$ &
$M_{2d}=1$ & $E_{\nu}\sim 0.21 E_\pi = 2.1\cdot 10^{-2} E_p$\\
\multicolumn {1}{l}{$\pi^-\rightarrow \mu^-+\bar{\nu}_{\mu}$} \\
& & & \\
& & & \\

\multicolumn {1}{l}{2d')} & & & \\
\multicolumn {1}{l}{$\mu^+\rightarrow e^++\nu_e+\bar{\nu}_{\mu}$}
& $P_{2d'}\simeq 1$
& $M_{2d'}=1$ & $E_{\nu}\sim 0.26 E_\pi = 2.6\cdot 10^{-2} E_p$ \\
\multicolumn {1}{l}{$\mu^-\rightarrow e^-+\bar{\nu}_e+\nu_\mu$} \\
& & & \\
& & & \\

\multicolumn {1}{l}{3d)} & & & \\
\multicolumn {1}{l}{$\nu_{\mu}+\bar{\nu}_{\mu_r}\rightarrow
Z^*\rightarrow 2 p+X$} &
$P_{\nu\nu}=\sigma_{\nu_{\mu}\bar{\nu}_{\mu}}n_{\nu_r}l_g\sim
 10^{-3}$ & & \\
\multicolumn {1}{l}{$\bar{\nu}_{\mu}+\nu_{\mu_r}\rightarrow
Z^*\rightarrow 2 p+X$} & & $M_{3d}=2$ & $E_p\sim
\frac{E_{\nu_\mu}}{80}\sim 2.6\cdot 10^{-4}
E_p$\\
& & & \\
& & & \\

\multicolumn {1}{l}{3d')} & & & \\
\multicolumn {1}{l}{$\nu_{\mu}+\bar{\nu}_{\mu_r}\rightarrow
Z^*\rightarrow 2 p+X $} &
$P_{\nu\nu}=\sigma_{\nu_{\mu}\bar{\nu}_{\mu}}n_{\nu_r}l_g\sim
8\cdot 10^{-4}$ & & \\
\multicolumn {1}{l}{$\bar{\nu}_{\mu}+\nu_{\mu_r}\rightarrow
Z^*\rightarrow 2 p+X $} & & $M_{3d'}=2$ & $E_p\sim
\frac{E_{\nu_\mu}}{83}\sim 3.15\cdot 10^{-4} E_p$\\
& & & \\
& & & \\
& & & \\
\multicolumn {1}{l}{finally: eq.(1d)} & \fbox{$^d P_{tot}^Z=\Pi_i
M_i P_i\sim 4 \cdot 10^{-3}$} & & \fbox{$E_p^Z\sim 10^{24}~eV$}\\
 & \fbox{$^{d'} P_{tot}^Z=\Pi_i M_i P_i\sim 8\cdot
10^{-3}$} & & \\
& & & \\
& & & \\
& & & \\

\multicolumn {4}{l}{\underline{Additional reactions not included
in the present
analysis:}}\\
& & & \\
& & & \\

\multicolumn {4}{l}{Z channels if there is neutrino flavor mixing
$\nu_\mu \leftrightarrow \nu_\tau$}\\
& & & \\
& & & \\

\multicolumn {1}{l}{3'd)} & & & \\
\multicolumn {1}{l}{$\bar{\nu}_\tau+\nu_{\tau_r}\rightarrow
Z^*\rightarrow 2 p+X$} & &
& \\
\multicolumn {1}{l}{$\nu_\tau+\bar{\nu}_{\tau_r}\rightarrow
Z^*\rightarrow 2 p+X$} & &
& \\
& & & \\
& & & \\

\multicolumn {4}{l}{Z channel whose efficiency is however
suppressed by lower $\nu_e, \bar{\nu}_e$ energies and $n_{\nu_e},
n_{\bar{\nu}_e}$ densities} \\
& & & \\
& & & \\
& & & \\

\multicolumn {1}{l}{3''d)} & & & \\
\multicolumn {1}{l}{$\bar{\nu}_e+\nu_{e_r}\rightarrow Z^*\rightarrow 2 p+X$} & & & \\
& & & \\
& & & \\
\hline
& & & \\

\multicolumn {4}{l}{$^a$ This multiplicity refers only to charged pions}\\
\multicolumn {4}{l}{$^b$ In calculating the probability for the
$\nu\nu$ interaction we assumed here: $m_{\nu_r} =0.154
eV$,(nearly twice the} \\
\multicolumn {4}{l}{observed atmospheric neutrino mass splitting)
$n_{\nu_r}= 2\cdot10^{3}~cm^{-3}$ (a number density contrast
above} \\
\multicolumn {4}{l}{cosmic background  $\simeq 40$),the distance
integral inside the halo volume $l_g\sim 10^{25}~cm$. } \\
\multicolumn {4}{l}{The $\sigma_{\nu_\mu\bar{\nu}_\mu}$ value has
been obtained from the corresponding cross section and center of
mass energy }

\end{tabular}
\end{center}
  \caption{Energy peak and energy fluence for different
decay channels as described in the text; the energy of the proton
and of the pion are respectively the averaged  ones observed in
LEP data for Z decay .} \label{tab:table2}
\end{table}

\normalsize

%\begin{table}[h]
%\begin{center}
%\begin{tabular}{|c|c|c|} \hline
%  % after \\ : \hline or \cline{col1-col2} \cline{col3-col4}...
%  $W_{decay}$ & E (eV) & $\frac{dN}{dE}E^2$ (eV) \\ \hline
%  p & $2.7 \cdot 10^{20}$ & 2 \\ \hline
%$\gamma$ & $4.5 \cdot 10^{19}$ & 6 \\ \hline
% $e_{\pi}$ & $2.4 \cdot 10^{19}$ & 3 \\ \hline
%   $e_{prompt}$ & $5 \cdot 10^{21}$ & 1 \\ \hline
%    $e_{\mu}$ & $1.6 \cdot 10^{21}$ & 0.5 \\ \hline
%     $e_{\tau}$ & $1.6 \cdot 10^{21}$ & 0.1 \\ \hline
%\end{tabular}
%\end{center}
%\caption{} \label{3}
%\end{table}

UHE electrons and photons are mostly produced by charged and
neutral pions. From Tab.\ref{tab:table2}, one can see that, on
average, the ratio $E_{e}^{\pi^{\pm}} / E_{\gamma}^{\pi^0} \sim
2$. While the $dN/dE \times E^2$ for photons( by W (or Z) decay)
is almost one order of magnitude greater than the corresponding
nucleon spectra.

The gamma (MeV - TeV) galactic background could present a
component due to UHE electron interactions in our galactic halo.
The initial energy of the ultra-relativistic particle is gradually
converted in gamma photons through different steps involving
radiative processes as synchrotron radiation and electron pair
production (see Fig.\ref{fig:fig7}-Fig.\ref{fig:fig8}).\\ An
electron lifetime in the Galaxy is approximately given by
$$
  \tau_e \simeq \frac{2.8 \cdot 10^{12}}{\gamma} \,yr
$$

that means for electron energy $E_e \sim 10^{16} \,eV$ a characteristic time scale:  $\tau \sim 10^{-4} \,yr$.\\

We considered as final electron production channel the reaction
$\nu_e \nu_{\tau}$ (Tab.\ref{tab:table1})  through W exchange
whose probability is one order of magnitude higher than channel
$\nu_{\mu} \nu_{\mu} \rightarrow W^+ W^-$ introduced  as the
mechanism able to produce Fly's Eye event at $3.2 \cdot 10^{20}
\,eV$. In Tab.\ref{tab:table2} we indicate the chain leading to
electrons through W W decay. This channel leads to electrons with
energies $E_e \sim 10^{19} \div 10^{20} \,eV$
%%%%%%%%%%%%%%%%%%%%%%%%%%%%%%%%%%%%%%%%%%%%%%%%%%%Tokio3

%%%%%%%%%%%%%%%%%%%% fig 2 %%%%%%%%%%%%%%%%%%%%%%%%%

%%%%%%%%%%%%%%%%%%%%%%%%%%%%%%%%%%%%%%%%%%%%%%%%%%%%%%%%%%%%%%%%%%%%
%%%%%%%%%%%%%%%%%%%%%%%%%%%%%%%%%%%%%%%%%%%%%%%%%%%%%%%%%%%%%%%%%%
%%%%%%%%%%%%%%%%%%%%%%%%%%%%%%%%%%%%%%%%%%%%%%%%%%%%%%%%%%%%%%%%%%%%%
%%%%%%%%%%%%%%%%%%%%%%%%%%%%%%%%%%%%%%
%%%%%%%%%%%%%%%%%%%%%%%%%%%%%%%%%%%%%%%%%%%%%%%5

%%%%%%%%%%%%%%%%%%%%%%%%%%%%%%%%%%%%%%%%%%%%%%%%%%%%%%%%%%%%%%%%%%%

\subsection{TeV tails from UHE electrons in Z-Showers}

\begin{figure}[h]
\begin{center}
 \includegraphics[width=0.8\textwidth] {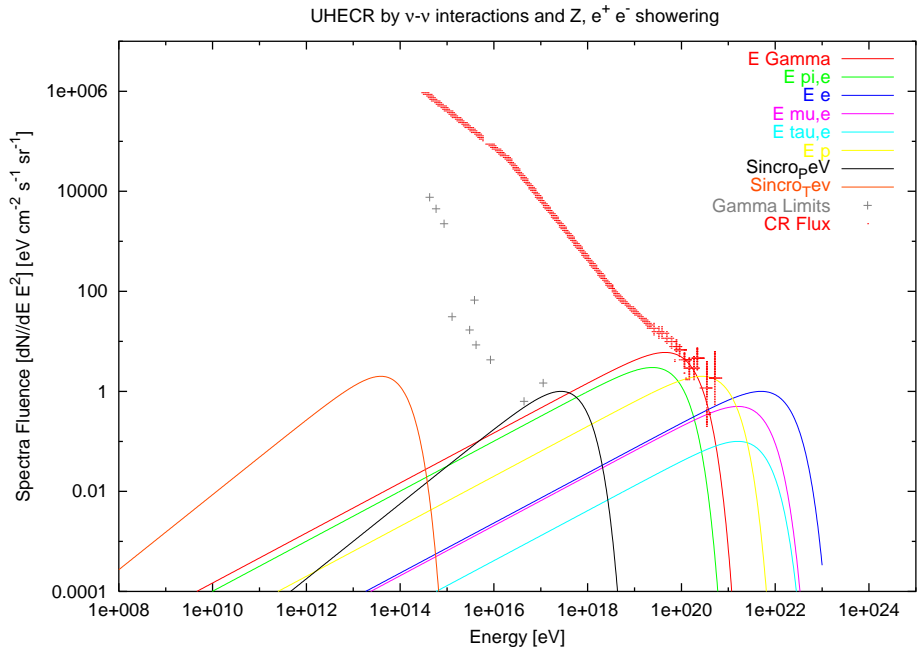}
%  \epsfigure{file=Z2.eps,width=0.45\textwidth}
\end{center}
\caption{Energy fluence by Z showering as in figures above and the
consequent $e^+ e^-$ synchrotron radiation } \label{fig:fig7}
\end{figure}

\begin{figure}[h]
\begin{center}
 \includegraphics[width=0.8\textwidth] {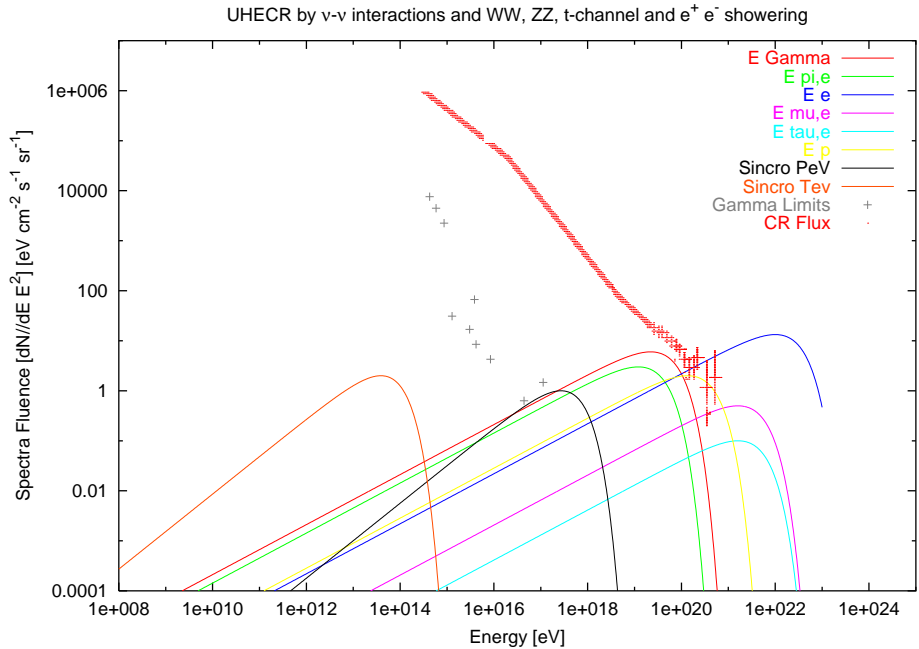}
%  \epsfigure{file=WW2.eps,width=0.45\textwidth}
\end{center}
\caption{Energy fluence by WW, ZZ, t-channel showering and the
consequent $e^+ e^-$ synchrotron radiation. The lower energy Z
showering is not included to make spectra more understandable.}
\label{fig:fig8}
\end{figure}

 As it is shown in Tab.\ref{tab:table2} and Fig.\ref{fig:fig7}-\ref{fig:fig8},  each electron
(positron) energy due to $\pi^{\pm}$ decays are around $E_e \sim 2
\cdot 10^{19} \, eV$ for an initial $E_Z \sim 10^{22} \, eV $ (
and incoming UHE neutrino energy $E_{\nu} \sim 10^{22} \, eV $)
assuming a nominal target neutrino mass $m_{\nu} \simeq 0.4 eV$.
Such electron pairs while not radiating efficiently in
extra-galactic magnetic fields will be interacting with the
galactic magnetic field ($B_G \simeq 10^{-6} \,G $) leading to
direct TeV photons:
\begin{displaymath}
  E_{\gamma}^{sync} \sim \gamma^2 \left( \frac{eB}{2\pi m_e } \right)
   \sim
\end{displaymath}
\begin{equation}\label{4b}
  \sim 27.2 \left( \frac{E_e}{2 \cdot10^{19}
  \,eV} \right)^2 \left( \frac{m_{\nu}}{0.4 \, eV} \right)^{-2} \left( \frac{B}{\mu G} \right)\,TeV.
\end{equation}
The same UHE electrons will radiate less efficiently with extra-
galactic magnetic field ($B_G \simeq 10^{-9} \,G $)  leading also
to  $27.2$ GeV photons direct peak.
   The spectrum of these photons is characterized by a law $dN
/dE dt \sim E^{-(\alpha + 1)/2} \sim E^{-1.25}$ where $\alpha$ is
the power law of the electron spectrum, and it is showed in
Fig.\ref{fig:fig2}-\ref{fig:fig6} above. As regards the prompt
electrons at higher energy ($E_e \simeq 10^{21}\, eV$), in
particular in the t-channels, their interactions with the
extra-galactic field first and galactic magnetic fields later are
sources of another kind of synchrotron emission around  energies
$E^{sync}_{\gamma}$ of tens PeV:

\begin{equation}\label{2}
 % E^{sync}_{\gamma}
 \sim
  6.8 \cdot 10^{13} \left( \frac{E_e}{10^{21}\,eV} \right)^2
  \left( \frac{m_{\nu}}{0.4 \, eV} \right)^{-2} \left( \frac{B}{nG}
  \right) \, eV
\end{equation}
\begin{equation}\label{2b}
 % E^{sync}_{\gamma}
 \sim
  6.8 \cdot 10^{16} \left( \frac{E_e}{10^{21}\,eV} \right)^2
  \left( \frac{m_{\nu}}{0.4 \, eV} \right)^{-2} \left( \frac{B}{\mu G}
  \right) \, eV
\end{equation}
The corresponding rate of energy loss is \cite{Ka}
\begin{equation}\label{3}
\left( \frac{1}{E} \frac{dE}{dt} \right)^{-1} = 3.8 \times \left(
\frac{E}{10^{21}} \right)^{-1} \left( \frac{B}{10^{-9} G}
\right)^{-2} \, kpc.
\end{equation}
For the first case the interaction length is few Kpcs while in the
second one in few days light flight (see
Fig\ref{fig:fig7}-Fig.\ref{fig:fig8}). Again one has the same
power law characteristic of a synchrotron spectrum with index
$E^{-(\alpha + 1 / 2)} \sim E^{-1.25}$.
 Photons at $10^{16} \div 10^{17}$ eV scatter onto
low-energy photons from isotropic cosmic background ($\gamma + BBR
\rightarrow e^+ e^-$) converting their energy in electron pairs.
 The expression for the pair production cross-section is:
\begin{equation} \sigma (s) = \frac{1}{2} \pi r_0^2 (1 - v^2) [
(3 - v^4) \ln \frac{1 + v}{1 - v} - 2 v (2 - v^2) ]
\end{equation}
where $v = (1 - 4m_e^2 / s)^{1/2}$,  $s = 2 E_{\gamma} \epsilon (1
- \cos \theta)$ is the square energy in the center of mass frame,
$\epsilon$ is the target photon energy, $r_0 $  is the classic
electron radius, with a peak cross section value at
\[ \frac{4}{137}\times \frac{3}{8\pi} \sigma_T \ln 183 = 1.2 \times
10^{-26} \,cm^2 \] Because the corresponding attenuation length
due to the interactions with the microwave background is around
ten kpc, the extension of the halo plays a fundamental role in
order to make this mechanism efficient. As it is shown in
 the contribution to  gamma signals of tens PeV by Z (or
W) hadronic decay, could be compatible with actual experimental
limits fixed by CASA-MIA detector on such a range of energies.
Considering a halo extension $l_{halo} \gtrsim 100 kpc$, the
secondary electron pair creation becomes efficient, leading to a
suppression of the signal of tens PeV . So electrons at $E_e \sim
3.5 \cdot 10^{16} \,eV$ loose again energy through additional
synchrotron radiation\cite{Ka}, with maximum $E_{\gamma}^{sync}$
around
\begin{equation}\label{3b}
  \sim 79 \left( \frac{E_e}{10^{21}
  \,eV} \right)^4 \left( \frac{m_{\nu}}{0.4 \, eV} \right)^{-4}
   \left( \frac{B}{\mu G} \right)^3 \, MeV.
\end{equation}

Since the relevant signal piles up at TeV it is not able to
pollute significantly the MeV-GeV region.

%%%%%%%%%%%%%%%%%%%%%%%%%%%%%%%%%%%%%%%%%%%%%%%%%%%%%%%%%%%%%%%%%%%%%%%%%%%%%%%5
Gamma rays with energies up to 20 TeV have been observed by
terrestrial detector only from nearby sources like Mrk 501 (z =
0.033) or very recently from MrK 421. This is puzzling because the
extra-galactic TeV spectrum should be, in principle, significantly
suppressed by the $\gamma$-rays interactions with the
extra-galactic Infrared background, leading to electron pair
production and TeV cut-off. The recent calibration and
determination of the infrared background by DIRBE and FIRAS on
COBE have inferred severe constrains on TeV propagation. Indeed,
as noticed by Kifune \cite{Kifune}, and Protheroe and
Meyer\cite{Meyer} we may face a severe infrared background - TeV
gamma ray crisis. This crisis imply a distance cut-off,
incidentally, comparable to the GZK one. Let us remind also an
additional evidence for IR-TeV cut-off is related to the possible
discover of tens of TeV counterparts of BATSE $GRB970417$,
observed by Milagrito\cite{Milagrito}, being most GRBs very
possibly at cosmic edges, at distances well above the IR-TeV
cut-off ones. In this scenario it is also important to remind the
possibilities that the Fly's Eye event has been correlated to TeV
pile up events in HEGRA \cite{Horns}. The very recent report
(private communication 2001) of the absence of the signal few
years  later at HEGRA may be still consistent with a bounded
Z-Showering volume and a limited UHE TeV tail activity.
 To solve the IR-TeV cut-off one may alternatively invoke unbelievable extreme hard intrinsic
spectra or exotic explanation as gamma ray superposition of
photons or sacrilegious  Lorentz invariance violation.
%%%%%%%%%%%%%%%   Figure 7       %%%%%%%%%%%%%%%%%   FFFFFFFFFFFFFFFFFFFFFFFFFFFFFFFFFFFFFF
%%%%%%%%%%%%%%%   Figure 7END   %%%%%%%%%%%%%%%%%   FFFFFFFFFFFFFFFFFFFFFFFFFFFFFFFFFFFFFF

%%%%%%%%%%%%%%%   Figure 8       %%%%%%%%%%%%%%%%%   FFFFFFFFFFFFFFFFFFFFFFFFFFFFFFFFFFFFFF
%%%%%%%%%%%%%%%   Figure 8END   %%%%%%%%%%%%%%%%%   FFFFFFFFFFFFFFFFFFFFFFFFFFFFFFFFFFFFFF

%%%%%%%%%%%%%%%%%%%%%%%%%%%%%%%%%%%%%%%%Tokio%%%%%%%
%%%%%%%%%%%%%%%%%%%%%%%%%%%%%%%%%%%%%
%%%%%%%%%%%%%%%%%%%%%%%%%%%%%%%%%%%%
Let us remind that  UHE neutrinos are un-effected by magnetic
fields and by BBR screening; they may reach us from far cosmic
edges without significant absorption. The UHE Z-shower in its
ultra-high energy nucleonic secondary component may be just the
observed final UHECR event on Earth. This possibility has been
reinforced by very recent correlations (doublets and triplets
events) between UHECR directions with brightest Blazars sources at
cosmic distances (redshift $\geq 0.1$) quite beyond ($\geq 300
Mpc$) any allowed GZK cut-off \cite{Gorbunov Tinyakov Tkachev
Troitsky}, \cite{AGASA 1999}, \cite{Takeda et all}. Therefore
there might be a role for GZK neutrino fluxes, either at very high
fluence as primary in the Z-Showering scenario or, at least, as
(but at lower intensities) necessary secondaries of all those
UHECR primary absorbed in cosmic BBR radiation fields by GZK cut
of. Naturally other solutions as topological defects or primordial
relics decay may play a role as a source of UHECR, but the
observed clustering \cite{AGASA 1999},
\cite{Tinyakov-Tkachev2001}, \cite{Takeda et all}, seems to favor
compact sources possibly overlapping far BL Lac sources \cite{
Gorbunov Tinyakov Tkachev Troitsky}. The most recent evidence for
self-correlations clustering at $10^{19}$, $2\cdot10^{19}$,
$4\cdot10^{19}$ eVs energies observed by AGASA (Teshima, ICRR26
Hamburg presentation 2001) maybe a first reflection of UHECR
Z-Showering secondaries: $p, \bar{p}, n, \bar{n}$ \cite{Fargion et
all. 2001b}. A very recent solution
    beyond the Standard Model (but within Super-Symmetry) consider
    Ultra High Energy Gluinos as the neutral particle bearing UHE
    signals interacting nearly as a hadron in the terrestrial atmosphere
\cite{Berezinsky 2002};
    this solution has a narrow window  for
    gluino masses allowable (and serious problems in production
    bounds), but it is an alternative that deserves attention. To
    conclude this brief Z-Shower model survey one finally needs to
    scrutiny the UHE $\nu$ astronomy and to test the GZK solution
    within Z-Showering Models by any independent search on Earth for
    such UHE neutrinos traces above PeV reaching even EeV-ZeV
    extreme energies.
%%%%%%%%%%%%%%%%%%%%%%%%%%%%%%%%%%%%%%%%%%%%%%%%%%%%%%%
%%%%%%Tokio%%

%\newpage

%%%%%%%%%%%%%%%%%%%% section 5 %%%%%%%%%%%%%%%%%%%%%%%%%

 \section{UHE $\nu$ Astronomy by $\tau$ air-shower: the ideal amplifier}

\begin{figure}
\centering
\includegraphics[width=13cm]{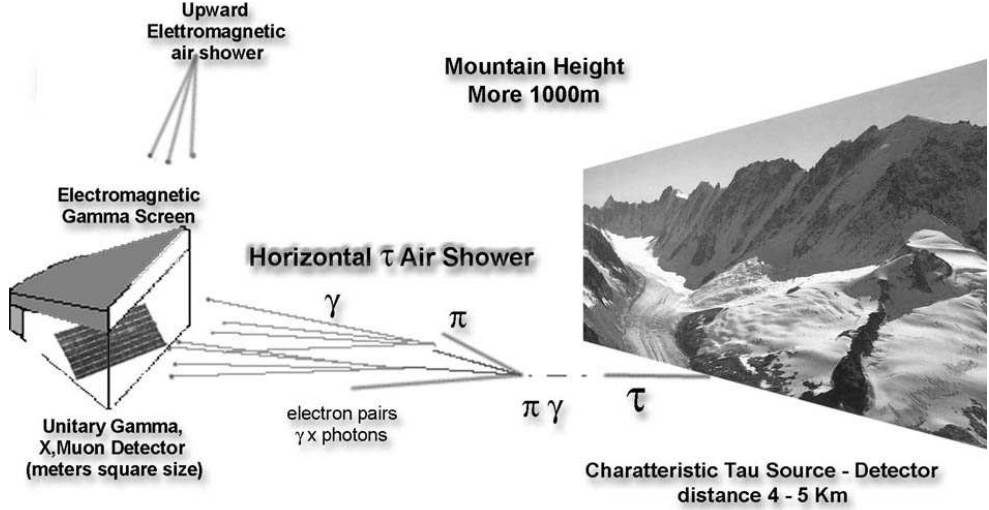}
\vspace{0.5cm} \caption {The Horizontal Tau on front of a Mountain
Chain; different interaction lengths will reflect in different
events rate  \cite{Fargion et all 1999}, \cite{Fargion
2000-2002}.} \label{fig:fig9}
\end{figure}

\begin{figure}
\centering
\includegraphics[width=13cm]{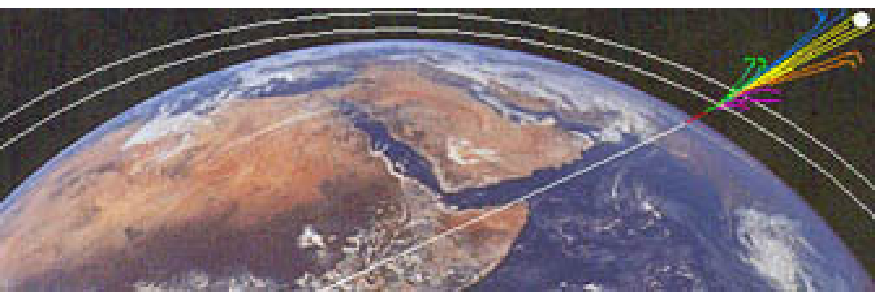}

\vspace{0.5cm} \caption {The Upward Tau air-shower UPTAUs and its
open fan-like jets due to geo-magnetic bending at high quota. The
gamma Shower is pointing to an orbital satellite detector as old
GRO-BATSE or very recent Integral \cite{Fargion 2000-2002},
\cite{Fargion2001a},  \cite{Fargion2001b}.} \label{fig:fig10}
%\end{figure}
%
%%%%%%%%%%%%%%%%%%%% end fig. 5 %%%%%%%%%%%%%%%%%%%%%%%%%
%
%%%%%%%%%%%%%%%%%%%% fig. 6 %%%%%%%%%%%%%%%%%%%%%%%%%
%\begin{figure}
\vspace{0.5cm}
\centering
\includegraphics[width=13cm]{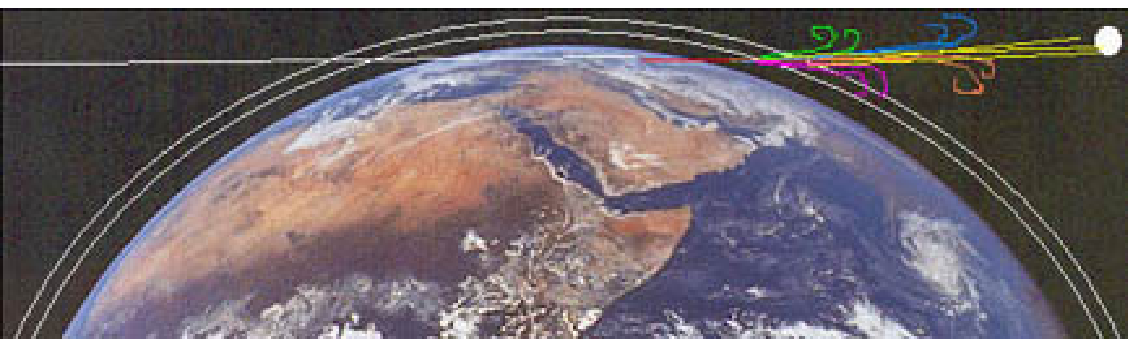}
\vspace{0.5cm} \caption {The Horizontal-Shower HORTAU and its open
fan-like jets due to geo-magnetic bending at high quota. The gamma
Shower is pointing to an orbital satellite detector as old
GRO-BATSE or very recent Integral just at the horizons
\cite{Fargion 2000-2002}, \cite{Fargion2001a},
\cite{Fargion2001b}.} \label{fig:fig11}
\end{figure}

\begin{figure}
\centering
\includegraphics[width=13cm]{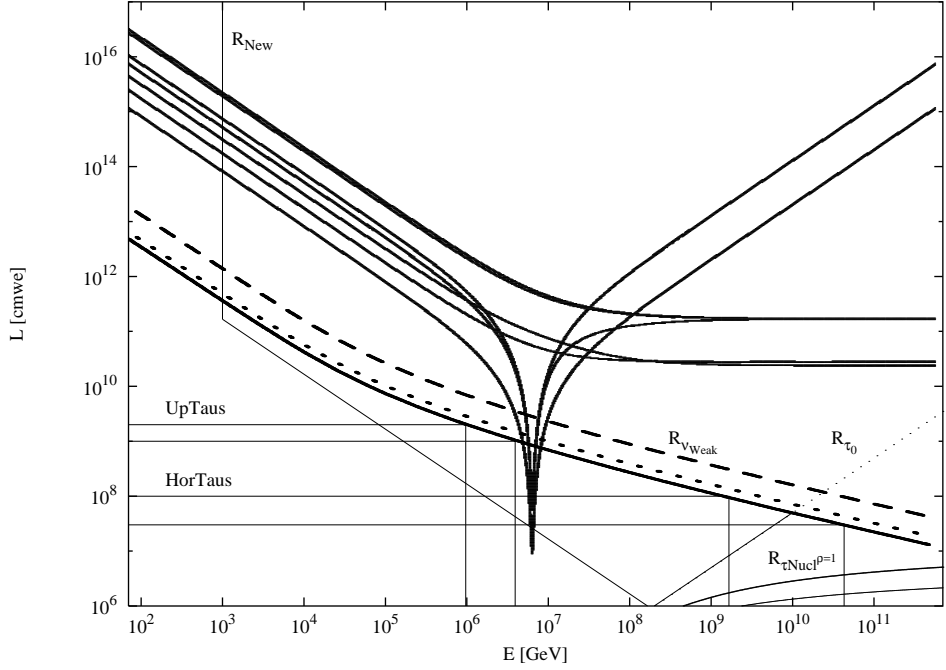}
\caption {Different interaction lengths for Ultra High Energy
Neutrinos, reflecting in differences of event rates either for
Horizontal Shower from Mountain Chains as well as from Upward and
Horizontal ones from Earth Crust. A severe decrease of  the
neutrino interaction length, $R_{New}$, is due to a   New TeV
gravity interaction. This interaction increases by  $2-3$ order of
magnitude the neutrino birth probability (in respect to known
interactions), leading to expected tens of thousand events a year
(in respect to a hundred a year). Here it has been  considered a
ten-km length Array detector on front of a Mountain Chain, see
Fig.\ref{fig:fig9},  assuming a cosmic neutrino fluence of $10^3
eV cm^{-2}s^{-1}$ \cite{Fargion 2000-2002}, \cite{Yoshida 2001}.}
\label{fig:fig12}
\end{figure}

    Recently
\cite{Fargion et all 1999}, \cite{Fargion 2000-2002}
     a new competitive UHE $\nu$ detection has been proposed.
    It is based on ultra high energy $\nu_{\tau}$
    interaction in matter and its consequent secondary $\tau$ decay
(different $\tau$ decay channels are given in
Tab.\ref{tab:table3})
    in flight while escaping from the rock (Mountain Chains, Earth
    Crust) or water (Sea,Ice)  in air leading to UPward or HORizontal
    TAU air-showers (UPTAUs and
    HORTAUs),
\cite{Fargion2001a}, \cite{Fargion2001b}.
    In a pictorial
    way one may compare the UPTAUs and HORTAUs as the double bang
    processes expected in $km^3$ ice-water volumes \cite{Learned
    Pakvasa 1995} : the double bang is due first to the UHE
$\nu_{\tau}$ interaction in matter and secondly by its consequent
$\tau$ decay in flight. Here we consider  a (hidden) UHE $\nu$-N
Bang $in$ (the rock-water within a mountain or the Earth Crust)
and a $\tau$ bang $out$ in air, whose shower is better observable
at high altitudes. A similar muon double bang amplifier is not
really occurring because of the extremely large decay length of
ultra-relativistic ($\gtrsim 10^{13}$ eV) muons. The main power of
the UPTAUs and HORTAUs detection is the huge amplification of the
UHE neutrino signal, which may deliver almost all its energy in
numerous secondaries traces (Cherenkov light, gamma, X photons,
electron pairs, collimated muon bundles) in a wider cone volume.
Indeed the multiplicity in $\tau$ air-showers secondary particles,
$N_{opt} \simeq 10^{12} (E_{\tau} / PeV)$, $ N_{\gamma} (<
E_{\gamma} > \sim  10 \, MeV ) \simeq 10^8 (E_{\tau} / PeV) $ ,
$N_{e^- e^+} \simeq 2 \cdot 10^7 ( E_{\tau}/PeV) $ , $N_{\mu}
\simeq 3 \cdot 10^5 (E_{\tau}/PeV)^{0.85}$ facilitates the
UPTAUs-HORTAUs discovery.
 These HORTAUs, also named Skimming neutrinos \cite{Feng et al 2002},
 studied also in peculiar approximation in the frame of AUGER
 experiment, in proximity of Ande Mountain Chains (see Fig.\ref{fig:fig9})
    \cite{Fargion et all 1999},
\cite{Bertou et all 2002},
    may be also originated on front of large volcano
\cite{Fargion et all 1999}, \cite{Hou Huang 2002}
    either by $\nu_{\tau}N$, $ \bar\nu_{\tau}N$ interactions as
well as by $ \bar\nu_{e} e \rightarrow W^{-} \rightarrow
\bar\nu_{\tau} \tau$. Also UHE $\nu_{\tau}N$, $ \bar\nu_{\tau}N$
at EeV may be present  in rare AGASA Horizontal Shower (one single
definitive event observed was used as un upper bound on EeV
neutrino flux \cite{Yoshida 2001}) facing Mountain Chain around
the Akeno Array (see Fig.\ref{fig:fig9}). This new UHE
$\nu_{\tau}$ detection is mainly based on the oscillated UHE
neutrino $\nu_{\tau}$ originated by more common astrophysical
$\nu_{\mu}$, secondaries of pion-muon decay at PeV-EeV-GZK
energies. The existence of these oscillations can be provided  by
Super Kamiokande evidences for flavor mixing within GeVs
atmospheric neutrino data \cite{Fukuda:1998mi} as well as by
recent evidences of complete solar neutrino mixing observed by
SNO detector \cite{Bellido et all 2001} and by most recent claims
on the disappearance of neutrino fluence from nearby nuclear
reactors \cite{Kamland2002}.
    Let us remind that HORTAUs (see Fig.\ref{fig:fig9})
    from mountain chains must nevertheless occur, even
for no flavor mixing, as being inevitable $\bar\nu_{e}$
secondaries of common pion-muon decay chains ($\pi^{-}\rightarrow
\mu^{-}+\bar\nu_{\mu}\rightarrow e^{-}+\bar\nu_{e}$) near the
astrophysical sources at PeV energies. These PeV $\bar\nu_{e}$ are
mostly absorbed by the Earth and are only rarely arising as
UPTAUs (see Fig.\ref{fig:fig10}-\ref{fig:fig11} and cross-section
in Fig.\ref{fig:fig12}). Their Glashow resonant interaction allow
them to be observed as HORTAUs only within very narrow and nearby
crown edges at horizons (not to be discussed here).

%%%%%%%%%%%%%%%%%%%% fig. 4 %%%%%%%%%%%%%%%%%%%%%%%%%

%%%%%%%%%%%%%%%%%%%% end fig. 4 %%%%%%%%%%%%%%%%%%%%%%%%%

%%%%%%%%%%%%%%%%%%%% fig. 5 %%%%%%%%%%%%%%%%%%%%%%%%%

%%%%%%%%%%%%%%%%%%%% end fig. 6 %%%%%%%%%%%%%%%%%%%%%%%%%
 At wider energies
windows ($10^{14}eV- 10^{20}eV$) only neutrino $\nu_{\tau}$,
$\bar{\nu}_{\tau}$ play a key role in UPTAUs and HORTAUs. These
Showers might be easily detectable looking downward the Earth's
surface from mountains, planes, balloons or satellites observer.
 Here the Earth itself acts as a "big mountain" or a wide beam dump target (see
Fig.\ref{fig:fig10}-\ref{fig:fig11}). The present upward $\tau$ at
horizons should not be confused with an independent and well
known, complementary (but rarer) Horizontal Tau air-shower
originated inside the same terrestrial atmosphere: we may refer
to it as to the Atmospheric Horizontal Tau air-shower. These rare
events are responsible for very rare double bang in air. Their
probability to occur, as derived  in  detail in next paragraph
(summarized in the last column of Tab.\ref{tab:table4} below,
labeled as R, being the ratio between the events caused  by
HORTAUs  in the ground and such events occurring in the air) is
more than two order magnitude below the event rate of HORTAUs.
The same UPTAUs (originated in Earth Crust) have a less
competitive upward showering due to $\nu_{e}$ $\bar\nu_{e}$
interactions within atmosphere, showering in thin upward air
layers \cite{Berezinsky 1990}: this atmospheric Upward Tau
presence is a very small additional contribute, because rocks are
more than $3000$ times denser than the air, see
Tab.\ref{tab:table4}. Therefore at different heights we need to
estimate (See for detail next paragraph) the UPTAUs and HORTAUs
event rate occurring along the thin terrestrial crust below the
observer, keeping care of their correlated  variables: from a
very complex sequence of functions we shall be able to define and
evaluate the effective HORTAUs volumes keeping care of the thin
shower beaming angle, atmosphere opacity and detector thresholds.
At the end of the study, assuming any given neutrino flux, one
might be easily able to estimate at each height $h_{1}$ the
expected event rate and the ideal detector size and sensibility
for most detection techniques (Cherenkov, photo-luminescent,
gamma rays, X-ray, muon bundles). The Upcoming Tau air-showers
and Horizontal ones may be already recorded as Terrestrial Gamma
Flashes (see TGF Recorded Data in Tab.\ref{fig:fig13} ) as shown
by their partial Galactic signature shown in Fig.\ref{fig:fig14}
(over EGRET celestial background) and in Fig.\ref{fig:fig15}(over
EeV anisotropy found by AGASA).

%%%%%%%%%%%%%%%%%%%% fig. 7 %%%%%%%%%%%%%%%%%%%%%%%%%

%%%%%%%%%%%%%%%%%%%% end fig. 7 %%%%%%%%%%%%%%%%%%%%%%%%%

%%%%%%%%%%%%%%%%%%%%end section 5 %%%%%%%%%%%%%%%%%%%%%%%%%

%%%%%%%%%%%%%%%%%%%% section 6 %%%%%%%%%%%%%%%%%%%%%%%%%

\subsection{Ultra High Energy $\nu_{\tau}$ astronomy by Upward  and Horizontal
$\tau$,UPTAUs-HORTAUs, detection}

The $\tau$ airshowers are observable at different height $h_{1}$
 leading to different underneath observable terrestrial
areas and crust volumes. HORTAUs in deep valley are also related
to the peculiar geographical morphology and composition
\cite{Fargion 2000-2002}. We remind in this case the very
important role of UHE  $ \bar\nu_{e}e \rightarrow
W^{-}\rightarrow \bar\nu_{\tau}\tau^{-} $ channels which may be
well observable even in absence of any $\nu_{\tau}$, $
\bar\nu_{\tau}$ UHE sources or any neutrino flavor mixing: its
Glashow peak resonance makes these neutrinos unable to cross all
the Earth across but it may be observable beyond mountain chain
\cite{Fargion 2000-2002}; while testing $\tau$ air-showers beyond
a mountain chain one must consider the possible amplification of
the signal because of possible New TeV Physics (see cross-section
in Fig.\ref{fig:fig12})\cite{Fargion 2000-2002}. In the following
we shall consider in general the main $\nu_{\tau}N$, $
\bar\nu_{\tau}N$ nuclear interactions on the Earth crust. It
should be kept in mind also that UPTAUs and in particular HORTAUs
are showering at very low densities and their geometrical
escaping opening angle from Earth (here assumed at far distances
$\theta\sim 1^o$ for rock and $\theta\sim 3^o$ for water) is not
in general conical (like common down-ward showers) but their
ending tails are more shaped in thin fan-like twin Jets . These
showers will be opened in a characteristic twin fan-jet ovals
looking like the $8$-shape, bent and split in two thin elliptical
beams by the geo-magnetic fields. These fan shapes are not widely
open by the Terrestrial Magnetic Fields while along the
North-South magnetic field lines. These UPTAUs-HORTAUs duration
times are also much longer than common down-ward showers because
their showering occurs at much lower air density and they are
more extended: from microsecond in the case of  UPTAUs  reaching
from mountains to millisecond in the case of  UPTAUs and HORTAUs
originated on Earth and observed from satellites. Indeed the GRO
observed upcoming Terrestrial Gamma Flashes which are possibly
correlated with the UPTAUs \cite{Fargion 2000-2002}; these events
show the expected millisecond duration times. In order to
estimate the rate and the fluence of UPTAUs and HORTAUs one has
to estimate the observable crown terrestrial crust mass, facing a
complex chain of questions, leading for each height $h_{1}$, to
an effective observable surface and volume from where UPTAUs and
HORTAUs might be originated. From this effective volume it is
easy to estimate the observable rates, assuming a given incoming
UHE $\nu$ flux model for galactic or extragalactic sources. Here
we shall only refer (see Appendices A-B) to the Masses estimate
unrelated to any UHE $\nu$ flux models.

\subsection{Tau air showers to discover UHE $\nu$: UHE $\tau$ decay
channels} The $\tau$ air-shower morphology would reflect the rich
and structured behaviors of $\tau$ decay modes. Indeed let us
label the main "eight finger" UHE decay channels (hadronic or
electromagnetic) and the consequent air-shower imprint, with
corresponding probability ratio as shown in the Tab.\ref{tab:table3}. %\\

%%%%%%%%%%%%%%%%%%% end fig. 5 %%%%%%%%%%%%%%%%%%%%%%%%%
%\newline \\
 This complex air-shower modes would exhibit different
interaction lengths in air at 1 atmosphere ($\sim 300$ meters for
electromagnetic interaction length, 500 meters for hadronic
interaction length or, more precisely 800 meters for $tau$ pions
secondaries). The consequent air-shower statistics will reflect
these imprint multi-channel modes also in its energy and
structured time arrival to detectors beyond mountains, on planes,
balloons or satellites. Possibly these channels may also be
reflected in observed terrestrial Gamma Flashes.
\begin{table}[h]

\begin{center}\footnotesize
\begin{tabular}{cccc}

 \hline\\
 {\bfseries Decay} & {\bfseries Secondaries} & {\bfseries Probability} &
{\bfseries Air-shower} \\
& & & \\
 \hline\\

& & & \\

  $\tau \rightarrow \mu^- \bar{\nu_{\mu}} \nu_{\tau}$ & $\mu^-$ & $\sim 17.4 \%$ & Unobservable \\
  $\tau \rightarrow e^- \bar{\nu_{e}} \nu_{\tau}$ & $e^-$ & $\sim 17.8 \%$
  &1 Electromagnetic \\
  $\tau \rightarrow \pi^- \nu_{\tau}$ & $\pi^-$ & $\sim 11.8 \%$ & 1
  Hadronic \\
  $\tau \rightarrow \pi^- \pi^0 \nu_{\tau}$ & $\pi^-, \, \pi^0 \rightarrow 2 \gamma$
  & $\sim 25.8 \%$ & 1 Hadronic, 2 Electromagnetic \\
  $\tau \rightarrow \pi^- 2\pi^0 \nu_{\tau}$ & $\pi^-, \, 2\pi^0 \rightarrow 4 \gamma$
  & $\sim 10.79 \%$ & 1 Hadronic, 4 Electromagnetic \\
  $\tau \rightarrow \pi^- 3\pi^0 \nu_{\tau}$ & $\pi^-, \, 3\pi^0 \rightarrow 6 \gamma$
  & $\sim 1.23 \%$ & 1 Hadronic, 6 Electromagnetic \\
  $\tau \rightarrow \pi^- \pi^- \pi^+ \nu_{\tau}$ & $2\pi^-, \, \pi^+$
  & $\sim 10 \%$ & 3 Hadronic \\
  $\tau \rightarrow \pi^- \pi^+ \pi^- \pi^0 \nu_{\tau}$ & $2\pi^-, \, \pi^+ , \, \pi^0 \rightarrow
  2 \gamma$ & $\sim 5.18 \%$ & 3 Hadronic, 2 Electromagnetic \\
& & & \\
 \hline\\

\end{tabular}
\end{center}
\caption{Tau decay channels, their Shower tail nature in each
case and their corresponding probabilities to occur. The
probabilities for pion secondaries take into account the
contribution of  $K$ mesons originated from $\tau$ decay.
\cite{Fargion 2000-2002}} \label{tab:table3}
\end{table}

%%%

\section{UPTAUs  and HORTAUs connection with Terrestrial Gamma Flash}

 The steps linking simple terrestrial spherical
geometry and its different geological composition and high energy
neutrino physics and UHE $\tau$ interactions leading to Tau
air-showers are not straightforward. The same UHE $\tau$ decay in
flight and its air-showering physics at various  quota (and air
density) behave differently. Detector physics threshold and
background noises, signal rates should also been kept in mind
\cite{Fargion 2000-2002}. Let us remind that a few  Tau decay
modes in muon channel  do not  lead to any observable air-shower.
%%%%%%%%%%%%%%%%%%%%%%%%%%%%%%%%%%%%%%%%%%%%%%%%%%%%%%%%%%%
\subsection{UPTAUs and HORTAUs  toward satellite : TGF events in
BATSE data} Let us estimate the UPTAUs (and HORTAUs) possible role
to trigger a Terrestrial Gamma Flash (TGF). These  short
(millisecond)  $\gamma$ ray burst , upcoming from the Earth,  have
been rarely ($78$ events in $10$ year of records) been observed
by the most sensible $\gamma$ experiment: BATSE in GRO satellite
($1991-2000$). Their interpretation have been first related to
upward lightening.The present rate of observed TGF at best (low
threshold and hard channel trigger set up) is much lower ($\sim$
factor ten) then predicted one \cite{Fargion 2000-2002} for an
incoming neutrino flux $\simeq 10^3 eV cm^{-2}s^{-1}sr^{-1}$: it
may be well possible that the usual BATSE threshold trigger is
suppressing and hiding this rate; otherwise tens of PeV UHE
$\nu_\tau$  are the TGF events source at BATSE sensibility edges.
The other possibility is that the real neutrino fluence is only
$\simeq 2\cdot 10^2 eV cm^{-2}s^{-1}sr^{-1}$. Nevertheless  a
small (factor $3 \div 5$) suppression, may reduce the $N_{ev} $
to the observed TGF rate on an expected higher  $\simeq 10^3 eV
cm^{-2}s^{-1}sr^{-1}$ fluence.
%%%%%%%%%%%%%%%%%%%%%%%%%%%%%%%%
\begin{table}
\centering
\includegraphics[width=13cm]{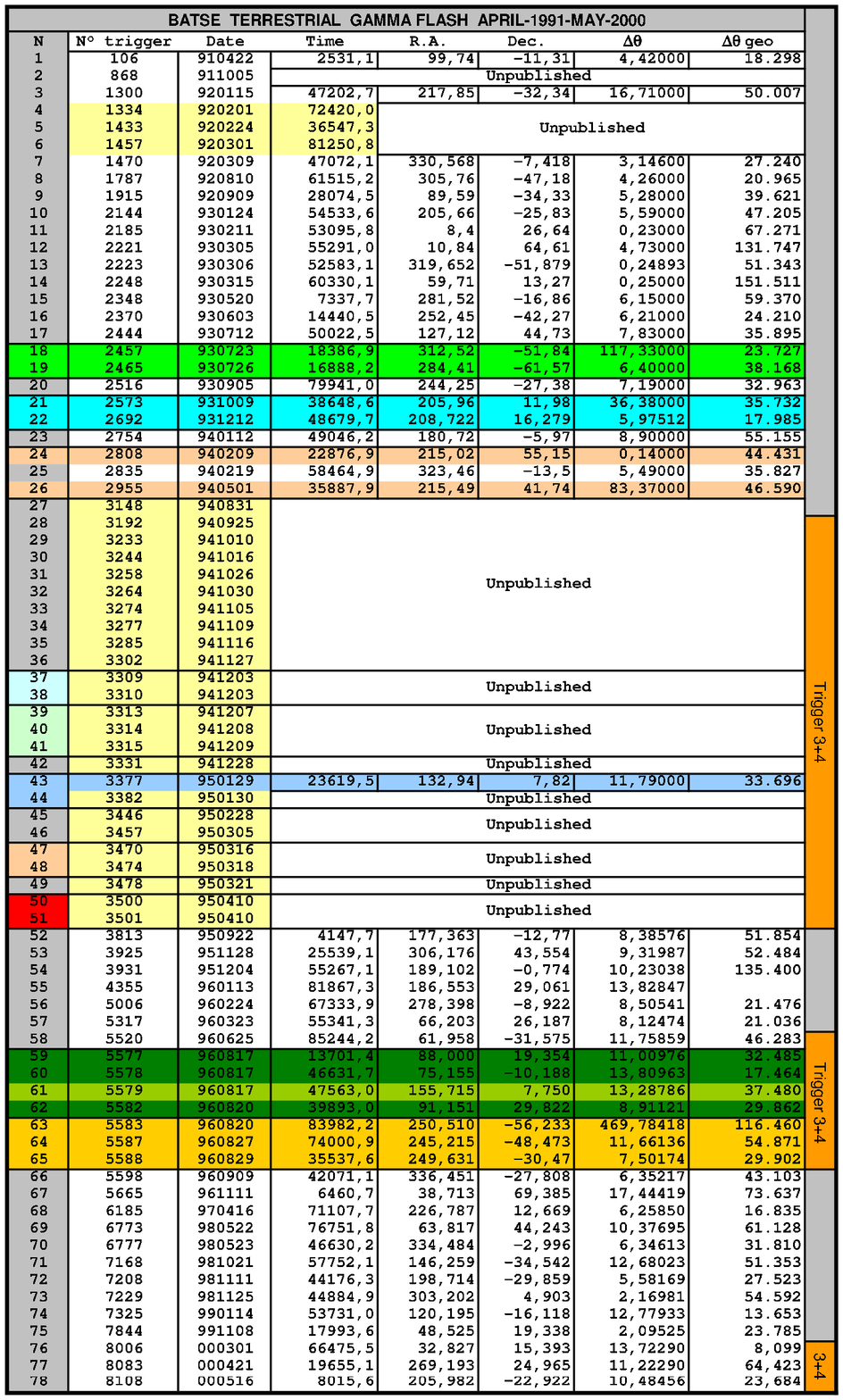}
 \caption {All BATSE terrestrial gamma burst
   data $1991-2000$. The colored
    TGF events associate common arrival directions (as Galactic
   Center, A.G.Center) associated also in time clustering; the
    date,time, celestial coordinate, error bar, and TGF-Earth Center
  angle are listed below; Hard Trigger set up Trigger periods
   (channel 3+4) have a colored orange side label.
   They mark a higher rate TGF activity  visibly correlated  (with two different   plateau in
      corresponding to higher TGF acceptance).  }\label{fig:fig13}
   \end{table}

\begin{figure}
\centering
\includegraphics[width=16cm]{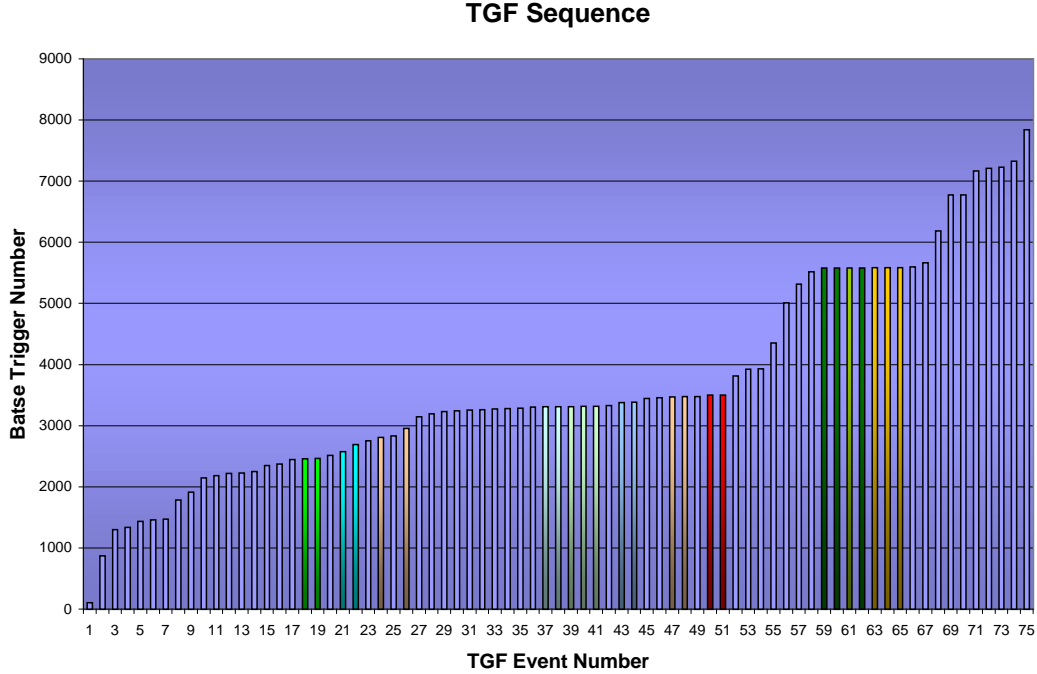}
 \caption {All BATSE terrestrial gamma burst
   data $1991-1999$. The colored
    TGF events associate both common arrival directions (as Galactic
   Center, A.G.Center)as well as their time clustering; the
    date,time, celestial coordinate, error bar, and TGF-Earth Center
  angle are listed in Table above;  Trigger periods
   (channel 3+4) have a higher and more prolific TGF activity.
   They mark  two different   plateau in
      corresponding to higher TGF acceptance. Last  Trigger 8006,8083,8108 on last year 2000, not included
      in the present table, confirmed and didn't change
   the general result above.}\label{fig:fig13b}
   \end{figure}

%%%%%%%%%%%%%%%%%%%%

Moreover HORTAUs at tens EeV \cite{Fargion 2000-2002} may also
lead to rare upward events at a rate comparable to TGF events.
The last secondaries from a $\tau$ air shower from  $ 3PeV $
neutrinos are mainly hard ($10^5 eV $ or above) bremsstrahlung
gamma photons produced by electron pairs whose approximated
number flux is comparable to
\begin{equation}
  N_{\gamma} \simeq \frac{E}{<E_{\gamma}>} \sim 3 \cdot
  10^{10} \ffrac{E}{3\,PeV}
\end{equation}
The atmosphere opacity may reduce the final value at least by 1/3:
$N_{\gamma} \sim 10^{10}$. The expected $X$ - $\gamma$ flux at
large 500 Km distances, is diluted even within a beamed angle
$\Delta \Omega \simeq 2 \cdot 10^{-5}$, leading to nearly
$\Phi_{\gamma}\sim 10^{-2}$ph/cm$^2$. The characteristic $\gamma$
burst duration is roughly defined by L/c $\sim$ few milliseconds
in agreement with the observed TGF events (See Appendix A). The
consequent TGF $\gamma$ burst number flux on BATSE is estimated
to be $\Phi_{\gamma} S \sim 10^{2}$ events, what is just
comparable with  observed TGFs  fluence,(see Appendix B). The
HORTAUs while being more energetic ($\sim 10^3$ factor for same
$\nu$ energy fluence) are rarer by the same factor (for equal
$\nu$ energy fluence) and by smaller arrival angle (two order of
magnitude) as well as diluted by the  longer tangential distances
( factor $\sim 25$); however most of these suppressions are well
recovered by much higher and efficient $\nu$ cross-sections at
GZK energies, longer $\tau$ tracks, possible rich primary spectra
and higher HORTAUs $\Phi_{\gamma}$ fluence, making them
complementary or even comparable to UPTAUs signals. The
bremsstrahlung spectra are hard, as the observed TGF ones. The
possible air shower time structure may reflect the different
eight $\tau$ decay channels (mainly involving hadronic and/or
electromagnetic decay products). The complex interplay between
UHE $\nu$ interaction with nuclear matter overimposed on
$\bar{\nu_e} e$ interactions is shown in (Fig.\ref{fig:fig12}).
The extremely narrow energy window where $\bar{\nu_e} e$ rate is
comparable to $\nu_{\tau} N$ while being transparent to Earth
makes UPTAUs-HORTAUs-TGF connection unrelated to $\bar{\nu_e} e$
resonant $W^-$ events possible only in HORTAUs beyond a mountain.
The characteristic interaction regions responsible for UPTAUs and
HORTAUs are within a narrow energy band shown in
(Fig.\ref{fig:fig12}). Peculiar $\nu_{\tau} N$ interaction
(Fig.\ref{fig:fig12}) departing from parton model, would lead to a
less restrictive UHE $\nu_{\tau}$ - Earth opacity, and a more
abundant vertical UPTAUs-TGF event rate at higher energies; the
TGF data do not support such a large flux variability and
therefore it might moderately favor the narrow energy window (PeV
- few tens PeV) constrained by parton model
 or the EeV energy window for HORTAUs. Indeed the TGF
data, collected by NASA BATSE archive and described in Tab.3 are
located in celestial map
 with their corresponding error boxes .  They are better readable,
 after few error bar calibration, in a galactic map over the diffused
  GeV - EGRET gamma background signal. One notes the surprising clustering
  of TGF  sources in the galactic plane center at maximal EGRET
  fluence; their correlations with important known TeV   are displayed.
   Let us remark that the last discovered TeV source ,  1ES1426+428,
associated with BL Lac object at redshift $z= 0.129$, do find also
a correlated event in the TGF BATSE Trigger, 2955, source, making
more plausible the TGF astrophysical nature than any random
terrestrial lightening origin (Fig.\ref{fig:fig15}). One should
foresee that UPTAUs must  be correlated to geological sites of
higher terrestrial densities,(rock over sea). The HORTAUs
location birth is correlated with higher terrestrial crust
elevation, (Mountain chains and Earth Crust discontinuity )
because target matter has a more deep penetrating probability.
All these air-showers may be better opened (and observed) at the
places with highest terrestrial magnetic field. Additional
remarkable correlations occur with AGASA UHECR non-homogeneities
at EeV energy band as shown in Fig.\ref{fig:fig15}, as well as
with most COMPTEL $\gamma$ sources toward the l = $18^o$ in
galactic plane. Some important locations of known galactic and
extra-galactic source (as nearby QSRs 3C273 and 3C279)  are
displayed in Fig.\ref{fig:fig15} over the EeV AGASA map. Very
recent and rarest UHECR AGASA triplet clustering, near or above
GZK energies, pointing toward BL Lac $1ES 0806+524$, finds,
surprisingly, a corresponding TGF event, within its error box:
BATSE (Trigger 2444). Also two (among four) additional UHECR
AGASA doublets (2EG J0432+2910 and TEX 1428+370) are  well
correlated to TGF events (Trigger 5317,2955). The present
TGF-$\tau$ air-shower identification could not   be produced by
UHE $\bar{\nu}_e$ charged current  at ($E_{\bar{\nu_{e}}} = M^2_W
/ 2m_e = 6.3 \cdot 10^{15}$ eV); therefore it stands for the UHE
$\nu_{\tau} \bar{\nu_{\tau}}$ presence. Consequently it could be
the result of  flavor mixing $\nu_{\mu}\leftrightarrow
\nu_{\tau}$  from far PSRs or AGNs sources toward the Earth. The
TGF-$\tau$ air-shower connection may be soon verified and
reinforced  (or partially mystified) by the BATSE-GRO publishing
of 28 missing TGFs data (as well as future GLAST data):  we
foresee that BATSE-TGF hide additional directional imprint of UHE
$\nu_{\tau}$ sources, (maybe the missing Mrk 421 and Mrk 501
extragalactic sources).

%%%%%%%%%%%%%%%%%%%%  fig. 8 %%%%%%%%%%%%%%%%%%%%%%%%%

\begin{figure}
\centering
\includegraphics[width=13cm]{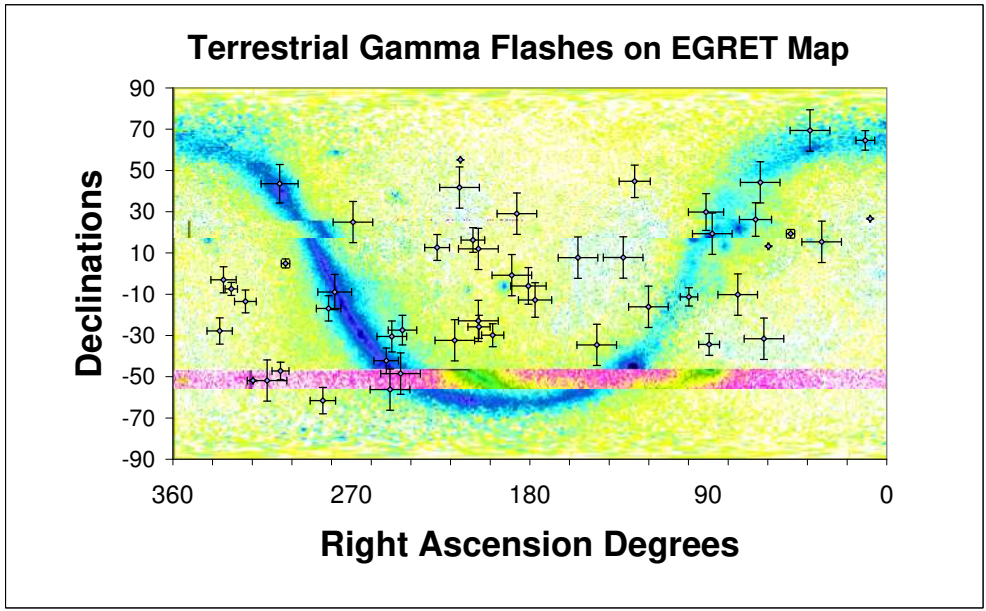}
\caption {The Terrestrial Gamma Flash arrival map over the EGRET
(hundred MeV-GeV) data in celestial coordinates. It manifests the
partial galactic signature and the clustering of repeater events
toward the Galactic Center. Also some relevant repeater events are
observed toward anti-galactic direction and to  well known
extragalactic sources (see next map and
    \cite{Fargion 2000-2002}).
}\label{fig:fig14}
\end{figure}

%%%%%%%%%%%%%%%%%%%% end fig. 8 %%%%%%%%%%%%%%%%%%%%%%%%%

%%%%%%%%%%%%%%%%%%%%  fig. 9 %%%%%%%%%%%%%%%%%%%%%%%%%

\begin{figure}
\centering
\includegraphics[scale=0.5]{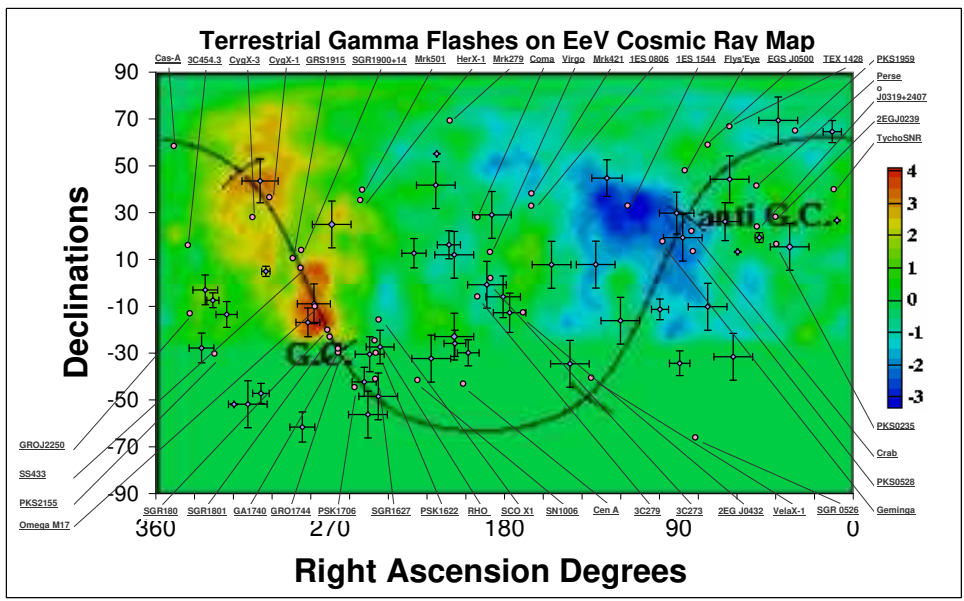}
\caption {The Terrestrial Gamma Flash arrival map over the EeV
anisotropic map  ($ 10^{18} eV$) data in celestial coordinates.
Some relevant X-$\gamma$-TeV sources are also shown in the same
region; see \cite{Fargion 2000-2002}. }\label{fig:fig15}
\end{figure}

%%%%%%%%%%%%%%%%%%%% end fig. 9 %%%%%%%%%%%%%%%%%%%%%%%%%

%%%%%%%%%%%%%%%%%%%% section 7  %%%%%%%%%%%%%%%%%%%%%%%%%

\section{Skin crown Earth volumes as a function
of  observation height $h$}

Below we define, list and estimate the sequence of the key
variables whose dependence (shown below or derived in Appendices)
leads to the desired HORTAUs volumes (useful to estimate the UHE
$\nu$ prediction rates) summarized in Tab.\ref{tab:table4}  and in
Conclusions. These Masses estimates are somehow only lower bounds
that ignore an additional contribution by more penetrating or
regenerated $\tau$ \cite{Halzen1998}.  Let us show the main
functions whose interdependence with the observer altitude leads
to an estimate of UPTAUs and HORTAUs equivalent detection
Surfaces, (See Fig.\ref{fig:fig16}-\ref{fig:fig17}), Volumes and
Masses (see Tab.\ref{tab:table4}).
\begin{enumerate}
  \item The horizontal distance $d_{h}$ at given height $h_{1}$
  toward the horizons (see Fig.\ref{fig:fig17}):

 % \begin{displaymath}
%d_{h} =  \sqrt{( R_{\oplus} + h_1)^2 - (R_{\oplus})^2}=
 % \end{displaymath}

 \begin{equation}
 d_{h}= \sqrt{( R_{\oplus} + h_1)^2 - (R_{\oplus})^2}= 113\sqrt{ \frac{h_1}{km} }\cdot {km}\sqrt{1+\frac{h_1}{2R_{\oplus}} }
 \end{equation}

The corresponding horizontal edge angle $\theta_{h}$ below the
horizons ($\pi{/2}$) is (see Fig.\ref{fig:fig17}):

\begin{equation}
 {\theta_{h} }={\arccos {\frac {R_{\oplus}}{( R_{\oplus} + h_1)}}}\simeq 1^o \sqrt{\frac {h_{1}}{km}}
\end{equation}

(All the approximations here and below hold for height
$h_{1}\ll{R_{\oplus}}$ )

%%%%%%%%%%%%%%%%%%%% fig. 10 %%%%%%%%%%%%%%%%%%%%%%%%%

\begin{figure}
\centering
\includegraphics[width=14cm]{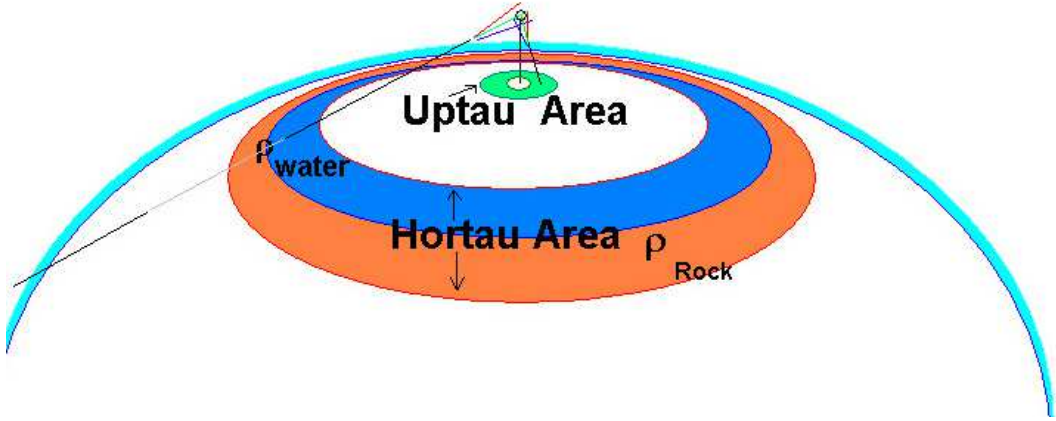}
\caption {The Upward Tau air-shower Ring or Crown Areas, labelled
UPTAUs and the Horizontal Tau air-shower Ring Area, labelled by
HORTAU, where the $\tau$ is showering and flashing toward an
observer at height $h_1$. The HORTAU Ring Areas are described both
for water and rock matter density.} \label{fig:fig16}
\end{figure}

%%%%%%%%%%%%%%%%%%%% end fig. 10 %%%%%%%%%%%%%%%%%%%%%%%%%

%%%%%%%%%%%%%%%%%%%% fig. 11 %%%%%%%%%%%%%%%%%%%%%%%%%

 \begin{figure}
\centering
\includegraphics[width=13cm]{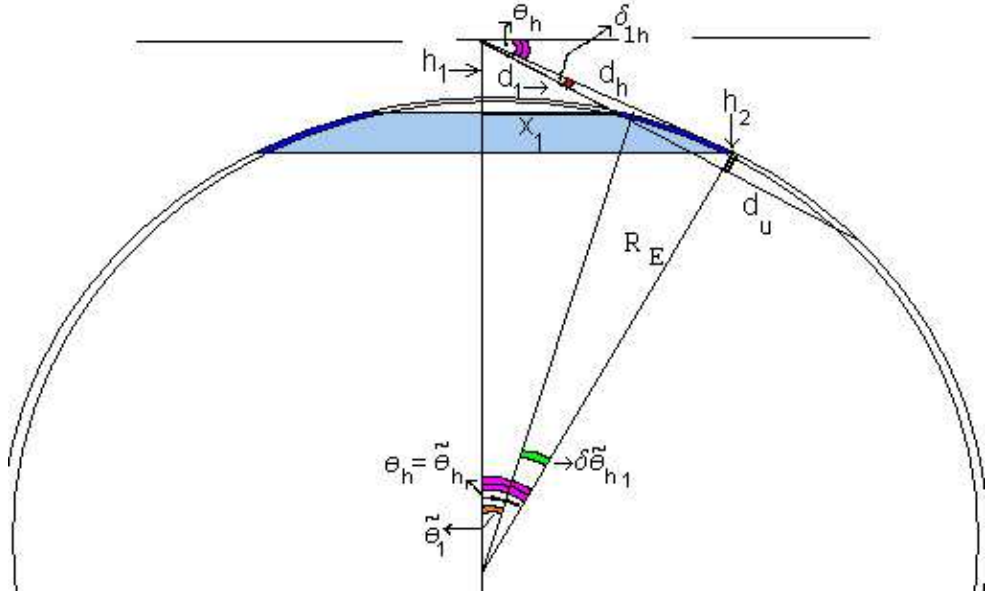}
\caption {The geometrical disposal   defining the UPTAUs and
HORTAUs Ring (Crown or Coronas) Areas as in the text; the
distances are exaggerated for simplicity.} \label{fig:fig17}
\end{figure}
%%%%%%%%%%%%%%%%%%%% end fig. 11 %%%%%%%%%%%%%%%%%%%%%%%%%

  \item
  The consequent characteristic  energy
  $E_{\tau_{h}}$ $\tau$ lepton decaying in flight from  $d_{h}$ distance just nearby
  the source:

\begin{displaymath}
 {E_{\tau_{h}}}=  {\left(\frac {d_{h}}{c\tau_{0}}\right)}m_{\tau} c^2
 \simeq{2.2\cdot10^{18}eV}\sqrt{\frac {h_{1}}{km}}
 %\sqrt{1 + \frac
%{h_{1}}{2R_{\oplus}}
\end{displaymath}

%\begin{equation}
%%%%%%%%%%%%%%%%%%%%%%
%%%%%%%%%%%%%%%%%%%%%% Tabl 1 %%%%%%%%%%%%%%%%%%%%%%%%%%%%%
\begin{table}[b]
\begin{center}
\caption{The Table of the main parameters leading to the
    effective HORTAUs Mass  from the observer height $h_1$, the
    corresponding $\tau$ energy $E_{\tau}$ able to let the $\tau$
    reach him from the horizons. The observed total Area $A_{TOT}$ underneath
    the observer is shown as well as  the corresponding $\tau$ propagation lenght in
    matter $l_{\tau}$. The opening angle seen by the observer  toward the terrestrial Crown crust
    $\delta\tilde{\theta}_{h_1}$
    and  $l_{\tau}$ has to be projected orthogonally
    in the Earth Crust, $l_{\tau_{\downarrow}}=l_{\tau}\cdot\sin\delta{\tilde{\theta}_{h_{1}}}$, are  also
    evalueted. The final Ring Areas for both
    (water,rock) densities $A_R$ at each characteristic high altitudes $h_1$ are shown with the consequent
     observable  Volume $\Delta V$,(made by $\Delta V = A_R \cdot
     l_{\tau_{\downarrow}}= A_R \cdot l_{\tau}\cdot\sin\delta{\tilde{\theta}_{h_{1}}}$).
 Finally the consequent effective Volume and  Mass
 $\Delta V_{eff.}= \Delta V \frac{\Delta\Omega}{4\pi}$, $\Delta M_{eff.} =\Delta V_{eff.}\cdot\varrho$,
(within the very narrow $\tau$ air-shower solid
    angle $\frac{\Delta\Omega}{4\pi} \simeq 2.5\cdot 10^{-5}$ ) as a function of density $\rho$ and height $h_1$. In the
    last Row the Ratio R $= \Delta M_{eff.}/\Delta M_{Atm}$ defines the ratio of
    HORTAUs produced within the Earth Crown Skin over the atmospheric
    ones: this ratio nearly reflects the matter density over the air one and
    it reaches nearly two order of magnitude and describes the larger
    probability to observe a HORTAU over the probability to observe
    a double bang in the air at EUSO or OWL detectors.}
\vspace{0.5cm}

\footnotesize
\begin{tabular} {||c|c|c|c|c|c|c|c|c|c||c|c||}\hline
& & & & & & & & & & &   \\ $\rho$&$h_1$&$E_{th}$ & $A_{TOT}$ &
$l_{\tau}$ & $\delta\widetilde{\theta}_{h_1}$ &
$l_{\tau_{\downarrow}}$  &$ A_{R}$ & $\Delta V$ &$ \Delta
V_{eff}$& $\Delta M_{eff.}$ &
$R$\\
&$(km)$ & $(eV) $ & $({km}^2)$ & $(km)$ &  &
 $(km)$ &
$({km}^2)$& $({km}^3)$&
$({km}^3)$&$\left(\frac{{km}^3}{\rho}\right)$ &
\\ \hline\hline$1$ & $2$&$ 3.12*10^{18}$
&$8*10^4$
&$21.7$&$1.31^o$&$0.496$&$7.9*10^4$&$3.95*10^4$&$0.987$&$0.987$&
$49.6$
\\ \hline$2.65$&$2$&$3.12*10^{18}$&$8*10^4$&$11$&$0.97^o$&$0.186$&$7.2*10^4$&$1.34*10^4$
&$0.335$&$0.89$& $49.2$
\\ \hline\hline$1$ &
$5$&$4.67*10^{18}$&$2*10^5$&$24.3$&$1.79^o$&$0.76$&$1.9*10^5$&$1.45*10^5$&$3.64$&$3.64$&
$75$
\\ \hline$2.65$
&$5$&$4.67*10^{18}$&$2*10^5$&$12.1$&$1.07^o$&$0.225$&$1.45*10^5$&$3.2*10^4$&$0.82$&$2.17$
& $59.6$
\\ \hline\hline
$1$&$25$&$8*10^{18}$&$10^6$&$27.5$&$2.36^o$&$1.13$&$7.16*10^5$&
$8.12*10^5$&$20.3$&$20.3$& $113$
\\ \hline$2.65$&$25$&$8*10^{18}$&$10^6$&$13.1$&$1.08^o$&$0.247$&$3.83*10^5$&
 $9.5*10^4$&$2.4$&$6.3$& $65.45$
\\ \hline\hline
$1$&$500$&$1.08*10^{19}$&$1.8*10^7$&$29.4$&$2.72^o$&$1.399$&$4.3*10^6$&$6*10^6$&$150.6$
&$150.6$& $140$
\\ \hline$2.65$
&$500$&$1.08*10^{19}$&$1.8*10^7$&$13.8$&$1.07^o$&$0.257$&$1.75*10^6$&$4.5*10^5$&$11.3$
&$30$& $68$
\\ \hline
\end{tabular}
%\caption {Hortau Energies, Areas, Volumes and Masses for
%different height h1 of observation}
\label{tab:table4}
\end{center}
\end{table}

\normalsize
%%%%%%%%%%%%%%%%%%
%\end{equation}
For each low quota ($h_1 \leq$ a few kms) there exists a
characteristic air depth, which is necessary for Tau decay  to
develop a detectable shower. The corresponding distance
 is $d_{Sh}$ $\sim 6 kms \ll d_h$. More precisely at low quota
($h_1\ll h_o$, where $h_o$ is the air density decay height$ =
8.55$ km.) one finds:
\begin{equation}
d_{Sh}\simeq 5.96km[ 1+ \ln {\frac{E_{\tau}}{10^{18}eV}})] \cdot
e^{\frac{h_1}{h_o}}
\end{equation}
Because these distances are usually much shorter than the far
horizons distances, we may neglect them. However at high altitude
($h_1\geq h_o$) this is no longer the case (see Appendix A).
Therefore  we shall introduce from here and in next steps a
small, but important modification , whose physical motivation is
just to include  the air dilution role at highest quota: $ {h}_1
\rightarrow \frac {h_{1}}{1 + h_1/H_o} $, where , as in Appendix
A, $H_o= 23$ km. Then previous definition (at height $h_{1} <
R_{\oplus}$) becomes:
\begin{equation}
 {E_{\tau_{h}}}\simeq{2.2\cdot10^{18}eV}\sqrt{\frac {h_{1}}{1 + h_1/H_o}}
\end{equation}
This procedure, applied tacitly everywhere, guarantees that there
we may extend our results to those HORTAUs at altitudes where the
residual air density  must exhibit a sufficient slant depth. For
instance, highest $\gg 10^{19} eV$ HORTAUs will not be easily
observable because their ${\tau}$ life distance exceedes (usually)
the horizons air depth length.
%\item
The parental UHE $\nu_{\tau}$,$\bar\nu_{\tau}$ or $\bar\nu_{e}$ of
energies $E_{\nu_{\tau}}$ able to produce such UHE $E_{\tau}$ in
matter are:
\begin{equation}
 E_{\nu_{\tau}}\simeq 1.2 {E_{\tau_{h}}}\simeq {2.64
 \cdot 10^{18}eV \cdot \sqrt{\frac {h_{1}}{km}}}
\end{equation}
\item The neutrino (underground) interaction lengths  at the
corresponding energies is $L_{\nu_{\tau}}$:
\begin{displaymath}
L_{\nu_{\tau}}= \frac{1}{\sigma_{E\nu_{\tau}}\cdot N_A\cdot\rho_r}
%\end{displaymath}
%\begin{displaymath}
= 2.6\cdot10^{3}km\cdot \rho_r^{-1}{\left(\frac{E_{\nu_h}}{10^8
\cdot GeV}\right)^{-0.363}}
\end{displaymath}
\begin{equation}
{\simeq 304 km \cdot
\left(\frac{\rho_{rock}}{\rho_r}\right)}\cdot{\left(\frac{h_1}{km}\right)^{-0.1815}}
\end{equation}
 For more details see \cite{Gandhi et al 1998}, \cite{Fargion 2000-2002}.
 It should be remind that here we ignore the $\tau$ multi-bangs \cite{Halzen1998}
 that reduce the primary $\nu_{\tau}$ energy and pile up the lower
 energies HORTAUs (PeV-EeV).

%\item
The maximal neutrino depth $h_{2}(h_{1})$, see Fig.\ref{fig:fig17}
-Fig.\ref{fig:fig19} above, under the chord along the UHE
neutrino-tau trajectory of length $L_{\nu}(h_{1})$ has been found:
\begin{displaymath}
h_{2}(h_{1}) = {\frac{L_{\nu_{h}}^2}{2^2\cdot{2}(R-h_{2})}
\simeq\frac{L_{\nu_{h}}^2}{8R_{\oplus}}}\simeq 1.81\cdot km
\cdot{\left(\frac{h_1}{km}\right)^{-0.363}\cdot
\left(\frac{\rho_{rock}}{\rho_r}\right)^2}
\end{displaymath}
%\begin{equation}

%\end{equation}
See Fig.\ref{fig:fig17}, for more details. Because the above $h_2$
depths are in general not  too deep respect to the Ocean depths,
we shall consider respectively either sea (water) or rock (ground)
materials as Crown matter density. \item The corresponding opening
angle observed from height $h_{1}$, $\delta_{1h}$ encompassing the
underground height  $h_{2}$ at horizons edge (see
Fig.\ref{fig:fig17}) and the nearest UHE $\nu$ arrival directions
$\delta_{1}$ is:
\begin{displaymath}
{{\delta_{1h}}(h_{2})}={2\arctan{\frac{h_{2}}{2 d_{h}}}}=
2\arctan\left[\frac{{8\cdot
10^{-3}}\cdot{(\frac{h_{1}}{km}})^{-0.863}\left(\frac{\rho_{rock}}{\rho_r}\right)^2}{{\sqrt{1+{\frac{h_{1}}{2R_{\oplus}}}}}}\right]
\end{displaymath}
\begin{equation}
{\simeq 0.91^{o}
\left(\frac{\rho_{rock}}{\rho_r}\right)^2}\cdot{(\frac{h_{1}}{km}})^{-0.863}
\end{equation}
\item The underground chord $d_{u_{1}}$ (see Fig.\ref{fig:fig17}-\ref{fig:fig19}) where
UHE $\nu_{\tau}$ propagates and the nearest distance $d_{1}$ for
$\tau$ flight (from the observer toward Earth) along the same
$d_{u_{1}}$ direction, within the angle $\delta_{1h}$ defined
above, angle below the horizons (within the upward UHE neutrino
and HORTAUs propagation line) is:
\begin{equation}
d_{u_{1}}=2\cdot{\sqrt{{\sin}^{2}(\theta_{h}+\delta_{1h})(R_{\oplus}+{h_{1}})^{2}-{d_{h}}^2}}
\end{equation}
 Note that by definition  and by construction $ L_{\nu} \equiv d_{u_{1}}$.
The nearest HORTAUs distance corresponding to this horizontal
edges still transparent to UHE $\tau$ is:
\begin{equation}
{d_{1}(h_{1})}=(R_{\oplus}+h_{1})\sin(\theta_{h}+\delta_{1h})-{\frac{1}{2}}d_{u_{1}}
\end{equation}
 Note also that for the height $h_{1}\geq km$ : $$ \frac{d_{u_{1}}}{2}\simeq{(R_{\oplus}+{h_{1}})\sqrt{\delta_{1h}\sin{2\theta_{h}}}}\simeq
{158\sqrt{\frac{\delta_{1h}}{1^o}}\sqrt{\frac{h_{1}}{km}}}km.$$

%%%%%%%%%%%%%%%%%%%%%%%%%%%%%%%%%%%%%%%%%%%%%%%%%%%%%%%%%%%%%%%%%%%%%%%%%%%
%\begin{figure}\centering\includegraphics[width=8cm]{Fig04.eps}
%\caption {Distances from the observer to the Earth ($d_1$) or to
%the Horizons ($d_h$)} \label{fig:fig4}
%\end{figure}
%%%%%%%%%%%%%%%%%%%%%%%%%%%%%%%%%%%%%%%%%%%%%%%%%%%%%%%%%%%%%%%%%%%%%%%%%%%%%

%%%%%%%%%%%%%%%%%%%% fig. 12 %%%%%%%%%%%%%%%%%%%%%%%%%

\begin{figure}
\centering
\includegraphics[width=14cm]{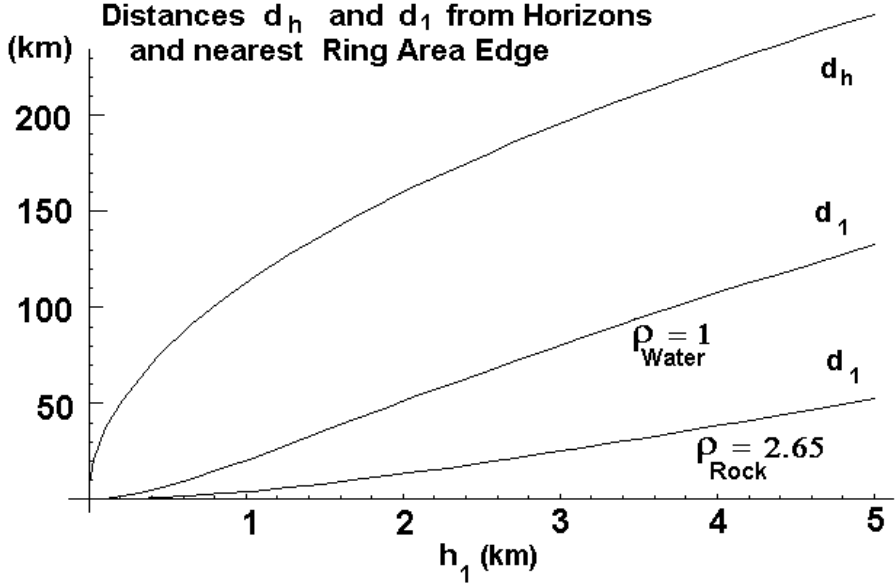}
\caption {Distances from the observer to the Earth ($d_1$) for
    different matter densities and to the Horizons ($d_h$) for low
    altitudes.} \label{fig:fig18}
\end{figure}

%%%%%%%%%%%%%%%%%%%% end fig. 12 %%%%%%%%%%%%%%%%%%%%%%%%%

%%%%%%%%%%%%%%%%%%%%%%%%%%%%%%%%%%%%%%%%%%%%%%%%%%%%%%%%%%%%%%%%%%%%%%%%%%%%%%%
\item The same distance projected cord $x_{1}(h_{1})$ along the
horizontal line (see Fig.\ref{fig:fig17}) is:
\begin{equation}
x_{1}(h_{1})=d_{1}(h_{1})\cos({\theta_{h}+\delta_{1h}})
\end{equation}

The total terrestrial Area $A_{T}$, underneath any observer at
height $h_{1}$, is  (see
Fig.\ref{fig:fig16}-\ref{fig:fig17}-\ref{fig:fig20}):
$$ A_{T} =2\pi{R_{\oplus}}^{2}(1-\cos{\tilde{\theta}_{h}})
=2\pi{R_{\oplus}}h_{1}\left({\frac{1}{1+\frac{h_{1}}{R_{\oplus}}}}\right)
A_{T}=4\cdot{10}^{4}{km}^{2}{\left({\frac{h_{1}}{km}}\right)}{\left({\frac{1}{1+{\frac{h_{1}}{R_{\oplus}}}}}\right)},$$

where $\tilde{\theta}_{h}$ is the opening angle from the Earth
 center (see Fig.\ref{fig:fig17}). At first sight one may be tempted to
consider all the Area  $A_{T}$ for UPTAUs and HORTAUs, but because
of the shower air opacity (HORTAUs) or for its small slant depth
 (UPTAUs) this is incorrect.  While for HORTAUs there is a more
complex Area estimated above and in the following, for UPTAUs the
Area Ring (or Disk) is quite simpler to derive  very similar
geometrical variables summarized in Appendix B.
 \item The
Earth Ring Crown crust area ${A_{R}}(h_{1})$ delimited by the
horizons distance $d_{h}$ and the nearest distance $d_{1}$ is
still transparent to UHE $\nu_{\tau}$ (see
Fig.\ref{fig:fig16}-\ref{fig:fig19}). The ring area
${A_{R}}(h_{1})$ is computed from the internal angles
$\delta{\tilde{\theta}_{h}}$ and $\delta{\tilde{\theta}_{1}}$
defined at the Earth center (Fig.\ref{fig:fig17})(note that
$\delta{\tilde{\theta}_{h}}={\delta{\theta_{h}}}$ but in general
$\delta{\tilde{\theta}_{1}}\neq{\delta{\theta_{1}}}$).

\begin{equation}
\hspace{-2.cm}A_{R}(h_{1})=2\pi{R_{\oplus}}^2(\cos{\tilde{\theta}_{1}}-\cos{\tilde\theta_{h}})
=2\pi{R_{\oplus}^{2}}{\left({{\sqrt{1-{\left({\frac{x_{1}({h_1})}{R_{\oplus}}}\right)^{2}}}}-{\frac{R_{\oplus}}{R_{\oplus}+{h_{1}}}}}\right)}
\end{equation}
 Here $x_{1}({h_1})$ is the cord defined above.
%%%%%%%%%%%%%%%%%%%%%%%%%%%%%%%%%%%%%%%%%%%%%%%%%%%%%%%%%%%%%%%%

%[height=0.95\textwidth,]

%%%%%%%%%%%%%%%%%%%% fig. 13 %%%%%%%%%%%%%%%%%%%%%%%%%

\begin{figure}
\centering
\includegraphics[width=13cm]{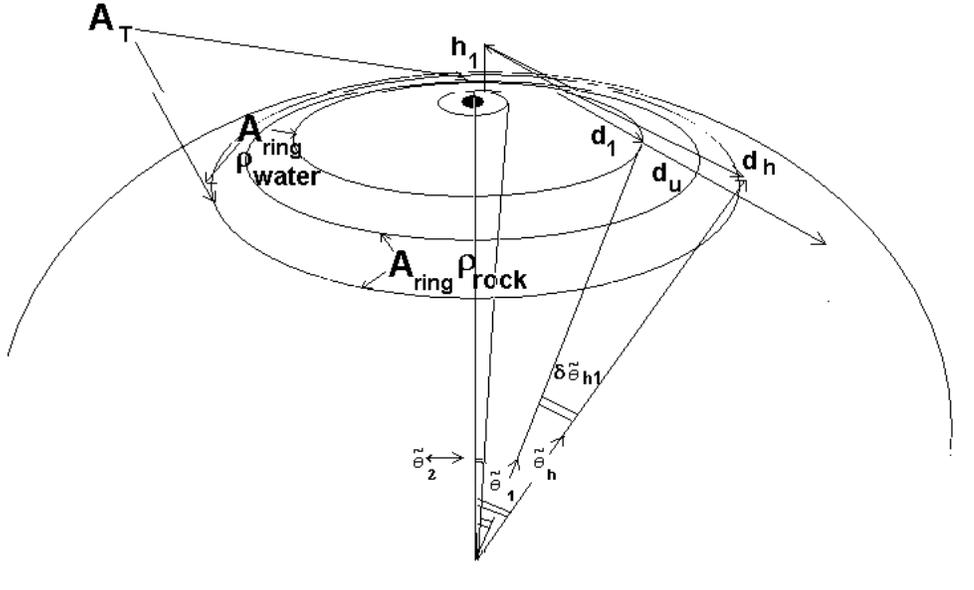}
\caption {Total and Ring (Crown) Areas and Angles for
    UPTAUs-HORTAUS observed at different heights.}
\label{fig:fig19}
\end{figure}

%%%%%%%%%%%%%%%%%%%% end fig. 13 %%%%%%%%%%%%%%%%%%%%%%%%%

%%%%%%%%%%%%%%%%%%%%  fig. 14 %%%%%%%%%%%%%%%%%%%%%%%%%

%%%%%%%%%%%%%%%%%%%%%%%%%%%%%%%%%%%%%%%%%%%%%%%%%%%%%%%%%%%%
\begin{figure}
\centering
\includegraphics[width=15.5cm]{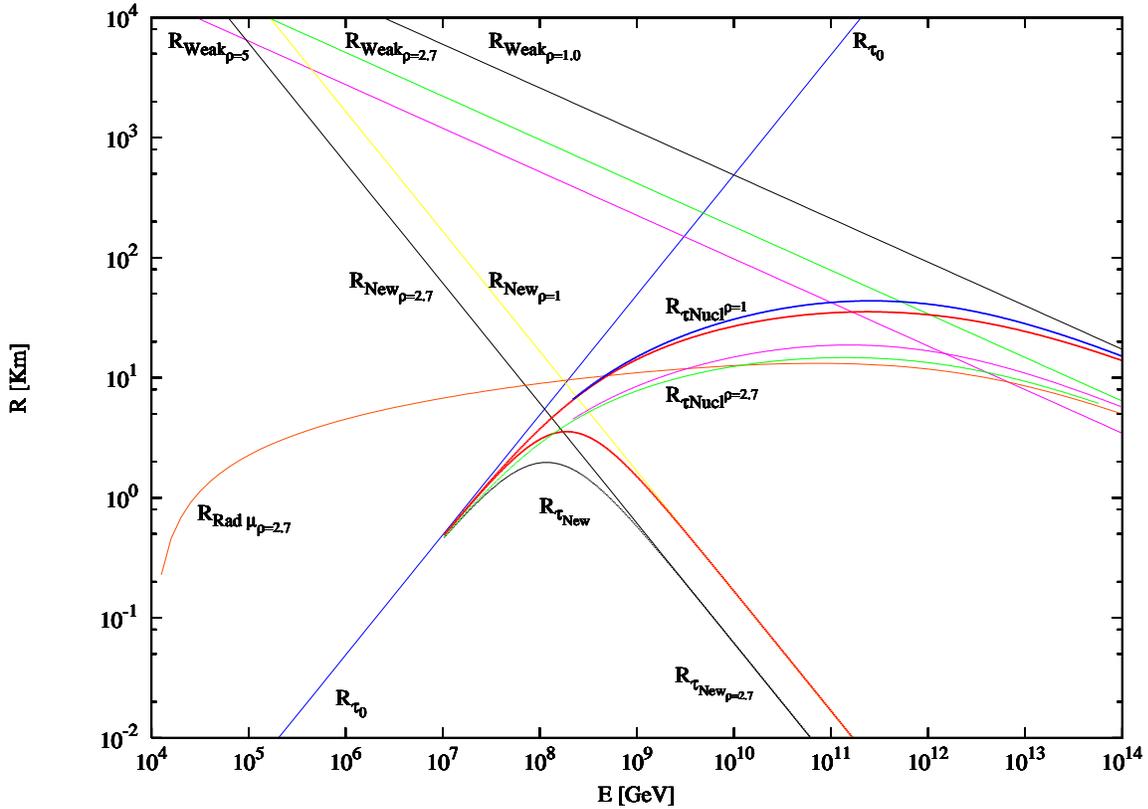}
\caption {Lepton $\tau$ (and $\mu$) Interaction Lengths for
different matter densities: $R_{\tau_{o}}$ is the free $\tau$
length,$R_{\tau_{New}}$ is the New Physics TeV Gravity interaction
range at corresponding densities,$R_{\tau_{Nucl}\cdot{\rho}}$
,\cite{Fargion 2000-2002}, see also \cite{Becattini Bottai 2001},
\cite{Dutta et al.2001}, is the combined $\tau$ Ranges keeping
care of all known interactions and lifetime and mainly the
photo-nuclear interaction. There are two slightly different split
curves (for each density) by two comparable approximations in the
interaction laws. $R_{Weak{\rho}}$ is the electro-weak Range at
corresponding densities (see also \cite{Gandhi et al 1998}),
\cite{Fargion 2000-2002}.} \label{fig:fig20}
\end{figure}

%%%%%%%%%%%%%%%%%%%% end fig. 14 %%%%%%%%%%%%%%%%%%%%%%%%%
%%%%%%%%%%%%%%%%%%%%  fig. 14 bis%%%%%%%%%%%%%%%%%%%%%%%%%

%%%%%%%%%%%%%%%%%%%%%%%%%%%%%%%%%%%%%%%%%%%%%%%%%%%%%%%%%%%%%%%%%%%%%%%%%%%%%%%%%%
\item The characteristic interaction tau lepton length $l_{\tau}$
is defined at the average $E_{\tau_{1}}$, from interaction in
matter (rock or water). These lengths have been derived by
analytical equations keeping care of the Tau lifetime, the
photo-nuclear losses, the electro-weak losses \cite{Fargion
2000-2002}. See Fig.\ref{fig:fig20} below. The tau length along
the Earth Skimming distance $l_{\tau_{2}}$ should be vertically
projected ( multiplied by $\sin(\delta\tilde{\theta}_{h_{1}})$)
in order to find the observable skin volume; the angle
$\delta\tilde{\theta}_{h_{1}}$ is given by:
\begin{equation}
\delta\tilde{\theta}_{h_{1}}\equiv \tilde\theta_{h}
-\arcsin{\left({\frac{x_{1}}{R_{\oplus}}}\right)}
\end{equation}

%%%%%%%%%%%%%%%%%%%%%%%%%%%%%%%%%%%%%%%%%%%%%%%%%%%%%%%%%%

%%%%%%%%%%%%%%%%%%%%  fig. 15 %%%%%%%%%%%%%%%%%%%%%%%%%

\begin{figure}
\centering
\includegraphics[width=13cm]{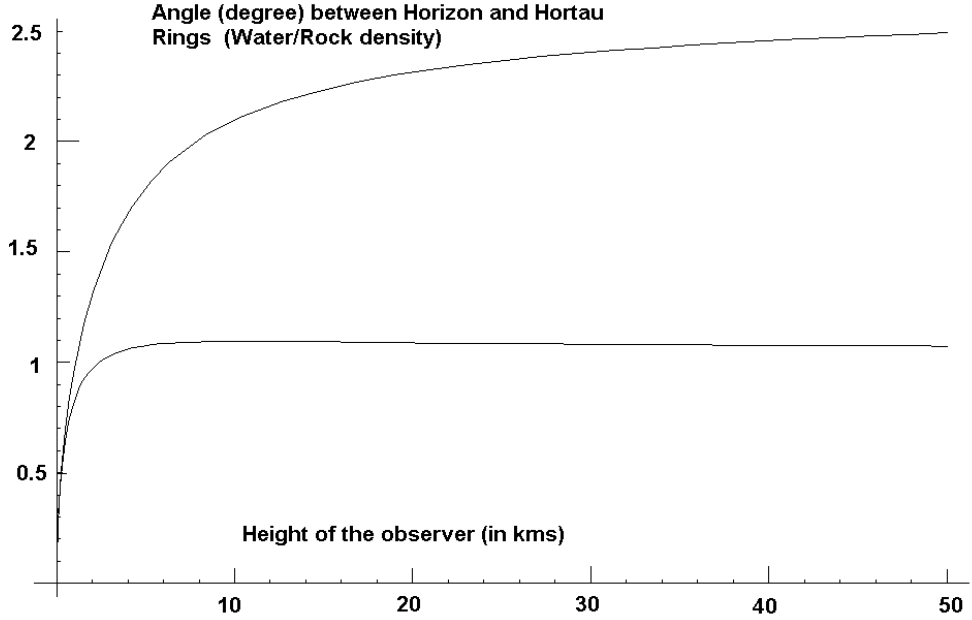}

\vspace{0.5cm} \caption {Opening angle
$\delta\tilde{\theta}_{h_{1}}$ toward Ring Earth Skin for density
$\rho_{water}$ and $\rho_{rock}$.(see Fig.\ref{fig:fig17}) }
\label{fig:fig21}
\end{figure}

%%%%%%%%%%%%%%%%%%%% end fig. 15 %%%%%%%%%%%%%%%%%%%%%%%%%

%%%%%%%%%%%%%%%%%%%%%%%%%%%%%%%%%%%%%%%%%%%%%%%%%%%%%
The same quantity $\tilde{\theta}_{h_{1}}$, See
Fig.\ref{fig:fig21},in a more direct approximation:

\begin{displaymath}
\sin\delta{\tilde{\theta}_{h_{1}}}\simeq\frac{L_{\nu}}{2R_{\oplus}}=\frac{{304}km}{2R_{\oplus}}{\left({\frac{\rho_{rock}}{\rho}}\right)}{\frac{h_{1}}{km}}^{-0.1815}.
\end{displaymath}
For highest ($h\gg H_o$=23km) altitude the exact approximation
reduces to:
$$\delta{\tilde{\theta}_{h_{1}}}\simeq{1}^o{\left({\frac{\rho_{rock}}{\rho}}\right)}\left({\frac{h_{1}}{500\cdot
km}}\right)^{-0.1815}$$ Therefore the penetrating $\tau$ skin
depth $l_{\tau_{\downarrow}}$ is
\begin{equation}
l_{\tau_{\downarrow}}=l_{\tau}\cdot\sin\delta{\tilde{\theta}_{h_{1}}}
\simeq{{0.0462\cdot
l_{\tau}{\left({\frac{\rho_{water}}{\rho}}\right)}}}{\frac{h_{1}}{km}}^{-0.1815}
\end{equation}

where the $\tau$ ranges in matter, $l_{\tau}$ have been calculated
and shown in Fig.\ref{fig:fig20}.
%%%%%%%%%%%%%%%%%%%%%%%%%%%%%%%%%%%%%%%%%%%%%%%%
  The consequent Observable Ring Areas at two different ranges of quota are displayed
  in Fig.\ref{fig:fig22}-Fig.\ref{fig:fig23}.
%%%%%%%%%%%%%%%%%%%%  fig. 16 %%%%%%%%%%%%%%%%%%%%%%%%%

\begin{figure}
\centering
\includegraphics[width=16cm]{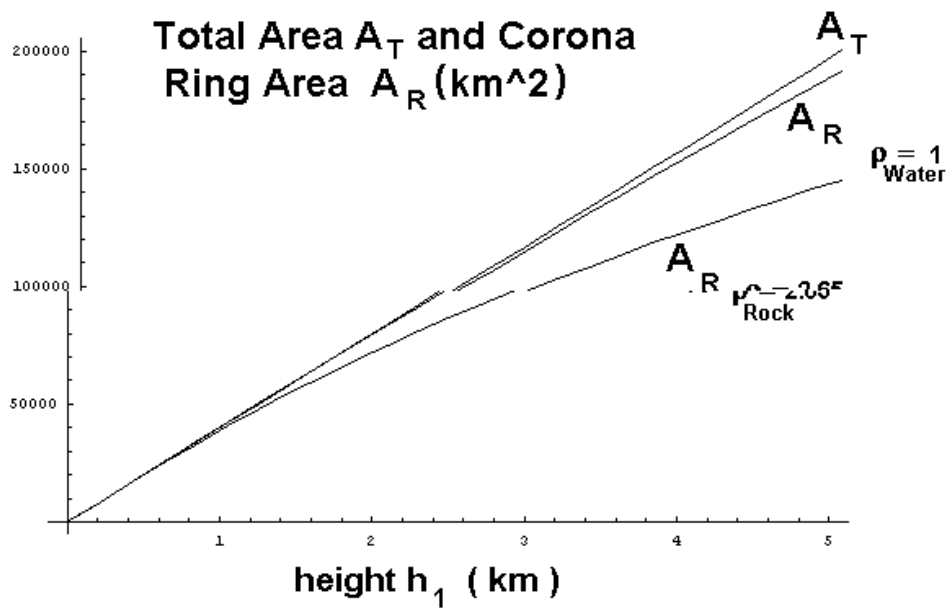}
\caption {Total Area $A_T$ and Ring ( Crown or Coronas) Areas for
two densities $A_R$ at low altitudes. }\label{fig:fig22}
\end{figure}

%%%%%%%%%%%%%%%%%%%% end fig. 16 %%%%%%%%%%%%%%%%%%%%%%%%%

%%%%%%%%%%%%%%%%%%%%%%%%%%%%%%%%%%%%%%%%%%%%%
%%%%%%%%%%%%%%%%%%%%%%%%%%%%%%%%%%%%%%%%%%%%%%%%

%%%%%%%%%%%%%%%%%%%% fig. 17 %%%%%%%%%%%%%%%%%%%%%%%%%

\begin{figure}
\centering
\includegraphics[width=16cm]{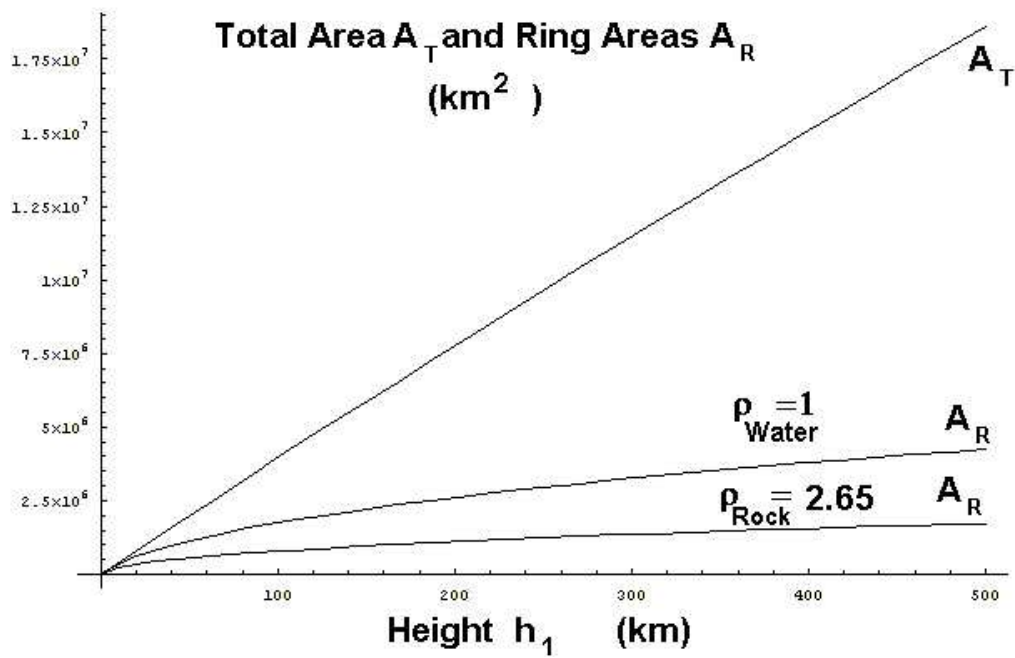}
\caption {Total Area $A_T$ and Ring (Crown) Areas for two
densities $A_R$ at high altitudes. } \label{fig:fig23}
\end{figure}

%%%%%%%%%%%%%%%%%%%% end fig. 17 %%%%%%%%%%%%%%%%%%%%%%%%%

%%%%%%%%%%%%%%%%%%%%%%%%%%%%%%%%%%%%%%%%%%%%%

\item
The final  analytical expression for the Earth Crust Skin Volumes
and Masses under the Earth Skin inspected by HORTAUs are derived
combining the above functions on HORTAUs Areas  with the previous
lepton Tau $l_{\tau_{\downarrow}}$ vertical depth:
\begin{equation}
{V_{h_{1}}}={A_{R}}(h_{1})\cdot l_{\tau_{\downarrow}};
\end{equation}
The detecting mass is directly proportional to this volume:
\begin{equation}
{M_{h_{1}}}={V_{h_{1}}}\cdot \rho
\end{equation}

\item A more approximated but easy to handle
 expression for Ring area for high altitudes ($h_1\gg 2km$ $h_1\ll R_{\oplus}$) may be
summirized as:
\begin{displaymath}
{A_R (h_1)}\simeq
2\pi{R_{\oplus}^2}\sin{\theta_{h}}{\delta{\tilde{\theta}_{{h}_{1}}}}\propto{\rho^{-1}}
\end{displaymath}
\begin{equation}
\simeq{2\pi{R_{\oplus}^{2}}{\sqrt{\frac{2h_{1}}{R_{\oplus}}}}\left({\frac{\sqrt{1+{\frac{h_{1}}{2R_{\oplus}}}}}{1+{\frac{h_{1}}{R}}}}\right)}{\left({\frac{L_{\nu}}{2R_{\oplus}}}\right)}
\end{equation}
 At high altitudes the above approximation, corrected
accordingly to the exact one, shown in Fig.\ref{fig:fig23},
becomes:
\begin{equation}
{A_R (h_1)}\simeq
2\pi{R_{\oplus}}{d_{h1}}{\delta{\tilde{\theta}_{{h}_{1}}}}
\simeq{4.65 \cdot 10^6{\sqrt {\frac{h_{1}}{500
km}}}}{\left({\frac{\rho_{water}}{\rho}}\right)}{km}^2
\end{equation}
 Within the above
approximation the final searched Volume ${V_{h_{1}}}$ and Mass
${M_{h_{1}}}$ from where HORTAUs may be generated  is:
\begin{equation}
{V_{h_{1}}}={\frac{\pi}{2}{\sqrt{\frac{2h_{1}}{R_{\oplus}}}}
\left({\frac{\sqrt{1+{\frac{h_{1}}{2R_{\oplus}}}}}{1+{\frac{h_{1}}{R_{\oplus}}}}}\right)}
{L_{\nu}^{2}}{l_{\tau}}\propto{\rho^{-3}}
\end{equation}
%%%%%%%%%%%%%%%%%%%%%%%%%%%%%%%%%%%%%%%%%%%%%%%%%%%%%%%%%%%%%%%%%%%%%%%%%%%%%%%%%%%%%%%%%%%
\begin{equation}
{M_{h_{1}}}={\frac{\pi}{2}{\sqrt{\frac{2h_{1}}{R_{\oplus}}}}\left({\frac{\sqrt{1+{\frac{h_{1}}{2R_{\oplus}}}}}{1+{\frac{h_{1}}{R_{\oplus}}}}}\right)}{L_{\nu}^{2}}{l_{\tau}}{\rho}\propto{\rho^{-2}}
\end{equation}
\item
The effective observable Skin Tau Mass $M_{eff.}(h_{1})$ within
the thin HORTAU or UPTAUs Shower angle beam $\simeq$ $1^o$ is
suppressed by the solid angle of view
${\frac{\delta\Omega}{\Omega}} \simeq 2.5\cdot 10^{-5}$.
\begin{equation}
{\Delta M_{eff.}(h_{1})={ V_{h_{1}}}\cdot\rho
{\frac{\delta\Omega}{\Omega}}}
\end{equation}
The lower bound Masses $M_{eff.}(h_{1})$  (summarized in the
Tab.\ref{tab:table4} and in Fig.\ref{fig:fig24}) are exactly
estimated  for different realistic high quota experiments.

%%%%%%%%%%%%%%%%%%%%%%% Section 8  %%%%%%%%%%%%%%%%%%%%%%%%%%%%%%%%%%%%%%%%%%%%%%%%%%%%%%%%%%%%%%%%%%

\subsection{Effective Volume in HORTAUs} The discovery of the
expected UHE neutrino astronomy is urgent and just behind the
corner. If the Terrestrial Gamma Flashes are indeed the traces of
such PeV-EeV UHE neutrino tau, then the neutrino flux is just at
the level of $\simeq  2\cdot 10^2 eV cm^{-2} s^{-1}sr^{-1}$ and
might be accessible by AMANDA in a near future. To make further
progress in studies of UHE neutrino astronomy huge volumes are
necessary. Beyond underground $km^3$ detectors a new generation
of UHE neutrino calorimeters lays on front of mountain chains and
just underneath our feet: The Earth itself offers huge Crown
Volumes as Beam Dump calorimeters observable via upward Tau Air
Showers, UPTAUs and HORTAUs. Their effective Volumes as a
function of the quota $h_1$ have been derived by an analytical
function variables in equations above and Appendix B. These
Volumes and Masses  are discussed below and summirized in the last
column of Tab.\ref{tab:table4}.
%at the end of the paper.
At a few tens meter altitude the UPTAUs and HORTAUs Ring are
almost overlapping. At low altitude $h_{1}\leq 2$ Km the HORTAUs
are nearly independent on the  matter density $\rho$:
%\begin{displaymath}
${\Delta M_{eff.}(h_{1}=2 Km)(\rho_{Water})}= 0.987 km^3$
%\end{displaymath}
%\begin{displaymath}
${\Delta M_{eff.}(h_{1}=2km)(\rho_{Rock})}= 0.89 km^3$
%\end{displaymath}
These volumes are the effective Masses expressed in Water
equivalent volumes. On the contrary at higher quotas, like highest
Mountain observations sites, Airplanes, Balloons and Satellites,
the matter density of the HORTAUs Ring (Crown) Areas plays more
and more dominant role asymptotically $proportional$ to
${\rho}^{-2}$:
%\begin{displaymath}
${\Delta M_{eff.}(h_{1}=5 km)(\rho_{Water})}= 3.64 km^3 $;
%\end{displaymath}
%\begin{displaymath}
${\Delta M_{eff.}(h_{1}=5 km)(\rho_{Rock})}= 2.17 km^3 $.
%\end{displaymath}
 From Air-planes or balloons the effective volumes $M_{eff.}$
increase and the density $\rho$ plays a relevant role.
%\begin{displaymath}
${\Delta M_{eff.}(h_{1}=25 km)(\rho_{Water})}= 20.3 km^3
%\end{displaymath}
$;
%\begin{displaymath}
${\Delta M_{eff.}(h_{1}=25 km)(\rho_{Rock})}= 6.3 km^3
%\end{displaymath}
$. Finally from satellite altitudes the same effective volumes
$M_{eff.}$ are reaching extreme values:
\begin{displaymath}
{\Delta M_{eff.}(h_{1}=500~ km)(\rho_{Water})}= 150.6~ km^3
\end{displaymath}
\begin{displaymath}
{\Delta M_{eff.}(h_{1}=500~ km)(\rho_{Rock})}=30~ km^3
\end{displaymath}
These masses must be compared with other proposed $km^3$
detectors, keeping in mind that these HORTAUs signals conserve the
original UHE $\nu$ direction information within a degree. One has
to discriminate  HORTAUS (only while observing from satellites)
from Horizontal High Altitude Showers (HIAS) \cite{Fargion2001b},
due to rare UHECR showering on high atmosphere. While wide (RICE)
one might also remind the UPTAUs (at PeV energies) volumes as
derived in Appendix B and in \cite{Fargion 2000-2002} whose values
(assuming an arrival angle$\simeq 45^o- 60^o$ below the horizons)
are nearly $proportional$ to the  density $\rho$:
\begin{displaymath}
{\Delta M_{eff.}(h_{1}=500~ km)(\rho_{Water})}= 5.9~ km^3
\end{displaymath}
\begin{displaymath}
{\Delta M_{eff.} (h_{1}=500~ km)(\rho_{Rock})}=15.6~ km^3
\end{displaymath}
These widest masses values, here estimated analytically for main
quota, are offering an optimal opportunity to reveal UHE $\nu$ at
PeV and EeV-GZK energies by crown array detectors (scintillation,
Cherenkov, photo-luminescent) facing vertically the Horizontal
edges, located at high mountain peaks or at air-plane low sides
and finally on  balloons and satellites. As it can be seen in the
last colums of Tab.\ref{tab:table4}, the ratio $ R $ between
HORTAUs events and Showers over atmospheric UHE $\nu$ interaction
is a greater and greater number with growing height, implying a
dominant role (above two orders of magnitude) of HORTAUs grown in
Earth Skin Crown over Atmospheric HORTAUs. These huge acceptance
may be estimated by comparison with other detector thresholds (see
Fig.\ref{fig:fig24} adapted to present Z-Shower GZK-$\nu$ models).

%%%%%%%%%%%%%%%%%%%%%%%%%%%%%%%%%%%%%%%%%%%%%%%%%%%%%%%%%%%%%%%%%%%%%%%%%%%%%%
%%%%%%%%%%%%%%%%%%%%%% section 9 %%%%%%%%%%%%%%%%%%%%%%%%%%%%%%%%%%%%%%%%%%%%%%%%

\subsection{Event rate of upward and horizontal Tau air-showers}

The event rates for HORTAUs are given at first approximation by
the following expression normalized to any given neutrino flux
${\Phi_{\nu}}$:
\begin{equation}
{\dot{N}_{year}}={\Delta
M_{eff.}}\cdot{\frac{\Phi_{\nu}}{{\Phi_{\nu}}_o}}\cdot{\dot{N_o}}\cdot\frac{\sigma_{E_{\nu
}}}{\sigma_{E_{\nu_o}}}
\end{equation}
where the ${\dot{N_o}}$ is  UHE neutrino rate estimated for the
water mass within the  $km^3$ volume,  at any given (unitary)
energy ${E_{\nu_o} }$, in absence of any Earth shadowing. In our
case we shall normalize our estimate at ${E_{\nu_o}=3}$ PeV energy
for standard electro-weak charged current in a standard parton
model \cite{Gandhi et al 1998} and we shall assume a
model-independent neutrino maximal flux ${\Phi_{\nu}}$ at a flat
fluence value of nearly ${\Phi_{\nu}}_o$ $\simeq 3\cdot 10^3 eV
cm^{-2}\cdot s^{-1}\cdot sec^{-1}\cdot sr^{-1}$ corresponding to a
characteristic Fermi power law in UHE $\nu$ primary production
rate decreasing as $\frac {dN_{\nu}}{dE_{\nu}}\simeq
{E_{\nu}}^{-2}$ just below present AMANDA bounds. The consequent
rate becomes:
\begin{displaymath}
{\dot{N}_{year}}= 29 {\frac{{\Delta M_{eff.}}}{\rho \cdot
km^{3}}\cdot\frac{{\Phi_{\nu}}}{{\Phi_{\nu}}_o}}\cdot\frac{\sigma_{E_{\nu
}}}{\sigma_{E_{\nu_o}}}
\end{displaymath}
\begin{equation}
=29\cdot{\left(\frac{E_{\nu}}{3 \cdot 10^6 \cdot
GeV}\right)^{-0.637}} {\frac{{\Delta M_{eff.}}}{\rho \cdot
km^{3}}\cdot\frac{{\Phi_{\nu}}}{{\Phi_{\nu}}_o}}
\end{equation}
For highest satellites and for a characteristic UHE GZK energy
fluence ${\Phi_{\nu}}_o \simeq $
\protect \newline $3 \cdot 10^3 eV cm^{-2}\cdot
s^{-1}\cdot sr^{-1}$(as the needed Z-Showering one), the
consequent event rate observable ${\dot{N}_{year}}$ above the Sea
is :
\begin{equation}
=24.76\cdot{\left(\frac{E_{\nu}}{10^{10} \cdot
GeV}\right)^{-0.637}} {\frac{{h}}{500
km}\cdot\frac{{\Phi_{\nu}}}{{\Phi_{\nu}}_o}}
\end{equation}
 This event rate is comparable to UPTAUs one (for comparable fluence) and it may be an
 additional source of Terrestrial Gamma Flashes already observed by GRO
 in last decade \cite{Fargion 2000-2002}. These event rates are
 considered as the detector thresholds for UPTAUs and HORTAUs and
 they are summarized in Tab.\ref{tab:table4} and in Fig.\ref{fig:fig24} with other present and future
 experimental thresholds.

\begin{figure}\centering\includegraphics[width=16cm]{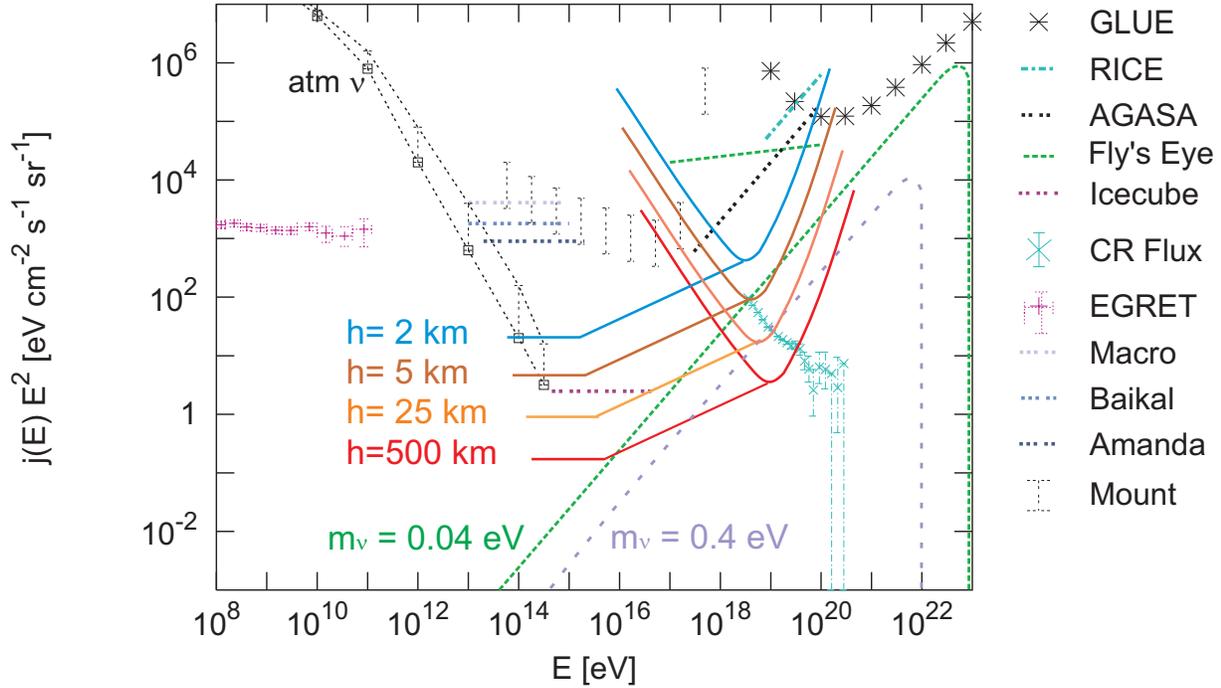}
\caption {UPTAUs (lower bound on the center) and HORTAUs (right
parabolic  curves)  sensibility at different observer heights h
($2,5,25,500 km $) assuming a $km^3$ scale volume (see Table
above)  adapted over a present neutrino flux estimate in Z-Shower
model scenario \cite{Kalashev:2002kx}, \cite{Fargion Mele Salis
1999} for light ($0.4-0.04$ eV) neutrino masses $m_{\nu}$; two
corresponding density contrast has been assumed \cite{Fargion et
all. 2001b}; the lower parabolic bound thresholds are at different
operation height, in Horizontal (Crown) Detector facing toward
most distant horizons edge; these limits are fine tuned (as
discussed in the text); we are assuming a duration of data records
of a decade comparable to the BATSE record data (a decade). The
parabolic bounds on the EeV energy range in the right sides  are
nearly un-screened by the Earth opacity while the corresponding
UPTAUs bounds  in the  center below suffer both of Earth opacity
as well as of a consequent shorter Tau interaction lenght in Earth
Crust, that has been taken into account. \cite{Fargion 2000-2002},
\cite{Fargion 2002b}, \cite{Fargion 2002c}.} \label{fig:fig24}
\end{figure}
\vspace{1 cm}
%%%%%%%%%%%%%%%%%%%%%%% Section 10 %%%%%%%%%%%%%%%%%%

%%%%%%%%%%%Union  EUSO%%%%%%%%%%%%%%%%%%
%%%%%%%%%%%%%%%%%%%%%%%%%%%%%%%%%%%%%%%%%%%%%%%%%%%%%%%%%%%%%%%%%

\section{  UHECR and UHE  $\nu$ observed by EUSO }
EUSO experiment, while monitoring the downward Earth atmosphere
layers, may observe among common Ultra High Energy Cosmic Rays,
UHECR, also High Energy Neutrino-Induced Showers either blazing
upward to the detectors at  high ($\sim$ PeV) energies or at much
higher GZK, $\sim E_{\nu}\geq 10^{19}$ eV energies, showering
horizontally in air or vertically downward.  A small fraction of
these upward, horizontal and vertical Shower maybe originated by a
direct astrophysical UHE neutrino interacting on terrestrial air
layers itself; however the dominant UHE neutrino signals are
Upward and Horizontal Tau air-showers, UPTAUSs and HORTAUs (or
Earth skimming $\nu$), born within widest Earth Crust Crown (Sea
or Rock) Areas, by UHE $\nu_{\tau} + Nuclei$ $\rightarrow \tau$
interactions, respectively at PeV and GZK energies: their rate and
signatures are shown in a neutrino fluence map for EUSO
thresholds versus other UHE air interacting neutrino signals and
backgrounds. The effective target masses originating HORTAUs, to
be observable by EUSO, may exceed (on sea) a wide and huge ring
volume $\simeq 2360$ $km^3$. The consequent HORTAUS event rate
(even at $10\%$ EUSO duty cycle lifetime) may deeply test the
expected Z-Burst models by at least a hundred of yearly events.
Even rarest but inescapable GZK neutrinos (secondary of photopion
production of observed cosmic UHECR) might be discovered in a few
(or a tens) horizontal shower events; in this view an extension
of EUSO detectability up to $\sim E_{\nu}\geq 10^{19}$eV
threshold is to be preferred. A wider collecting EUSO telescope
(3m diameter) might be considered.

%%% ----------------------------------------------------------------------
The very possible discover of the UHECR astronomy, the solution of
the GZK paradox, the very urgent rise of an UHE neutrino astronomy
are among the main goals of EUSO project. This advanced experiment
in a very near future will encompass AGASA-HIRES and AUGER and
observe for highest cosmic ray showers on Earth Atmosphere
recording their tracks from International Space Station by
Telescope facing dawn-ward the Earth. The recent doublets and
triplets clustering found by AGASA seem to favor compact object
(as AGN) over more exotic topological relic models, mostly fine
tuned in mass (GUT, Planck one) and time decay rate to fit all
the observed spectra. However the missing AGN within a GZK volume
is wondering. A possible remarkable correlation recently shows
that most of the UHECR event cluster point toward BL Lac sources
\cite{Gorbunov Tinyakov Tkachev Troitsky}. This correlation
favors a cosmic origination for UHECRs, well above the near GZK
volume. In this frame a relic neutrino mass \cite{Dolgov2002},
$m_{\nu} \simeq 0.4$ eV or ($m_{\nu} \simeq 0.1 \div 5$ eV) may
solve the GZK paradox \cite{Fargion Salis 1997} , \cite{Fargion
Mele Salis 1999},\cite{Weiler 1999},\cite{Yoshida et all
1998},\cite{Fargion et all. 2001b},\cite{Fodor Katz Ringwald
2002} overcoming the proton opacity being ZeV UHE neutrinos
transparent (even from cosmic edges to cosmic photon Black Body
drag) while interacting in resonance with relic neutrinos masses
in dark halos (Z-burst or Z-WW showering models). These light
neutrino masses do not solve the galactic or cosmic dark matter
problem but it is well consistent with old and  recent solar
neutrino oscillation evidences
\cite{Gallex92},\cite{Fukuda:1998mi},\cite{SNO2002} and most
recent claims by KamLAND \cite{Kamland2002} of anti-neutrino
disappearance  (all in agreement within a Large Mixing Angle
neutrino model and $\triangle {m_{\nu}}^2 \sim 6.9 \cdot
10^{-5}{eV}^2$) as well as these light masses are in agreement
with atmospheric neutrino mass splitting ($\triangle m_{\nu}
\simeq 0.07$ eV) and in possible fine tune with more recent
neutrino double beta decay experiment mass claim $m_{\nu} \simeq
0.4$ eV \cite{Klapdor-Kleingrothaus:2002ke}. In this Z-WW
Showering for light neutrino mass models large fluxes of UHE
$\nu$ are necessary,\cite{Fargion Mele Salis 1999},\cite{Yoshida
et all 1998}\cite{Fargion et all. 2001b}, \cite{Fodor Katz
Ringwald 2002},\cite{Kalashev:2002kx} or higher than usual
gray-body spectra of target relic neutrino or better clustering
are needed \cite{Fargion et all. 2001b}\cite{Singh-Ma}: indeed a
heaviest neutrino mass  $m_{\nu} \simeq 1.2-2.2$ eV while still
being compatible with known bounds, might better gravitationally
cluster leading to denser dark local-galactic halos and lower
neutrino fluxes\cite{Fargion et all. 2001b}\cite{Singh-Ma}. It
should remarked that in this frame the main processes leading to
UHECR above GZK are mainly the WW-ZZ and the t-channel
interactions \cite{Fargion Mele Salis 1999},\cite{Fargion et all.
2001b}. These expected UHE neutrino fluxes might and must  be
experienced  in complementary and independent tests.

\subsection{UHE $\nu$ Astronomy by the $\tau$ Showers and UHECRs in EUSO}
While longest ${\mu}$ tracks in
$km^3$ underground detector have been, in last three decades, the
main searched UHE neutrino signal, Tau Air-showers by UHE
neutrinos generated in Mountain Chains or within Earth skin crust
at PeV up to GZK ($>10^{19}$ eV) energies have been recently
proved to be a  new powerful amplifier in Neutrino Astronomy
\cite{Fargion et all 1999}, \cite{Fargion 2000-2002},\cite{Bertou
et all 2002},\cite{Hou Huang 2002},\cite{Feng et al 2002}. This
new Neutrino $\tau$ detector will be (at least) complementary to
present and future, lower energy, $\nu$ underground  $km^3$
telescope projects (from AMANDA,Baikal, ANTARES, NESTOR, NEMO,
IceCube). In particular Horizontal Tau Air shower may be naturally
originated by UHE $\nu_{\tau}$ at GZK energies crossing the thin
Earth Crust at the Horizon showering far and high in the
atmosphere \cite{Fargion 2000-2002},\cite{Fargion2001a},
\cite{Fargion2001b},\cite{Bertou et all 2002},\cite{Feng et al
2002}. UHE $\nu_{\tau}$ are abundantly produced by flavor
oscillation and mixing from muon (or electron) neutrinos, because
of the large galactic and cosmic distances respect to the neutrino
oscillation ones (for already known neutrino mass splitting).
Therefore EUSO may observe many of the above behaviors and it may
constrains among models and fluxes and it may also answer open
standing questions. I will briefly enlist, in this first
preliminary presentation, the main different signatures and rates
of UHECR versus UHE $\nu$ shower observable by EUSO at 10\% duty
cycle time within a 3 year record period, offering a first
estimate of their signals. Part of the results on UHECR are
probably  well known, nevertheless they are here re-estimated.
Part of the results, regarding the UPTAUs and HORTAUs, are new and
they rule the UHE $\nu$ Astronomy in EUSO.
 \vspace{2cm}
%%%%%%%%%%%%%%%%%%%% fig. 1a %%%%%%%%%%%%%%%%%%%%%%%%%
\vspace{2cm}
\begin{figure}\centering\includegraphics[width=12cm]{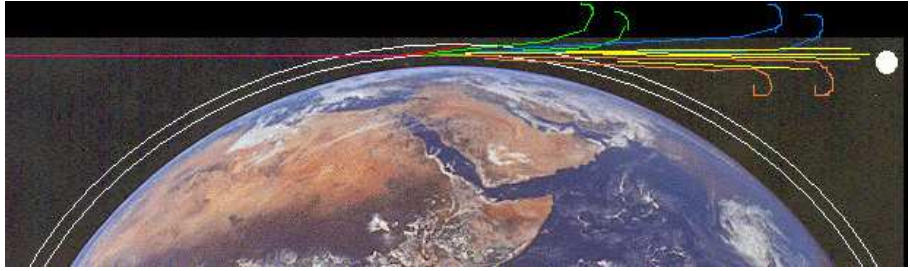}
\caption {A very schematic Horizontal High Altitude Shower (HIAS);
its fan-like imprint is due to geo-magnetic bending of charged
particles at high quota ($\sim 44 km$). It is similar to HORTAUs
event due to Earth Skimming neutrinos. The Shower may point to a
satellite as old gamma GRO-BATSE detectors or very recent
Beppo-Sax,Integral, HETE, Chandra or future Agile and Swift ones.
\cite{Fargion 2000-2002},\cite{Fargion2001a},\cite{Fargion2001b}.
The HIAS Showers are open and forked in  five (either three or at
least two main component): ($e^+,e^-,\mu^+,\mu^-, \gamma $, or
just positive-negative); these  multi-finger tails may be seen as
split tails  by EUSO. } \label{fig:fig25}
\end{figure}
%%%%%%%%%%%%%%%%%%%% end fig. 1a %%%%%%%%%%%%%%%%%%%%%%%%%

%%%%%%%%%%%%%%%%%%%%%%%%%%EUSO%%%%%%%%%%%
%%%%%%%%%%%%%%%%%%% fig.8 %%%%%%%%%%%%%%%%%%%%%%%%%

\begin{figure}\centering\includegraphics[width=13cm]{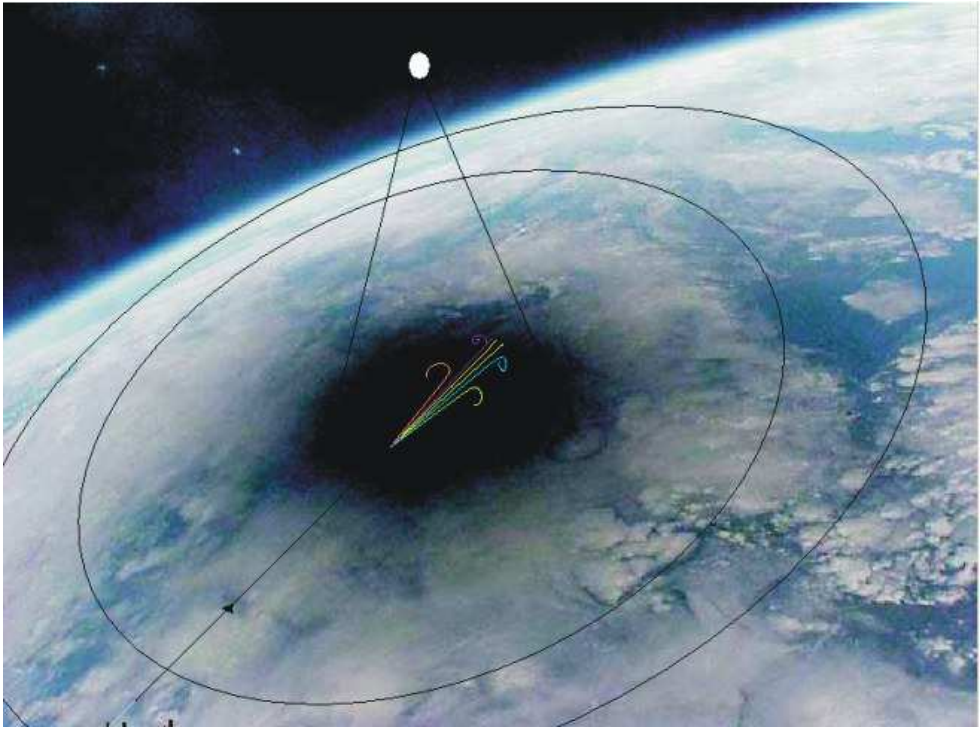}
\caption {A Horizontal High Altitude Shower or similar Horizontal
Tau air-shower (HORTAUs) and its open fan-like jets due to
geo-magnetic bending seen from a high quota by EUSO satellite. The
forked Shower is a multi-finger containing a inner $\gamma$ core
and external fork spirals due to $e^+  e^-$ pairs (first opening)
and  ${\mu}^+  {\mu}^-$ pairs \cite{Fargion 2000-2002},
\cite{Fargion2001a}, \cite{Fargion2001b}.} \label{fig:fig26}
\end{figure}
%%%%%%%%%%%%%%%%%%%% end fig. 8 %%%%%%%%%%%%%%%%%%%%%%%%%

%%%%%%%%%%%EUSO%%%%%%%%%%%%%%%%%%

 \vspace{2cm}
%%%%%%%%%%%%%%%%%%%% end fig. 3%%%%%%%%%%%%%%%%%%%%%%%%%
\subsection{Upward UHE $\nu$  Showering in Air observable by EUSO} Let us first
consider the last kind of Upward $\tau$ signals due to their
interaction in Air or in Earth Crust. The Earth opacity will
filter mainly  $10^{14}-10^{15}$eV upward events \cite{Gandhi et
al 1998},\cite{Halzen1998},\cite{Becattini Bottai
2001},\cite{Dutta et al.2001},\cite{Fargion 2000-2002}; therefore
only the direct upward $\nu$ shower in air or the UPTAUs around
PeV will be able to flash toward EUSO telescope facing downward
the Earth. The showers eject in a narrow beam ($2.5 \cdot 10^{-5}$
solid angle) jet an apparent total energy  corresponding to
$10^{19}-10^{20}$eV isotropically emitted energy. The shower will
be opened in a fan like shape and it will emerge from the Earth
atmosphere spread as a triplet or multi-dot signal aligned
orthogonal to a local terrestrial magnetic field lines. This
signature will be easily revealed. However the effective observed
air mass by EUSO is not $\ 10\%$ (because of duty cycle) of the
inspected air volume $\sim 150 km^3$, but because of the narrow
blazing shower cone it corresponds only to $3.72\cdot 10^{-3}$
$km^3$. The target volume  increases  for upward neutrino Tau
interacting vertically in Earth Crust in last matter layer
(either rock or water), making upward relativistic $\simeq PeV$
$\tau$ whose decay in the air born finally UPTAUs; in this case
the effective target mass is  (for water or rock at tau energy $3$
PeV) respectively $5.5\cdot 10^{-2}$$km^3$ or $1.5 \cdot10^{-1}$
$km^3$. These volume are not extreme. The consequent foreseen
thresholds are summarized for $3$ EUSO years of data recording in
Fig.\ref{fig:fig24}. The UPTAUs signal is nearly $15$ times larger
than the Air-Induced $\nu$ Shower. A more detailed analysis may
show an additional factor three (due to the neutrino flavors) in
favor of Air-Induced Showers, but the more transparent role of PeV
multi-generating upward $\nu_{\tau}$ while crossing the Earth,
makes the results summarized in figure. The much wider acceptance
of BATSE in respect to EUSO and the consequent better threshold
(in BATSE) is due to the wider angle view of the gamma detector,
the absence of any suppression factor as in EUSO duty cycle, as
well as the $10$ (for BATSE) over $3$ (for EUSO) years assumed of
record life-time. Any minimal neutrino  fluence
$\Phi_{\nu_{\tau}}$ of PeV neutrino
 $$ \Phi_{\nu_{\tau}}\geq 10^2 eV cm^{-2} s^{-1}$$ might be detectable by EUSO.
%%%%%%%%%%%%%%%%%%%% fig.4 %%%%%%%%%%%%%%%%%%%%%%%%%
\vspace{2cm}
\begin{figure}\centering\includegraphics[width=16cm]{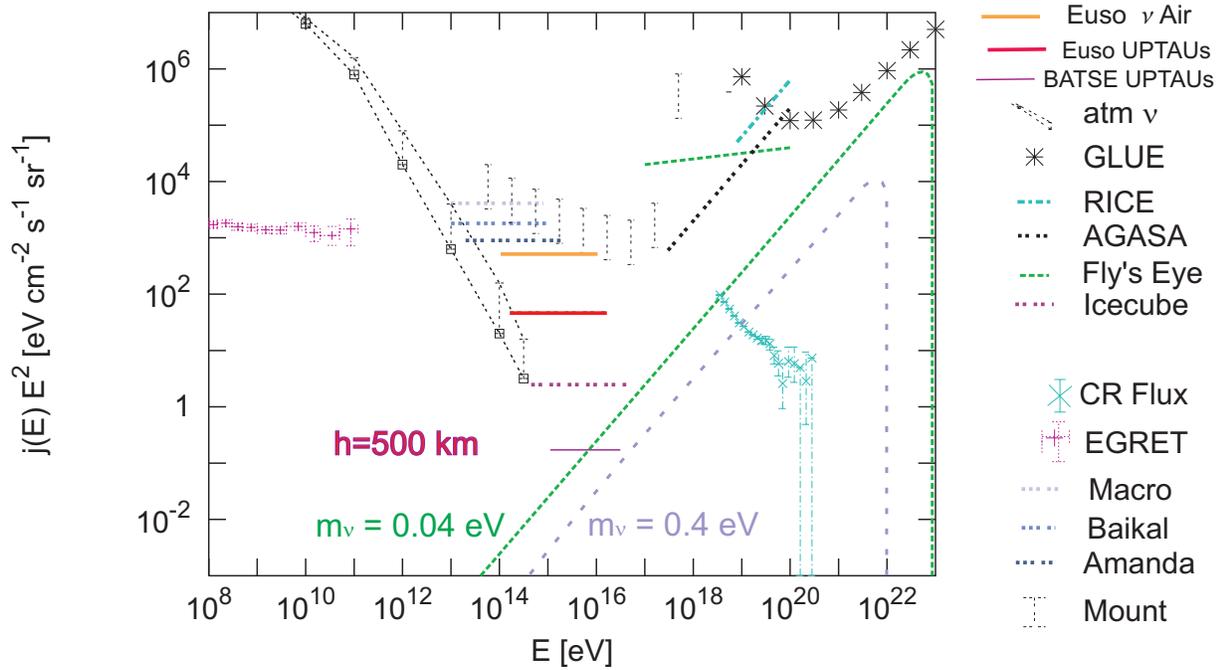}
\caption {Upward Neutrino air-shower and Upward Tau Air-shower,
UPTAUs, Gamma and Cosmic Rays Fluence Thresholds and bounds in
different energy windows for different past and future detectors.
The UPTAUs threshold for EUSO has been estimated for a three year
experiment lifetime. BATSE recording limit is also shown from
height $h = 500km$ and for ten year record. Competitive
experiments are also shown as well as the Z-Shower expected
spectra in light neutrino mass values $m_{\nu} = 0.4, 0.04$ eV.
\cite{Fargion 2000-2002},
\cite{Fargion2001a},\cite{Kalashev:2002kx}, \cite{Fargion et all.
2001b},\cite{Fargion 2002d}.} \label{fig:fig27}
\end{figure}

%%%%%%%%%%%%%%%%%%%% end fig. 4 %%%%%%%%%%%%%%%%%%%%%%%%%
\subsection{Downward and Horizontal UHECRs in EUSO}
%---------------------------------------------------
%The UHE  $\nu_{\tau}$ versus UHECR in EUSO
%---------------------------------------------------
Reconsider briefly the nature of common Ultra High Cosmic Rays
(UHECR) showers. Their  rate  will offer a useful test for any
additional UHE neutrino signals.
 Assume for a sake of simplicity a characteristic opening angle of
EUSO telescope of $30^o$ and a nominal satellite  height of $400$
km, leading to an approximate atmosphere area under inspection of
EUSO $\sim 1.5 \cdot 10^5 km^2$. Let us discuss the UHECR shower.
In this case the estimated rate is a $\sim 2\cdot10^{3}$
event/year above $3\cdot10^{19}$ eV. Among these "GZK" UHECR
(either proton, nuclei or $\gamma$) nearly $7.45\%\approx 150$
event/year will shower in Air Horizontally with no Cherenkov hit
on the ground. The critical angle of $\sim 6.7^o$ corresponding to
$7.45\%$ of all the events, is derived from first interacting
quota (here assumed for Horizontal Hadronic Shower near $44$ km
following \cite{Fargion
2000-2002},\cite{Fargion2001a},\cite{Fargion2001b}). Indeed the
corresponding horizontal edge critical angle $\theta_{h}$ $=$
$6.7^o$ below the horizons ($\pi{/2}$) is given, as in previous
eq. 17, but used  at an interacting height h near $44$ km: $
{\theta_{h} }={\arccos {\frac {R_{\oplus}}{( R_{\oplus} +
h_1)}}}\simeq 1^o \sqrt{\frac {h_{1}}{km}} $. These Horizontal
High Altitude Showers
(HIAS)\cite{Fargion2001a},\cite{Fargion2001b}, will be able to
produce a new peculiar showering, mostly very long (hundred
kilometers) and bent and forked (by few or several degrees) by
local geo-magnetic fields. The total UHECR above $3\cdot10^{19}$
eV will be $\sim 6000$ UHECR and $\sim 450$ Horizontal Showers
within 3 years; these latter horizontal signals are relevant
because they may mimic Horizontal  induced $\nu$ air-shower, but
mainly at high quota ($\geq 30-40 km$) and downward see
Fig.\ref{fig:fig28}. On the contrary UHE neutrino tau showering,
HORTAUs, to be discussed later, are also at high quota ($\geq 23
km$), but upward-horizontal. Their outcoming angle will be ($\geq
0.2^o-3^o$) upward. Therefore it may be necessary to have a good
angular ($\leq 0.2-0.1 ^o$) resolution to distinguish between the
two signals  as a key discriminator between HORTAUs and horizontal
UHECR. While Horizontal UHECR are an important piece of evidence
in the UHECR calibration and its GZK study , at the same time
they are a severe back-ground noise competitive with
Horizontal-Vertical GZK Neutrino Showers originated in Air, to be
discussed below. However Horizontal-downward UHECR are not
confused with upward Horizontal HORTAUs by UHE neutrinos to be
summarized in last section. Note that air-induced Horizontal UHE
neutrino as well as all down-ward air-induced UHE $\nu$ will
shower mainly at lower altitudes ($\leq 10 km$) ; however they
are respectively only a small ($\leq 2\% $, $\leq 8\%$) fraction
than HORTAUs showers to be discussed in the following. An
additional factor $3$ due to their three flavor over $\tau$
unique  may lead to respectively ($\leq 6\% $, $\leq 24\% $)
ratio of air induced $\nu$ shower in air over  all HORTAUs events:
a contribute ratio that may be in principle a useful test to
study the balanced neutrino flavor mixing.

%%%%%%%%%%%%%%%%%%%% fig. 5 %%%%%%%%%%%%%%%%%%%%%%%%%

\begin{figure}\centering\includegraphics[width=16cm]{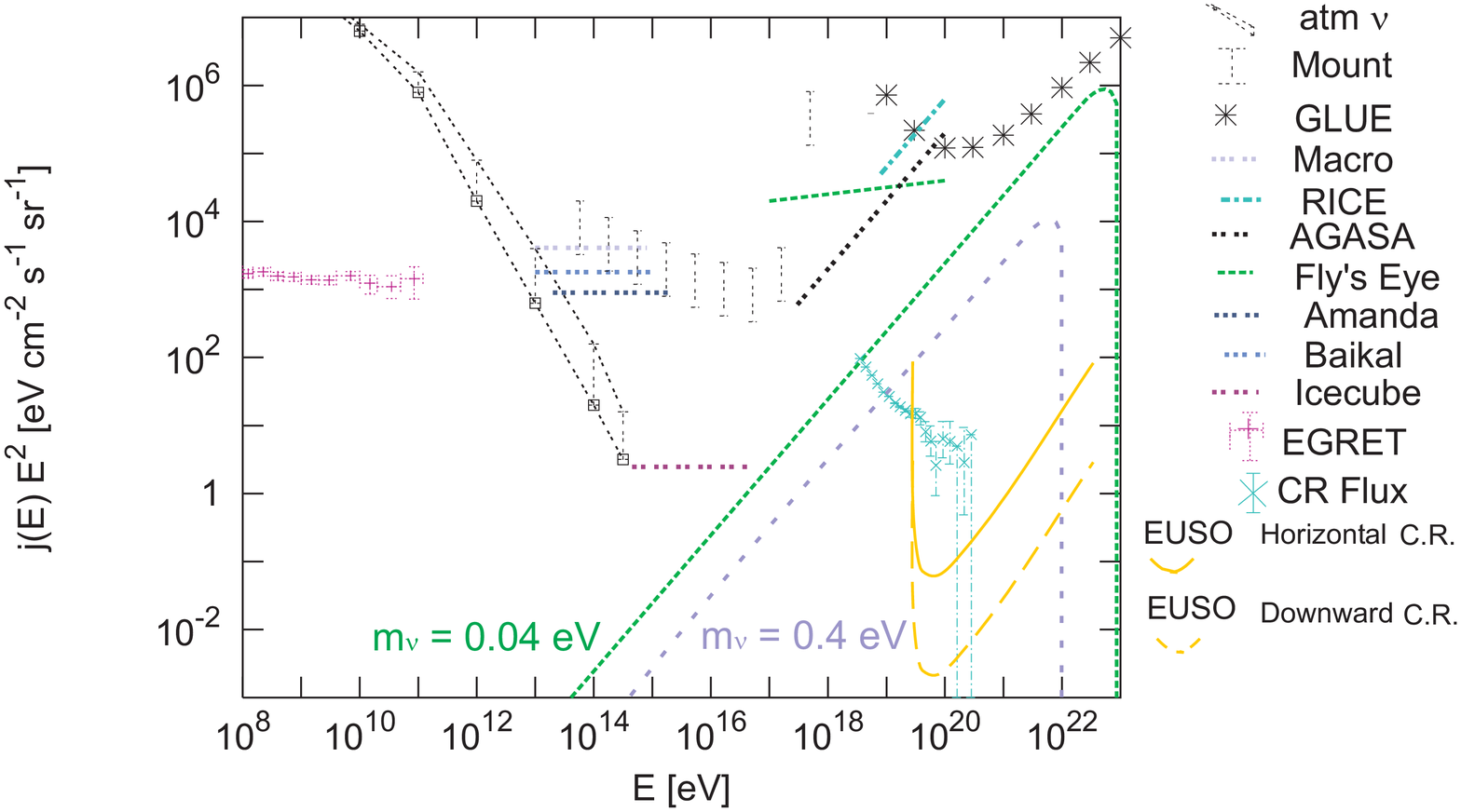}
\caption {Neutrino, Gamma and Cosmic Rays Fluence Thresholds and
bounds in different  energy windows. The Cosmic Rays Fluence
threshold for EUSO has been estimated  for a three year
experiments lifetime. The parabolic bound shape threshold may
differ upon the EUSO optics and acceptance. Competitive
experiment are also shown as well as the Z-Shower expected
spectra in light mass values. \cite{Fargion 2000-2002},
\cite{Fargion2001a},\cite{Kalashev:2002kx}, \cite{Fargion et all.
2001b},\cite{Fargion 2002d}.} \label{fig:fig28}
\end{figure}

%%%%%%%%%%%%%%%%%%%% end fig.5 %%%%%%%%%%%%%%%%%%%%%%%%%

\subsection{Air Induced UHE $\nu$   Shower }
 UHE $\nu$ may hit an air nuclei and shower
vertically or horizontally or more rarely nearly up-ward: its
trace maybe observable by EUSO preferentially in inclined or
horizontal case.  Indeed  vertical down-ward  ($\theta \leq 60^o$)
neutrino induced air showers  occur mainly
 at lowest quota and they will only partially shower their UHE $\nu$ energy
 because of the small slant depth ($\leq 10^3 g cm^{-2}$) in most vertical down-ward UHE $\nu$ shower.
 The observed  EUSO air mass ($1500 km^3$, corresponding to a $\sim 150$ $km^3$ for $10\%$ EUSO record time)
  is  the UHE neutrino calorimeter only ideally. Indeed  only the inclined ($\sim{\theta\geq 60^o }$) and horizontal
 air-showers ($\sim{\theta\geq 83^o }$) (induced by GZK UHE neutrino) may reach their maximum output  and
 their events maybe well observed; therefore only a
 small fraction ($\sim 30\%$ corresponding to $\sim 50$ $km^3$ mass-water volume for EUSO observation)
 of vertical downward UHE neutrino may be seen by EUSO. This
 signal may be somehow hidden (or masked) by the more common downward UHECR
 showers.  The key reading  signature will be the shower height
  origination: $(\geq 40 km)$ for most downward-horizontal UHECR, $(\leq 10 km)$ for most inclined-horizontal Air UHE
 $\nu$ Induced Shower. A corresponding smaller fraction ($\sim 7.45\%$)  of totally Horizontal
  UHE neutrino Air shower, orphan of their final Cherenkov flash, in competition
  with the horizontal UHECR, may be also clearly observed:
  their observable mass is only $V_{Air-\nu-Hor}$ $\sim 11.1$ $km^3$ for EUSO
  observation duty cycle. These masses reflect into a characteristic
  threshold behavior shown by bounds in Fig.\ref{fig:fig30}.
\subsection{UHE $\nu_{\tau}-\tau$  Double Bang Shower }
 A more  rare, but spectacular, double $\nu_{\tau}$-$\tau$ bang  in Air (comparable in principle to
 the PeV expected  "double bang" in water \cite{Learned Pakvasa 1995})
 may be exciting, but difficult to be observed; the EUSO effective calorimeter mass for such Horizontal event is only $10\%$ of the
 UHE $\nu$ Horizontal ones (($\sim 1.1$ $ km^3$)); therefore its event rate is nearly excluded
 needing too high neutrino fluxes see Fig.\ref{fig:fig29}; indeed it should be also noted that the EUSO energy threshold
($\geq 3\cdot 10^{19}$eV) implies such a very large  ${\tau}$
Lorentz boost distance; such large ${\tau}$ track  exceed (by more
than a factor three) the EUSO disk Area diameter ($\sim 450$km);
therefore the expected Double Bang Air-Horizontal-Induced ${\nu}$
Shower thresholds are suppressed by a corresponding factor as
shown in Fig.\ref{fig:fig29}. More abundant single event
Air-Induced ${\nu}$ Shower (Vertical or Horizontal) thresholds
are facing different Air volumes and  quite different visibility
as shown and summarized in Fig.\ref{fig:fig6}. It must be taken
into account an additional factor three (because of three light
neutrino states) in the Air-Induced ${\nu}$  Shower  arrival flux
respect to incoming $\nu_{\tau}$ (and $\bar{\nu_{\tau}}$ ),
making the Air target not totally a negligible calorimeter.
\subsection{UHE $\nu_{\tau}-\tau$  Air  Single Bang Shower }
There are also a sub-category  of $\nu_{\tau}$ - $\tau$ "double
 bang" due to a first horizontal UHE $\nu_{\tau}$ charged current interaction
 in air  nuclei (the first bang) that is lost from the EUSO view;
 their UHE  secondary $\tau$ fly and decay leading to a Second Air-Induced Horizontal Shower, within the EUSO
 disk area. These  horizontal "Double-Single $\tau$ Air Bang"  Showers
   are produced within a very wide Terrestrial Crown Air Area whose radius is exceeding $\sim 600- 800$ km
 surrounding  the EUSO Area of view. However it is easy to show
 that they will just double the  Air-Induced ${\nu}$  Horizontal Shower
 rate due to one unique flavor. Therefore the total Air-Induced Horizontal Shower (for
 all $3$ flavors and the additional $\tau$ decay in flight) are summirized and considered
 in Fig.\ref{fig:fig30} including also the present Single Bang Shower.
  The relevant UHE neutrino signals, HORTAUs, as discussed below, are
  originated within the (much denser) Earth
 Crust.
%%%%%%%%%%%%%%%%%%%% fig. 6 %%%%%%%%%%%%%%%%%%%%%%%%%

\begin{figure}\centering\includegraphics[width=16cm]{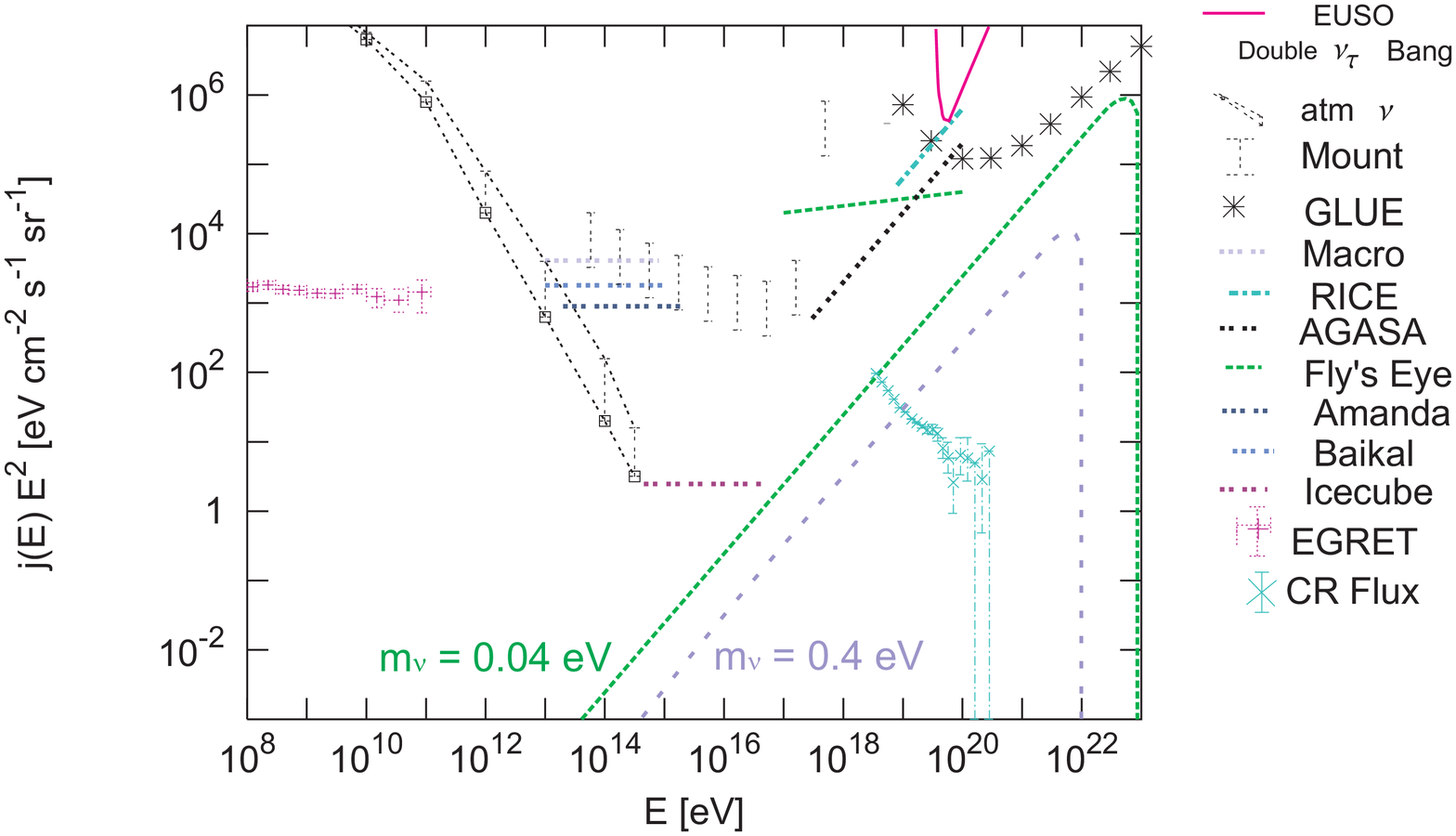}
\caption {EUSO threshold for Double bang $\tau$ Neutrino over
other $\gamma$, $\nu$ and Cosmic Rays (C.R.) Fluence  and bounds
in different energy windows. The  Fluence threshold for EUSO has
been estimated for a three year experiments lifetime. Competitive
experiment are also shown as well as the Z-Shower expected
spectra in most probable light neutrino mass values ($m_{\nu} =
0.04, 0.4$ eV). \cite{Fargion
2000-2002},\cite{Fargion2001a},\cite{Kalashev:2002kx},
\cite{Fargion et all. 2001b},\cite{Fargion 2002d}.}
\label{fig:fig29}
\end{figure}

%%%%%%%%%%%%%%%%%%%% end fig. 6%%%%%%%%%%%%%%%%%%%%%%%%%
%%%%%%%%%%%%%%%%%%%% fig.7 %%%%%%%%%%%%%%%%%%%%%%%%%

\begin{figure}\centering\includegraphics[width=16cm]{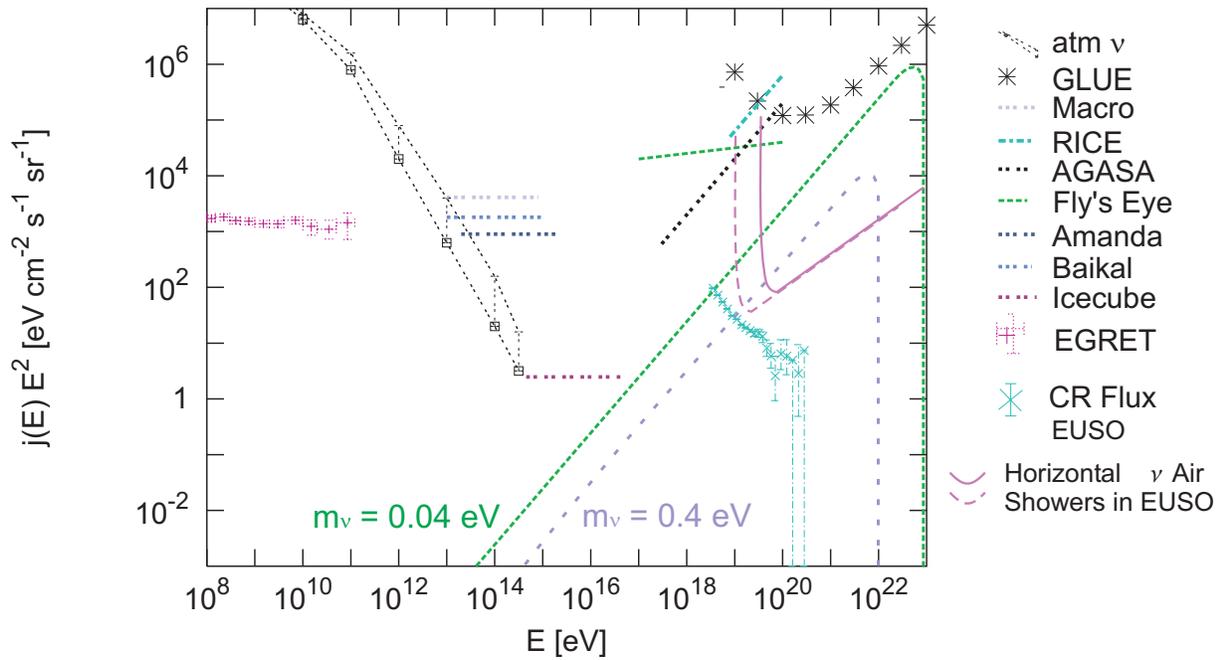}
\caption {EUSO thresholds for  Vertical Downward Neutrino Air
induced shower, at arrival angle  $> 60^o$ , $< 90^o$, over other
$\gamma$, $\nu$ and Cosmic Rays (C.R.) Fluence  and bounds. The
Fluence threshold for EUSO has been estimated for a three year
experiment lifetime. Competitive experiments are also shown as
well as the Z-Shower expected spectra in light neutrino mass
values ($m_{\nu} = 0.04, 0.4$ eV). \cite{Fargion 2000-2002},
\cite{Fargion2001a},\cite{Kalashev:2002kx}, \cite{Fargion et all.
2001b},\cite{Fargion 2002d}.} \label{fig:fig30}
\end{figure}

%%%%%%%%%%%%%%%%%%% end fig. 7%%%%%%%%%%%%%%%%%%%%%%%%%

\subsection{HORTAUs  in EUSO}
 As already mentioned the UHE $\nu$ astronomy may be greatly strengthened by
 $\nu_{\tau}$ appearance via flavor mixing and oscillations. The
  consequent scattering of $\nu_{\tau}$ on Mountains or into the
 Earth Crust may lead to Horizontal Tau air-showers: HORTAUs (or
 so called Earth Skimming Showers \cite{Fargion2001a},\cite{Fargion2001b},\cite{Fargion 2000-2002}\cite{Feng et al 2002}
 \cite{Fargion2002e}).
 Indeed UHE $\nu_{\tau}$ may skip below
 the Earth and escape as $\tau$ and finally decay in flight, within the air atmosphere,
 as well as inside  the Area of  view of EUSO, as shown in
 Fig.\ref{fig:fig26}. Any UHE-GZK Tau Air Shower induced event is approximately born within
a wide  ring  (whose radiuses extend between $R \geq 300$ km and
$R \leq 800$ km from the EUSO Area center). Because of the wide
area and deep $\tau$ penetration \cite{Fargion
2000-2002},\cite{Fargion 2002b},\cite{Fargion 2002d} the amount
of interacting matter where UHE $\nu$ may lead to $\tau$ is huge
($\geq 2 \cdot 10^5$ $km^3$) ;however only a tiny fraction of
these HORTAUs will beam and Shower within the EUSO Area within
EUSO. After carefully estimate (using also results in
\cite{Fargion 2000-2002},\cite{Fargion 2002b}, \cite{Fargion
2002c},\cite{Fargion 2002d}) we probed  a lower bound (in sea
matter) for these effective Volumes respectively at ($1.1 \cdot
10^{19}$eV) and  at ($3 \cdot 10^{19}$ eV) energy shown in
figures below. Therefore at GZK energies ($1.1 \cdot 10^{19}$eV)
the horizontal $\tau$s by HORTAUs are more than $6$ times
abundant than any corresponding Vertical observable Air Induced
neutrino Showers as shown in Fig.\ref{fig:fig31}.

%%%%%%%%%%%%%%%%%% fig.9 %%%%%%%%%%%%%%%%%%%%%%%%%
\begin{figure}\centering\includegraphics[width=16cm]{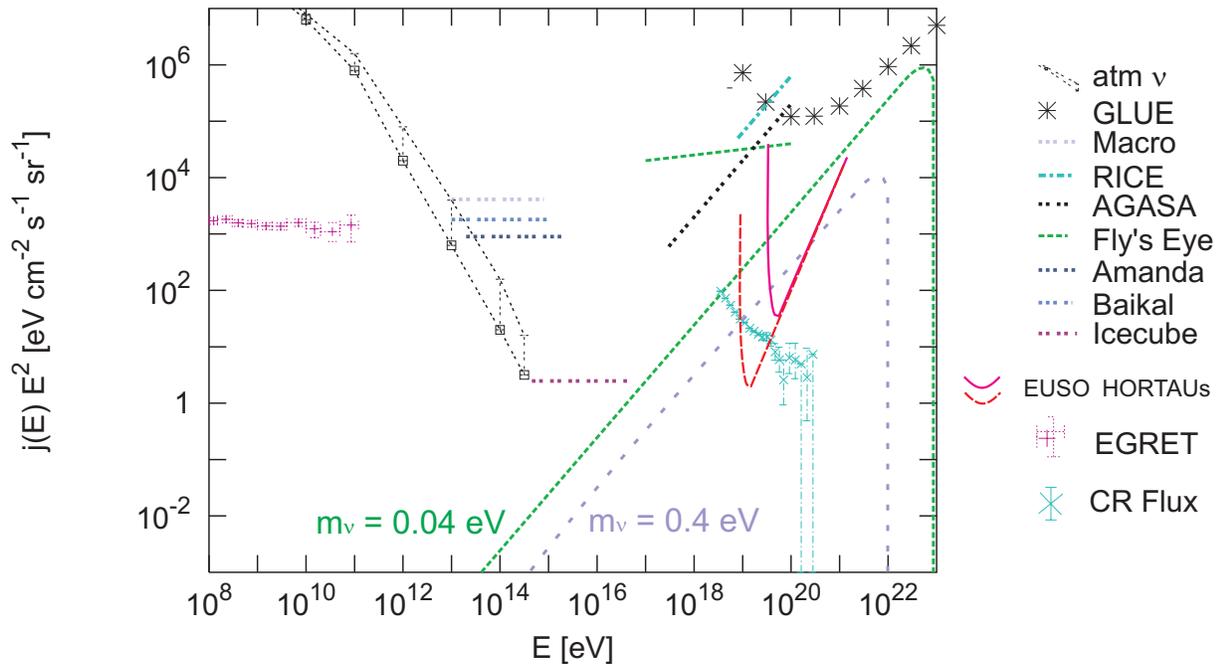}
\caption {EUSO thresholds for Horizontal Tau air-shower HORTAUs
 over other $\gamma$, $\nu$ and Cosmic
Rays (C.R.) fluence and bounds. Dash curves for HORTAUs  are
drawn assuming an EUSO threshold at $10^{19}$eV. Because the
bounded $\tau$ flight distance (due to the contained terrestrial
atmosphere height) the main signal is  better observable at $1.1
\cdot 10^{19}$eV than higher energies. The Fluence threshold for
EUSO has been estimated for a three year experiment lifetime.
Z-Shower or Z-Burst expected spectra in light neutrino mass
values ($m_{\nu} = 0.04, 0.4$ eV) are shown. \cite{Fargion
2000-2002}, \cite{Fargion2001a},\cite{Kalashev:2002kx},
\cite{Fargion et all. 2001b},\cite{Fargion 2002d}.}
\label{fig:fig31}
\end{figure}

%%%%%%%%%%%%%%%%%%%% end fig. 9 %%%%%%%%%%%%%%%%%%%%%%%%%

%%%%%%%%%%%%%%%%%%% fig.10 %%%%%%%%%%%%%%%%%%%%%%%%%
\begin{figure}\centering\includegraphics[width=16cm]{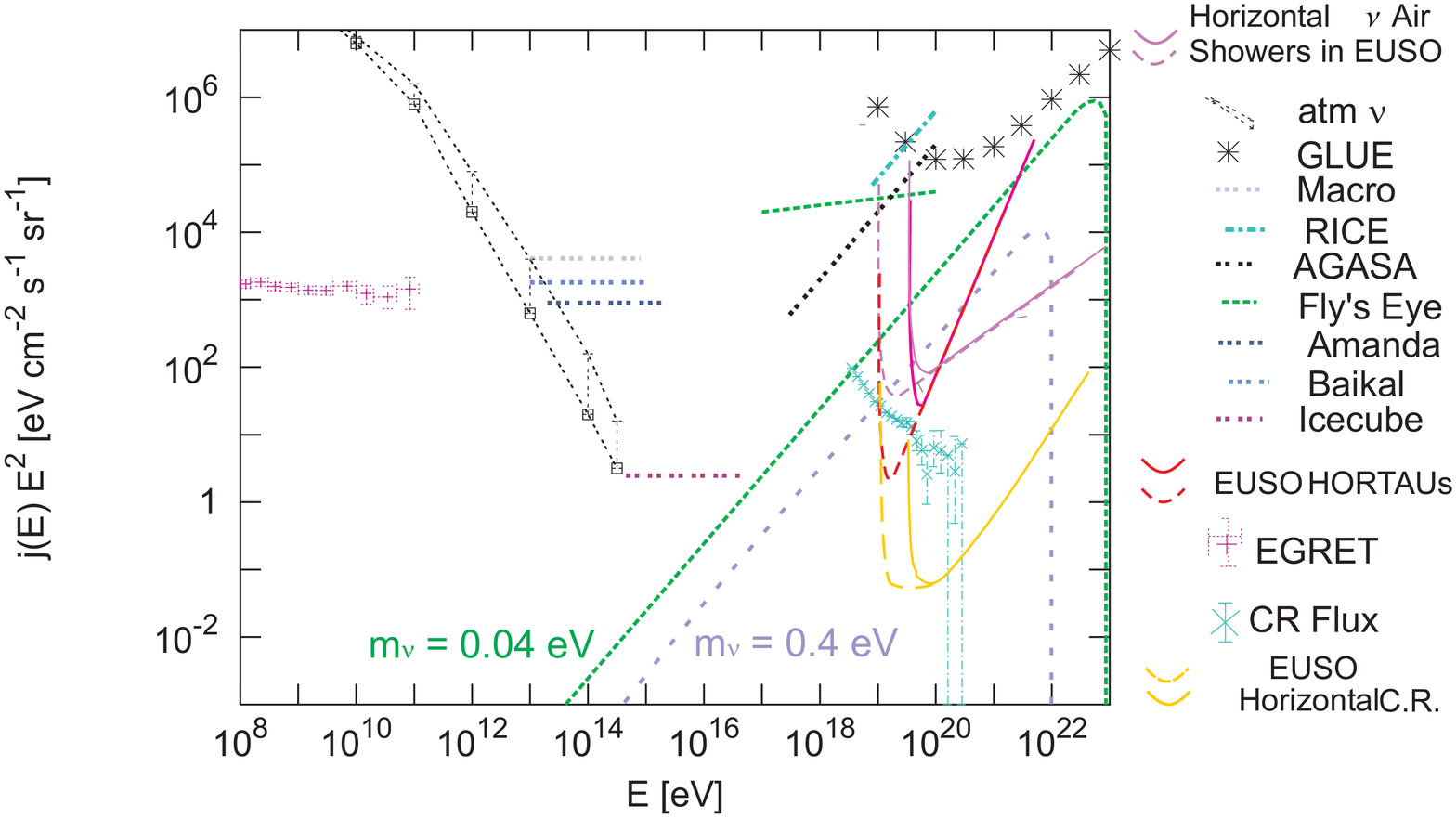}
\caption {EUSO thresholds for Horizontal Tau air-shower shower,
HORTAUs (or Earth Skimming Showers) over all other $\gamma$, $\nu$
and Cosmic Rays (C.R.) fluence and bounds. The fluence threshold
for EUSO has been estimated for a three year experiments lifetime.
Competitive experiment are also shown as well as the Z-Shower
expected spectra in two different light neutrino mass values
($m_{\nu} = 0.04, 0.4$ eV). As above dash curves for both HORTAUs
and Horizontal Cosmic Rays are drawn assuming an EUSO threshold at
$10^{19}$eV. \cite{Fargion 2000-2002},
\cite{Fargion2001a},\cite{Kalashev:2002kx}, \cite{Fargion et all.
2001b},\cite{Fargion 2002d}.} \label{fig:fig32}
\end{figure}
%%%%%%%%%%%%%%%%%%%% end fig. 10 %%%%%%%%%%%%%%%%%%%%%%%%%
%%%%%%%%%%%%%%%%%%%%%%%%%%%%%%%%%%%%%%%%%%%%%%%%%%%%%%%%%%%%%%%%%%%%%%%
\begin{figure}
\centering
\includegraphics[width=16.2cm]{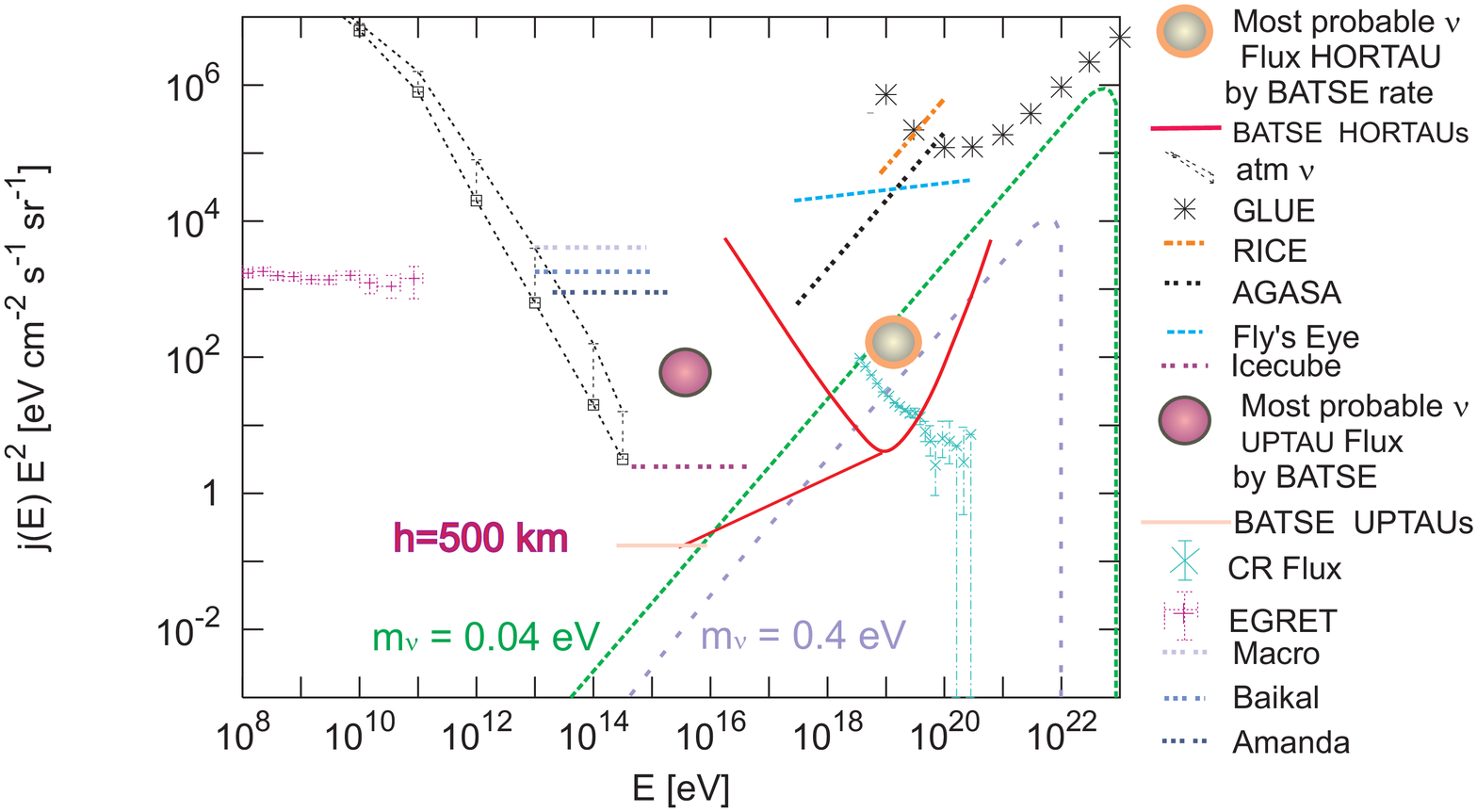}
\caption {Neutrino Flux derived by  BATSE Terrestrial Gamma
Flashes assuming them as $\gamma$ secondaries of upward Tau
Air-showers. These fluxes are estimated using the
Tab.\ref{tab:table4} normalized  during  their most active
trigger and TGF activities ($3+4$) as labeled  in last column. The
UPTAUs and HORTAUs rate are normalized assuming that the events at
geo-center angle above $50^o$ might be of HORTAU nature. }
\label{fig:fig33}
\end{figure}

%%%%%%%%%%%%%%%%%%%%%%%%%%%%%%%%%%%%%%%%%%%%%%%%%%%%%%%%%%%%%%%%%%%%%%%%%%%%%%%%

However the air-shower induced neutrino may reflect all three
light neutrino flavors, while HORTAUs are made only by
$\nu_{\tau}$,$\bar{\nu_{\tau}}$ flavor. Nevertheless the dominant
role of HORTAUs overcomes (by a factor $\geq 15$) all other
Horizontal EUSO neutrino events: their  expected event rates are,
at $\Phi_{\nu}\geq 3 \cdot 10^{3}$ eV $cm^{-2} s^{-1}$ neutrino
fluence (as in Z-Shower model in
Fig.\ref{fig:fig27}-\ref{fig:fig31}), a few hundred events a year
and they may already be comparable or even may exceed the
expected Horizontal CR rate. Dash curves for both HORTAUs and
Horizontal Cosmic Rays are drawn assuming the EUSO threshold at
$10^{19}$eV. Because of the relativistic $\tau$ flight distance
growth and because of the finiteness of the terrestrial
atmosphere, the main signal is better observable at $1.1 \cdot
10^{19}$eV than at higher energies (as emphasized in
Fig.\ref{fig:fig30},\ref{fig:fig31},\ref{fig:fig32} at different
threshold curves).
%%% ----------------------------------------------------------------------

\subsection{Visibility and Signatures of UHE neutrino in EUSO }

Highest Energy Neutrino signals may be well observable by next
generation satellite as EUSO: the main sources of such neutrino
traces are UPTAUs (Upward Tau blazing the telescope born in Earth
Crust) and mainly HORTAUs (Horizontal Tau air-showers originated
by Earth-Skimming UHE $\nu_{\tau}$). These showers will be opened
in a characteristic twin fan-jet ovals looking like the $8$-shape
horizontal cosmic rays showers observed on Earth. The UPTAUs will
arise mainly at PeV energies (because the Earth neutrino opacity
at higher energies and because the shorter $\tau$ boosted length
, at lower energies)\cite{Fargion 2000-2002}; UPTAUs will be
detected as a thin stretched multi-pixel event by EUSO, whose
orientation is polarized orthogonal to the local geo-magnetic
field. The EUSO sensibility (effective volume ($V_{eff}$$\sim 0.1
km^3$) for 3 years of detection) will be an order of magnitude
below present AMANDA-Baikal bounds. Horizontal Tau air-showers at
GZK energies will be better searched and revealed. They are
originated along huge Volumes around the EUSO Area. Their
horizontal skimming secondary $\tau$ decays occur far away $\geq
500$ km, at high altitude ($\geq 20-40$ km) and it will give
clear signals distinguished from downward horizontal UHECR.
HORTAUs are grown by UHE neutrino interactions inside huge
volumes ($V_{eff}$$\simeq 2360$ $km^3$); see for more details
\cite{Fargion 2002b},\cite{Fargion 2002c},\cite{Fargion 2002d}.
As summirized in last Fig.\ref{fig:fig32} the expected UHE fluence
$$\Phi_{\nu}\simeq 10^{3} eV cm^{-2} s^{-1}$$ needed in most
Z-Shower models (as well as in most topological relic scenarios)
to solve GZK puzzle, will lead to a hundred of horizontal events
comparable to UHECR ones. Even in the most conservative scenario
where a minimal GZK-$\nu$ fluence must take place at least at
$$\Phi_{\nu} \simeq 10 eV cm^{-2} s^{-1}$$ (just comparable to
well observed Cosmic Ray fluence), a few or a ten of such UHE
astrophysical neutrinos must be observed (respectively at $3
\cdot10^{19}$ eV and $10^{19}$ eV   energy windows). To improve
their visibility EUSO, on our opinion, may:\\
 a) Improve the fast pattern recognition of Horizontal Shower
Tracks with their forking signature.\\
 b) Enlarge the Telescope
Radius to embrace also lower $10^{19}$ eV showers.\\
 c) Consider a detection threshold at  angular $\Delta\theta$ and at height $\Delta h$ level within an
accuracy $\Delta\theta \leq 0.2^o$,$\Delta h \leq 2$ km.\\ Even
the above results have been derived carefully in  minimal
realistic scenarios they may be used within $10\%$ approximated
value due to present uncertainties in  EUSO detection
capabilities.
%%%%%%%%%%%%%%%%%%%%%%%%%%%%%%%%%%%%%%%%%%%%%%%%%%%%%%%%%%%%%%%%%%%%%%%%%%%%%%%%%%%%%
\section{Conclusions}
%%%%%%%%%%%%%%%%%%%%%%%%%%%%%%%%%%%%%%%%%%%%%%%%%%%
In the present article we have considered the connection between
Ultra High Energy Cosmic Ray Astronomy, relic neutrino masses and
UHE neutrino scattering or relic ones  (the Z-Shower solution) to
solve the GZK puzzles: indeed  because of the UHECR rigidity we do
expect the rise of a new UHE particle astronomy. But because UHECR
interactions on relic cosmic photons this UHECR astronomy must be
very local, within our nearby Universe. The absence of any defined
known structure (as our galactic plane and super-galactic group)
is puzzling. The absence of the GZK cut-off is also puzzling.
Because of photopion production along the UHECR path we must
expect also a secondary neutrino astronomy  at GZK energies. In
Z-Showering model, on the contrary, larger UHE neutrino fluxes
are the primary of UHECR. These UHE neutrino may hit relic ones
whose lightest masses are reflecting into highest Z-boosted
resonance peaks and are modulating highest un-explored UHECR
tails. UHE neutrino at ZeV energies may be needed. We have
investigated the interplay between neutrino masses and UHECR
spectra. We also studied the possibility to test part of this UHE
neutrino astronomy by Tau Air-Showers. In particular we have
shown the virtue of such tau astronomy at PeVs and at EeVs energy
window as Upward and Horizontal Tau air-showers (UPTAUs-HORTAUs).
Tau showering are able to leave a strong imprint in detectors
located either beyond a mountain chain or, better, above mountain
peak or an plane, balloons and satellites. In particular we
estimated the detectable mass via Tau air-shower observing by  any
quota. We had estimated the  terrestrial crust mass and the
consequent event rate for given neutrino flux. We also discussed
the possibility to observe the same horizontal tau air-showers
within the next generation telescope (EUSO like). We also
discussed the possibility that the same upward and horizontal
air-showers will cross upward the terrestrial atmosphere leading
to UPTAUs or HORTAUs. These events may be already recorded on
best gamma satellites as very sharp (millisecond) gamma bursts.
Indeed the last Gamma Ray Observatory did recorded , in BATSE
experiment, rare upward gamma flashes, the Terrestrial Gamma
Flashes. Their fluence and duration are well in agreement with
expected UPTAUs and HORTAUs power and shower dispersion at high
altitude. We reconsidered their maps in celestial coordinates in
correlation with known EGRET and AGASA anisotropies and sources.
We also correlated the observed TGF rate with the expected
neutrino flux threshold finding a first estimate of the UPTAUs
and HORTAUs fluxes.  Both their fluence  $\Phi_{\nu} \simeq 75 eV
cm^{-2} s^{-1} sr^{-1}$ are nearly one order of magnitude below
present bounds and might be confirmed in a very near future by
EUSO and other experiments.

%%%%%%%%%%%%%%%%%%%%%%%%%%%%%%%%%%%%%%%%%%%%%%%%%%%%%

\section{Appendix A: Influence of Atmosphere depth in HORTAUs}
 As soon as the altitude $h_1$ and the corresponding energy
$E_{\tau_{h_1}}$ increases the corresponding  air density
decreases. At a too high quota there is no more $X$ slant depth
for any air-showering to develop. Indeed its value is :
\begin{displaymath}
X=\int_{\frac{d_u}{2}+c\tau\gamma_t}^{d_1+\frac{d_u}{2}}{{n_0}e^{-\frac{R_{\oplus}}{h_0}{\left[{\sqrt{\left(1-\frac{h_2}{R_{\oplus}}\right)^2+\left({\frac{x}{R_{\oplus}}}\right)^2}}-1\right]}}{dx}}
\end{displaymath}
\begin{equation}
\simeq\int_{\frac{d_u}{2}+c\tau\gamma_t}^{d_1+\frac{d_u}{2}}{n_0e^{-\frac{x^2}{2R_{\oplus}h_0}}dx}\leq{n_0h_0}
\end{equation}
\end{enumerate}
In order to find this critical height $h_{1}$ where the maximal
energy HORTAU terminates  we remind our recent approximation. The
transcendental equation that defines the Tau distance $c\tau$
 has been more simplified in:
\begin{equation}\label{13}
  \int_{0}^{+ \infty} n_0 e^{-\frac{\sqrt{(c\tau+x)^2+R_\oplus^2} - R_\oplus}{h_0}}
   dx \cong n_0 h_0 A
\end{equation}
\begin{equation}\label{14}
  \int_{0}^{+ \infty} n_0 e^{-\frac{(c\tau+x)^2} {2h_0R_\oplus}}
   dx \cong n_0 h_0 A
\end{equation}
\begin{equation}\label{15}
%  c\tau = \sqrt{2R_\oplus h_0}
%  \sqrt{ ln \ffrac{R_\oplus}{c\tau} - ln A \left(1+\alpha \cdot ln
%  \ffrac{E}{E_{max}}\right)}
 c\tau = \sqrt{2R_\oplus h_0}
 \sqrt{ln \ffrac{R_\oplus}{c\tau} - ln A }
\end{equation}
Here $A=A_{Had.}$ or $A=A_{\gamma}$ are slow logarithmic
functions of values near unity; applying known empirical laws to
estimate this logarithmic growth (as a function of  the X slant
depth) we  derived respectively for hadronic and gamma UHECR
showers \cite{Fargion 2000-2002}, \cite{Fargion2001a}:
\begin{equation}\label{5}
 A_{Had.}=0.792 \left[1+0.02523 \ln\ffrac{E}{10^{19}eV}\right]
\end{equation}
\begin{equation}\label{5}
 A_{\gamma}=\left[1+0.04343\ln\ffrac{E}{10^{19}eV}\right]
\end{equation}
The solution of the above transcendental equation leads to a
characteristic maximal UHE $c\tau_{\tau}$ = $546 \;km$ flight
distance, corresponding to $E \leq 1.1\cdot 10^{19}eV$ energy
whose decay occurs at height $H_o= 23$ km; nearly 600 Km far from
the horizon it was originated from there on the HORTAUs begins to
shower. At higher quotas the absence of sufficient air density
lead to a suppressed development or to a poor particle shower,
hard to be detected. At much lower quota the same air opacity
filter most of the electromagnetic shower allowing only to muon
bundles and Cherenkov lights to survive at low a somehow $(\leq
10^{-3})$ level.
%%%%%%%%%%%%%%%%%%%%%%%%%%%%%%%%%%%%%%%%%%%%%%%%%%%%%%%%%%%%%%%%%%%%%%%%%%%%%%%%
\section{Appendix B: The UPTAUs area}
The Upward Tau air-showers, mostly at PeV energies, might travel a
minimal air depth before reaching the observer in order to
amplify its signal. The UPTAUS Disk Area $A_U$ underneath an
observer at height $h_1$ within a opening angle $\tilde{\theta}_2$
from the Earth Center is:
\begin{equation}
 A_{U}= 2\pi{R_{\oplus}}^2(1 - \cos{\tilde{\theta}_{2}})
\end{equation}
Where the $\sin{\tilde{\theta}_{2}= (x_2/{R_{\oplus}})}$ and
$x_2$ behaves like $x_1$ defined above for HORTAUs. In general the
UPTAUs area are constrained in a narrow Ring (because the mountain
presence itself or because the too near observer distances from
Earth are encountering a too short air slant depth for showering
or a too far and opaque atmosphere for the horizontal
UPTAUs)\cite{Fargion 2002b},\cite{Fargion 2002c}:
\begin{equation}
 A_{U}= 2\pi{R_{\oplus}}^2( \cos{\tilde{\theta}_{3}-\cos{\tilde{\theta}_{2}}})
 \end{equation}
An useful Euclidean approximation is:
\begin{equation}
 A_{U}= \pi {h_1}^2 ({\cot{\theta}_{2}}^2-{\cot{\theta}_{3}}^2)
 \end{equation}
 Where ${\theta}_{2}$, ${\theta}_{3}$ are the outgoing $\tau$
 angles on the Earth surface \cite{Fargion 2000-2002}.

 For  UPTAUs (around $3\cdot10^{15} eV$ energies) these
volumes have been estimated in \cite{Fargion 2000-2002}, assuming
an arrival values  angle$\simeq 45^o- 60^o$ below the horizons.
For two characteristic densities one finds respectively:
$${\Delta M_{eff.}(h_{1}=500 km)(\rho_{Water})}= 5.9~ km^3;$$
$${\Delta M_{eff.}.  (h_{1}=500 km)(\rho_{Rock})}=15.6~ km^3 $$
Their detection efficiency is displayed in last figure
(Fig.\ref{fig:fig27}), and it exceed by more than an order of
magnitude, the future ICE-CUBE threshold.

%%%%%%%%%%%%%%%%%%%%%%%%%%%%%%%%%%%%%%%%%%%%%%%%%%%%%%%%

%%% ----------------------------------------------------------------------

% ------------------------------------------------------------------------

\subsection*{Acknowledgment}
The authors wish to thank Prof. L. Scarsi, M.Teshima for inspiring
discussions and  C.Leto for technical support. The work of one of
us (M.Khlopov) was partially performed in the framework of
Russian State Contract 40.022.1.1.1106 and supported in part by
RFBR grant 02-02-17490 and grant UR.02.01.026.

\end{document}